\documentclass[prb,aps,twocolumn,preprintnumbers,amsmath,amssymb,superscriptaddress,showpacs]{revtex4-2}

\usepackage[colorlinks,bookmarks=true,citecolor=blue,linkcolor=red,urlcolor=blue,citecolor=blue]{hyperref}

\usepackage{graphicx}
\usepackage{subfigure}
\usepackage{amssymb,amsmath}
\allowdisplaybreaks
\usepackage{color}
\usepackage{ulem}
\usepackage{lineno}
\usepackage{tikz}

\definecolor{mygreen}{RGB}{51,180,76}
\definecolor{myblue}{RGB}{100,149,237}
\definecolor{myred}{RGB}{211,56,28}
\definecolor{myorange}{RGB}{255,170,0}

\newcommand{\Tensor}[4]{
\filldraw[draw=mygreen!100,thick,fill=white!20] (#1,#2) circle (#3) ;
\draw [mygreen, thick] (#1-#3*0.7071,#2+#3*0.7071) to (#1-#3*0.7071-#4,#2+#3*0.7071+#4) ; 
}

\newcommand{\Tensora}[4]{
\filldraw[draw=mygreen!100,thick,fill=white!20] (#1,#2) circle (#3) ;
\draw [mygreen, thick] (#1-#3*0.7071,#2+#3*0.7071) to (#1-#3*0.7071-#4,#2+#3*0.7071+#4) ; 
}

\newcommand{\Tensorb}[4]{
\filldraw[draw=myblue!100,thick,fill=white!20] (#1,#2) circle (#3) ;
\draw [myblue, thick] (#1-#3*0.7071,#2+#3*0.7071) to (#1-#3*0.7071-#4,#2+#3*0.7071+#4) ; 
}

\newcommand{\Tensorc}[4]{
\filldraw[draw=myred!100,thick,fill=white!20] (#1,#2) circle (#3) ;
\draw [myred, thick] (#1-#3*0.7071,#2+#3*0.7071) to (#1-#3*0.7071-#4,#2+#3*0.7071+#4) ; 
}
\newcommand{\TensorA}[4]{
\filldraw[draw=mygreen!100,thick,fill=white!20] (#1,#2) circle (#3) ;
\draw [mygreen, thick] (#1-#3*0.7071,#2+#3*0.7071) to (#1-#3*0.7071-#4,#2+#3*0.7071+#4) ; 
\draw (#1,#2-0.2) node [below] {\textcolor{mygreen}{$A$}} ;
}

\newcommand{\CenterB}[4]{
\filldraw[draw=myred!100,thick,fill=myred!100] (#1,#2) circle (#3) ;
\draw (#1+#3*2.8,#2) node [above] {$\boldsymbol{r}$} ;
\draw (#1+#3*2.6,#2) node [below] {$B$} ;
\draw [myred, thick] (#1-#3*0.7071,#2+#3*0.7071) to (#1-#3*0.7071-#4,#2+#3*0.7071+#4) ; 
}

\newcommand{\projector}[3]{
\filldraw[draw=myblue!100,thick,fill=myblue!100] (#1-#3/2,#2-#3*0.2887) -- (#1+#3/2,#2-#3*0.2887) -- (#1,#2+#3*0.5774) ;
\draw[myblue!100,thick] (#1-#3/2,#2-#3*0.2887) -- (#1-#3/2-0.1,#2-#3*0.2887+0.4);
\draw[myblue!100,thick] (#1+#3/2,#2-#3*0.2887) -- (#1+#3/2-0.1,#2-#3*0.2887+0.4);
\draw[myblue!100,thick] (#1,#2+#3*0.5774) -- (#1-0.1,#2+#3*0.5774+0.4) ;
}

\newcommand{\projectorA}[3]{
\filldraw[draw=myblue!100,thick,fill=myblue!100] (#1-#3/2,#2-#3*0.2887) -- (#1+#3/2,#2-#3*0.2887) -- (#1,#2+#3*0.5774) ;
\draw[myblue!100,thick] (#1-#3/2,#2-#3*0.2887) -- (#1-#3/2-0.1,#2-#3*0.2887+0.4);
\draw[myblue!100,thick] (#1+#3/2,#2-#3*0.2887) -- (#1+#3/2-0.1,#2-#3*0.2887+0.4);
\draw[myblue!100,thick] (#1,#2+#3*0.5774) -- (#1-0.1,#2+#3*0.5774+0.4) ;
\draw (#1,#2-0.2) node [below] {\textcolor{myblue}{$A$}} ;
}

\newcommand{\simplex}[3]{
\draw[myred!100,thick] (#1,#2) -- (#1-#3/2,#2+#3*0.2887) ;
\draw[myred!100,thick] (#1,#2) -- (#1+#3/2,#2+#3*0.2887) ;
\draw[myred!100,thick] (#1,#2) -- (#1,#2-#3*0.5774) ;
}

\newcommand{\simplexS}[3]{
\draw[myred!100,thick] (#1,#2) -- (#1-#3/2,#2+#3*0.2887) ;
\draw[myred!100,thick] (#1,#2) -- (#1+#3/2,#2+#3*0.2887) ;
\draw[myred!100,thick] (#1,#2) -- (#1,#2-#3*0.5774) ;
\draw (#1,#2+0.2) node [above] {\textcolor{myred}{$S$}} ;
}

\newcommand{\CenterA}[4]{
\filldraw[draw=mygreen!100,thick,fill=white!20] (#1,#2) circle (#3) ;
\draw (#1+#3*2.,#2) node [below] {$A$} ;
\draw [mygreen, thick] (#1-#3*0.7071,#2+#3*0.7071) to (#1-#3*0.7071-#4,#2+#3*0.7071+#4) ; 
}

\newcommand{\CenterAa}[4]{
\filldraw[draw=mygreen!100,thick,fill=white!20] (#1,#2) circle (#3) ;
\draw (#1+#3*2.5,#2) node [below] {\small $A_3$} ;
\draw [mygreen, thick] (#1-#3*0.7071,#2+#3*0.7071) to (#1-#3*0.7071-#4,#2+#3*0.7071+#4) ; 
}

\newcommand{\CenterAb}[4]{
\filldraw[draw=myblue!100,thick,fill=white!20] (#1,#2) circle (#3) ;
\draw (#1+#3*2.5,#2) node [below] {\small $A_2$} ;
\draw [myblue, thick] (#1-#3*0.7071,#2+#3*0.7071) to (#1-#3*0.7071-#4,#2+#3*0.7071+#4) ; 
}

\newcommand{\CenterAc}[4]{
\filldraw[draw=myred!100,thick,fill=white!20] (#1,#2) circle (#3) ;
\draw (#1+#3*2.5,#2) node [below] {\small $A_1$} ;
\draw [myred, thick] (#1-#3*0.7071,#2+#3*0.7071) to (#1-#3*0.7071-#4,#2+#3*0.7071+#4) ; 
}

\newcommand{\triangular}[3]{
\filldraw[draw=myblue!100,thick,fill=myblue!100] (#1-#3/2,#2-#3*0.2887) -- (#1+#3/2,#2-#3*0.2887) -- (#1,#2+#3*0.5774) ;
}

\newcommand{\kagomelattice}[1]{
\draw[dashed] (-2.2*#1,-#1*0.2887-#1*0.8660) -- (4.2*#1,-#1*0.2887-#1*0.8660);
\draw[dashed] (-2.2*#1,-#1*0.2887+#1*0.8660*1) -- (4.2*#1,-#1*0.2887+#1*0.8660*1);
\draw[dashed] (-2.2*#1,-#1*0.2887-#1*0.8660*3) -- (4.2*#1,-#1*0.2887-#1*0.8660*3);

\Dottedlinel{-1.5*#1}{-#1*0.2887}{1.4}{2.3}
\Dottedlinel{0.5*#1}{-#1*0.2887}{3.3}{2.3}
\Dottedlinel{2.5*#1}{-#1*0.2887}{3.3}{2.3}
\Dottedlinel{4.5*#1}{-#1*0.2887}{3.3}{-0.6}
\Dottedliner{-3.5*#1}{-#1*0.2887}{3.3}{-2.6}
\Dottedliner{-1.5*#1}{-#1*0.2887}{3.3}{1.4}
\Dottedliner{0.5*#1}{-#1*0.2887}{3.3}{2.3}
\Dottedliner{2.5*#1}{-#1*0.2887}{3.3}{2.3}
\Dottedliner{4.5*#1}{-#1*0.2887}{-0.6}{2.3}
}

\newcommand{\blackline}[4]{
\draw[gray,thick] (#1-#3*0.5,#2-#3*0.5*1.7321)--(#1+#4*0.5,#2+#4*0.5*1.7321);
}

\newcommand{\Dottedlinel}[4]{
\draw[dashed] (#1-#3*0.5,#2-#3*0.5*1.7321)--(#1+#4*0.5,#2+#4*0.5*1.7321);
}

\newcommand{\Dottedliner}[4]{
\draw[dashed] (#1+#3*0.5,#2-#3*0.5*1.7321)--(#1-#4*0.5,#2+#4*0.5*1.7321);
}

\begin{document}

\switchlinenumbers

\title{Dynamical Spectral Function of the Kagome Quantum Spin Liquid}

\author{Jiahang Hu}\email{These authors contributed equally to this work}
\affiliation{Beijing National Laboratory for Condensed Matter Physics and Institute of Physics, Chinese Academy of Sciences, Beijing 100190, China.}
\affiliation{School of Physical Sciences, University of Chinese Academy of Sciences, Beijing 100049, China.}

\author{Runze Chi}\email{These authors contributed equally to this work}
\affiliation{Beijing National Laboratory for Condensed Matter Physics and Institute of Physics, Chinese Academy of Sciences, Beijing 100190, China.}
\affiliation{School of Physical Sciences, University of Chinese Academy of Sciences, Beijing 100049, China.}
\affiliation{Division of Chemistry and Chemical Engineering, California Institute of Technology, Pasadena, California 91125, USA}

\author{Yibin Guo}
\affiliation{Beijing National Laboratory for Condensed Matter Physics and Institute of Physics, Chinese Academy of Sciences, Beijing 100190, China.}
\affiliation{School of Physical Sciences, University of Chinese Academy of Sciences, Beijing 100049, China.}

\author{B. Normand}\email{bruce.normand@psi.ch}
\affiliation{PSI Center for Scientific Computing, Theory and Data, Paul Scherrer Institute, CH-5232 Villigen-PSI, Switzerland}

\author{Hai-Jun Liao}\email{navyphysics@iphy.ac.cn}
\affiliation{Beijing National Laboratory for Condensed Matter Physics and Institute of Physics, Chinese Academy of Sciences, Beijing 100190, China.}

\author{T. Xiang}\email{txiang@iphy.ac.cn}
\affiliation{Beijing National Laboratory for Condensed Matter Physics and Institute of Physics, Chinese Academy of Sciences, Beijing 100190, China.}
\affiliation{School of Physical Sciences, University of Chinese Academy of Sciences, Beijing 100049, China.}

\begin{abstract}
Quantum spin liquids (QSLs) host exotic fractionalized magnetic and gauge-field excitations whose microscopic origins and experimental verification remain frustratingly elusive. In the absence of static magnetic order, the spin excitation spectrum constitutes the crucial probe of QSL behavior, but its theoretical computation is a serious challenge. Here we employ state-of-the-art tensor-network methods to obtain the full dynamical spectral function of the $J_1$-$J_2$ kagome Heisenberg model and benchmark our results by tracking their evolution across the magnetically ordered and QSL phases. Reducing $|J_2|/J_1$ causes increasingly strong spin-wave renormalization, flattening these modes then merging them into a continuum characteristic of deconfined spinons at all finite energies in the QSL. The low-energy continuum and the occurrence of gap closure at multiple high-symmetry points identify this gapless QSL as the U(1) Dirac spin liquid. These results establish a unified understanding of spin excitations in highly frustrated quantum magnets and provide clear spectral fingerprints for experimental detection in candidate kagome QSL materials.
\end{abstract}

\maketitle

\textit{Introduction---}Quantum spin liquids (QSLs) continue to attract extensive interest due to their unconventional properties, which are characterized by long-ranged quantum entanglement rather than conventional magnetic order at zero temperature, with accompanying fractional spin excitations~\cite{Anderson1973, Savary2016, Zhou2017}. QSLs offer exciting prospects for deepening our understanding of strongly correlated electronic systems, entanglement, topological order, and even error-protected quantum computation. The kagome lattice, with its frustrated triangular geometry and low coordination number, has long been a focal point for exploring the exotic states of quantum matter, and the $S = 1/2$ kagome model with only nearest-neighbor antiferromagnetic Heisenberg interactions is now widely accepted as hosting a QSL ground state.

This model presents one of the hardest problems in quantum magnetism~\cite{sindzingre2009,Sachdev1992,Yan2011,Savary2016,he2017,Lauchli2019} due to its extensive manifold of low-lying energy states. Introducing a next-neighbor coupling ($J_2$) is one of the more revealing ways to lower the degeneracy of this manifold and hence to track the emergence of QSL behavior \cite{gong2014, suttner2014, Iqbal2015, kolley2015, Liao2017, Zhu2019}. Nevertheless, an accurate characterization of the spin dynamics remains a fundamental challenge. Every numerical method faces significant limitations in this problem due to the complex interplay of quantum fluctuations, frustration, and finite system sizes; although computational studies have provided valuable insight into the ground-state properties, a quantitative description of the spectral function at all energies has not been possible. This is particularly important because the excitation spectrum can be measured directly by techniques including inelastic neutron scattering (INS) and tunneling spectroscopy, making it a critical link between theory and experiment.

In this Letter, we present a systematic investigation of the excitation spectrum in the $J_1$-$J_2$ kagome Heisenberg antiferromagnet (KHAF). We leverage our own recent developments in single-mode excited-state tensor-network methods that overcome previous limitations in order to resolve the full spectrum of low-energy excitations directly for the infinite system, meaning at all wave vectors. By controlling $J_2$, we track the evolution of the spectrum from the $q = 0$ and $\sqrt{3}$$\times$$\sqrt{3}$ magnetically ordered phases to the intermediate QSL phase, revealing how the higher-lying continuum is suppressed into the low-energy magnon branches, closing the spin gap at the emergence of U(1) Dirac QSL behavior. By identifying the fundamental characteristics of the KHAF, our work reveals the spectral signatures that will be crucial for the experimental identification of this QSL phase in quantum magnetic materials realizing the KHAF Hamiltonian. 

Theoretically, substantial efforts have been made to understand the ground-state properties of the KHAF by applying the methods of exact diagonalization~\cite{sindzingre2009, nakano2011, Lauchli2011, Lauchli2019, Prelovsek2021}, the density-matrix renormalization group (DMRG)~\cite{Jiang2008, Yan2011, Stefan2012, Jiang2012, Nishimoto2013, gong2014, sun2024, Zhu2019, he2017}, variational Monte Carlo~\cite{Ran2007, Iqbal2013, iqbal2014, Iqbal2015, Mei2015, Zhang2020, Ferrari2021}, and tensor networks~\cite{Liao2017, jiang2019, jahromi2020}. These studies have uncovered a multiplicity of candidate QSL phases, including gapped, nonchiral $\mathbb{Z}_2$ spin liquids~\cite{Jiang2008, Yan2011, Stefan2012, Jiang2012, Nishimoto2013, gong2014, mei2017}, chiral spin liquids (CSLs)~\cite{gong2014, Zhu2019, sun2024}, and gapless QSLs related to the U(1) Dirac spin liquid (DSL)~\cite{Ran2007, Iqbal2013, iqbal2014, Iqbal2015, Liao2017, he2017, jiang2019, Zhu2019, jahromi2020, Zhang2020}. This diversity underlines the sensitivity of KHAF physics and indicates that a definitive characterization will require working beyond the ground state. Numerical studies to date have examined the excitation spectrum of finite-sized systems~\cite{laeuchli2009, Lauchli2011, endo2018, Lauchli2019, Zhu2019, Ferrari2021, jiang2026}, with results highly dependent on system size and boundary conditions. Efforts to benchmark the spectrum from experiment remain complicated by the fact that no materials yet available provide a faithful representation of the KHAF, and we summarize this situation in the End Matter. A high-precision calculation of the spectral function in the infinite system is therefore required to clarify the nature of the kagome QSL.

\begin{figure*}[t]
\centering
\includegraphics[width=0.98\textwidth]{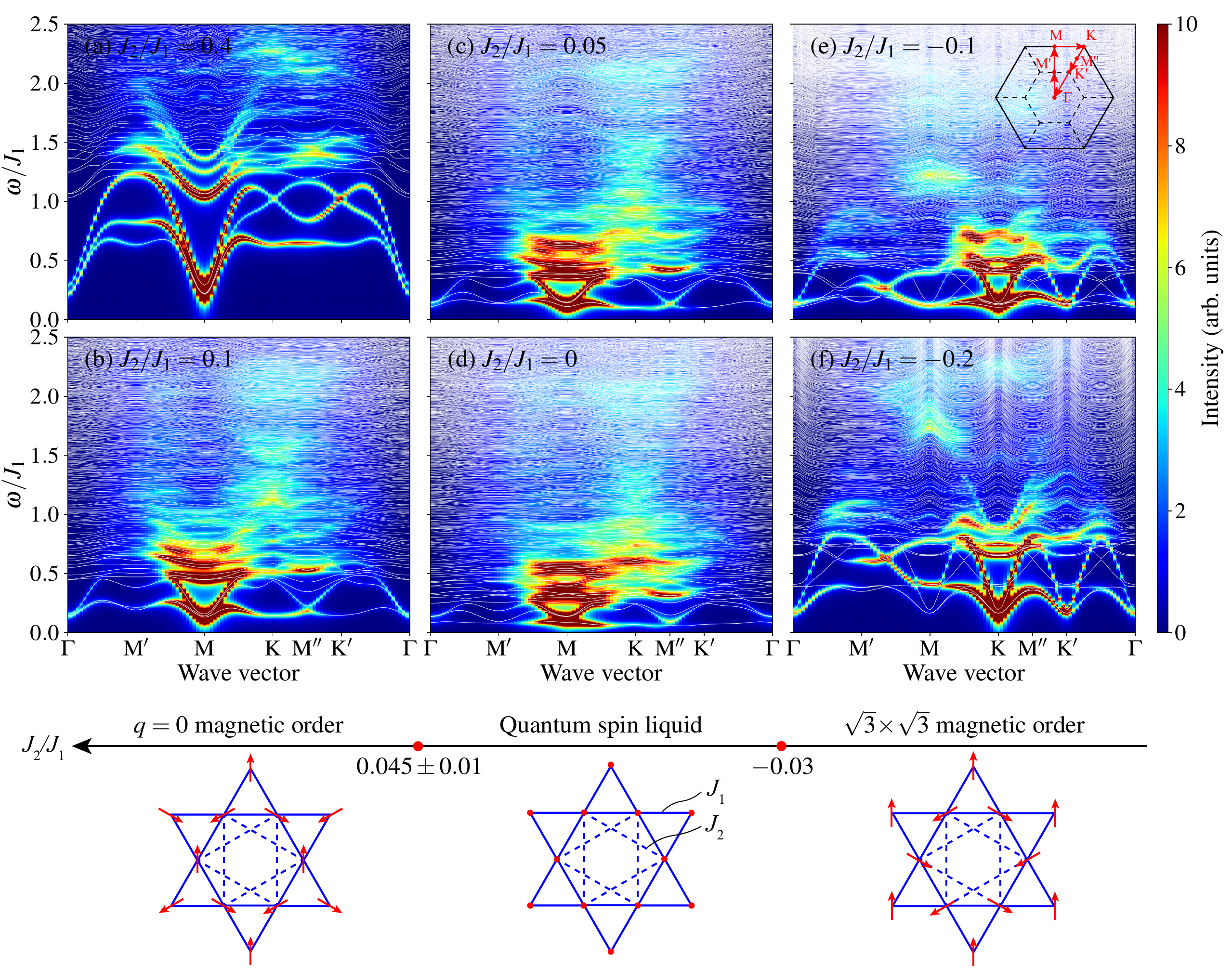}
\caption{{\bf Spin excitation spectra of the $J_1$-$J_2$ kagome Heisenberg antiferromagnet (KHAF).} (a)-(f) Example spectra in the $q = 0$ magnetically ordered phase (a),(b), the quantum spin-liquid (QSL) phase (c),(d), and the $\sqrt{3}$$\times$$\sqrt{3}$ ordered phase (e),(f). White lines show the energy levels obtained from iPEPS calculations. The inset in panel (e) shows the Brillouin zone and the high-symmetry path followed in our spectra; dashed and solid hexagons mark, respectively, the elementary and extended Brillouin zones. Below the spectra we represent the phase diagram determined by an iPEPS study of the ground states~\cite{Liao2017}. 
\label{fig1}}
\end{figure*}

\textit{Model and method---}We consider the model 
\begin{equation}
H = J_1 \sum_{\langle i j\rangle} \vec{S}_{i} \cdot \vec{S}_{j} + J_2 \sum_{\langle\langle i j\rangle\rangle} \vec{S}_{i} \cdot \vec{S}_{j}
\label{eq:Ham}
\end{equation}
on the kagome lattice with nearest- ($J_1$) and next-neighbor ($J_2$) Heisenberg interactions. A tensor-network study of the ground state~\cite{Liao2017} found a continuous transition from the ``$q = 0$'' ordered phase [Fig.~\ref{fig1}] to a gapless QSL at $J_2/J_1 = 0.045 \pm 0.01$ and a discontinuous transition into the ``$\sqrt{3}$$\times$$\sqrt{3}$'' ordered phase at $J_2/J_1 = - 0.03$. Tensor-network methods operate directly in the limit of infinite system size, which is a major advantage in capturing the key quantum fluctuation effects, and we stress that strikingly different phase boundaries are estimated by techniques restricted to finite systems~\cite{suttner2014, kolley2015, Iqbal2015, Zhu2019, jiang2026}. 

The single-mode excitation Ansatz for computing the spectral function~\cite{Ostlund1995, Haegeman2012, Vanderstraeten2015} within the framework of infinite projected entangled-pair states (iPEPS) is summarized in Sec.~S1 of the Supplemental Material (SM) \cite{sm}. Ground-state optimization and excited-state calculations are performed using automatic differentiation (AD)~\cite{Liao2019, chi2022, Ponsioen2022}. This method obtains excited states with definite momentum ({\bf k}) directly, eliminating the spatiotemporal Fourier transforms required in real-time evolution approaches, and thereby yielding excitation spectra with significantly higher resolution. The quantities we calculate are the equal-time spin structure factor,
 \begin{eqnarray}
    S(\mathbf{k}) & = & \sum_\alpha S^{\alpha\alpha} (\mathbf{k}) \; {\rm with} \;
    S^{\alpha\beta} (\mathbf{k}) = \langle 0 | S_{-\mathbf{k}}^{\alpha} \, S_{\mathbf{k}}^{\beta} |0 \rangle,
   \end{eqnarray}
where $\alpha, \beta = x, y, z$, and the spin excitation spectrum,
 \begin{eqnarray}
  S(\mathbf{k}, \omega) & = & \sum_\alpha S^{\alpha\alpha} (\mathbf{k}, \omega) \; {\rm with} \; \\
  S^{\alpha\beta} (\mathbf{k}, \omega) &=& \langle 0 | S_{-\mathbf{k}}^{\alpha} \, \delta(\omega - H + E_0) \, S_{\mathbf{k}}^{\beta} |0 \rangle.
 \end{eqnarray}
The discrete energy levels we obtain are shown by the white lines in Fig.~\ref{fig1} and in postprocessing our data we apply a Lorentzian broadening factor, $\eta$, for display purposes, as well as to model an experimental spectral function measured with finite energy resolution. 

For the present work, we have developed a number of methodological improvements that speed up our calculations by over 2 orders of magnitude. As described in Sec.~S2 of the SM \cite{sm}, by the reuse of CTMRG projectors \cite{chi2022,Ponsioen2022,tu2024} we achieve efficient parallelization, which we accelerate further by introducing a new fixed-point CTMRG method. This progress allows us to implement systematic spectral-function calculations over the entire Brillouin zone and over a wide range of $J_2/J_1$. The iPEPS truncation parameter is the tensor bond dimension, $D$, and finite-$D$ calculations always return a gapped spectrum, from which the intrinsic physics is deduced by extrapolation~\cite{Liao2017}. Our present calculations allow us to benchmark gapped spectra at fixed $D$ by comparing different $J_2$ values (below). At fixed $J_2$, we show a systematic comparison of finite-$D$ spectra in Sec.~S3 of the SM \cite{sm}. In Sec.~S4 \cite{sm} we demonstrate that the broadening $\eta$ cannot close gaps or shift spectral weight, but acts only to round the spectral features. In the figures to follow we use $D = 4$ and $\eta = 0.02J_1$ unless otherwise stated. 

\textit{Results---}Figure~\ref{fig1} illustrates the evolution of the spin excitation spectrum across the $J_2/J_1$ phase diagram (spectra for more $J_2$ values are shown in Sec.~S5 of the SM \cite{sm}). Deep in the $q = 0$ ordered phase [$J_2 = 0.4$, Fig.~\ref{fig1}(a)] we observe three spin-wave branches with a band width of order $J_1$, the lowest (the Goldstone mode) having a finite-$D$ gap at the M point. As $J_2$ is decreased toward the QSL regime [$J_2 = 0.1$, Fig.~\ref{fig1}(b)], the three branches not only exhibit a strongly downward energetic renormalization but the dense manifold of higher-lying states begins to cross the third branch. The lowest branch forms an almost flat band along K-K$^{\prime}$ and we remark on the persistence of the symmetry-protected crossings between the second and third spin-wave branches (also referred to as Dirac magnons~\cite{Fransson2016}) at the K and K$^{\prime}$ points. The ability to resolve these crossings, which are a consequence of reflection and time-reversal symmetries, highlights the accuracy of our tensor-network methods. 

At the onset of the QSL [$J_2 = 0.05$, Fig.~\ref{fig1}(c)], the spectrum continues its net downward shift and the dense level manifold has fully dispersed spectral weight, which is characteristic of a continuum. As this continuum merges with the upper edge of the low-energy branches, we observe that the minimum of the second-lowest branch approaches the lowest (flat) branch at the M$^{\prime\prime}$ point. This behavior is qualitatively different from spin-wave theory, in which the second branch approaches the third, merging with it at $J_2 = 0$ (Sec.~S7 of the SM \cite{sm}). 

The same trends continue at the nearest-neighbor kagome point [$J_2 = 0$, Fig.~\ref{fig1}(d)], with the continuum descending below $\omega \approx 0.25J_1$ at all wave vectors and the lowest branch becoming flat at an energy, $\omega < 0.1J_1$, below our finite-$D$ gap, which we benchmark below. A small negative $J_2$ compresses the spectrum yet more (Fig.~S7 of the SM \cite{sm}), but further reduction of $J_2$ drives a first-order transition into the $\sqrt{3}$$\times$$\sqrt{3}$ ordered phase \cite{Liao2017}. This causes a rapid reemergence of sharp spin-wave excitations [Fig.~\ref{fig1}(e)], with the Goldstone mode appearing at K rather than M, and a rapid recovery of their renormalized energy scale toward $J_1$ [Fig.~\ref{fig1}(f)].

\begin{figure}[t]
\centering
\includegraphics[width=0.5\textwidth]{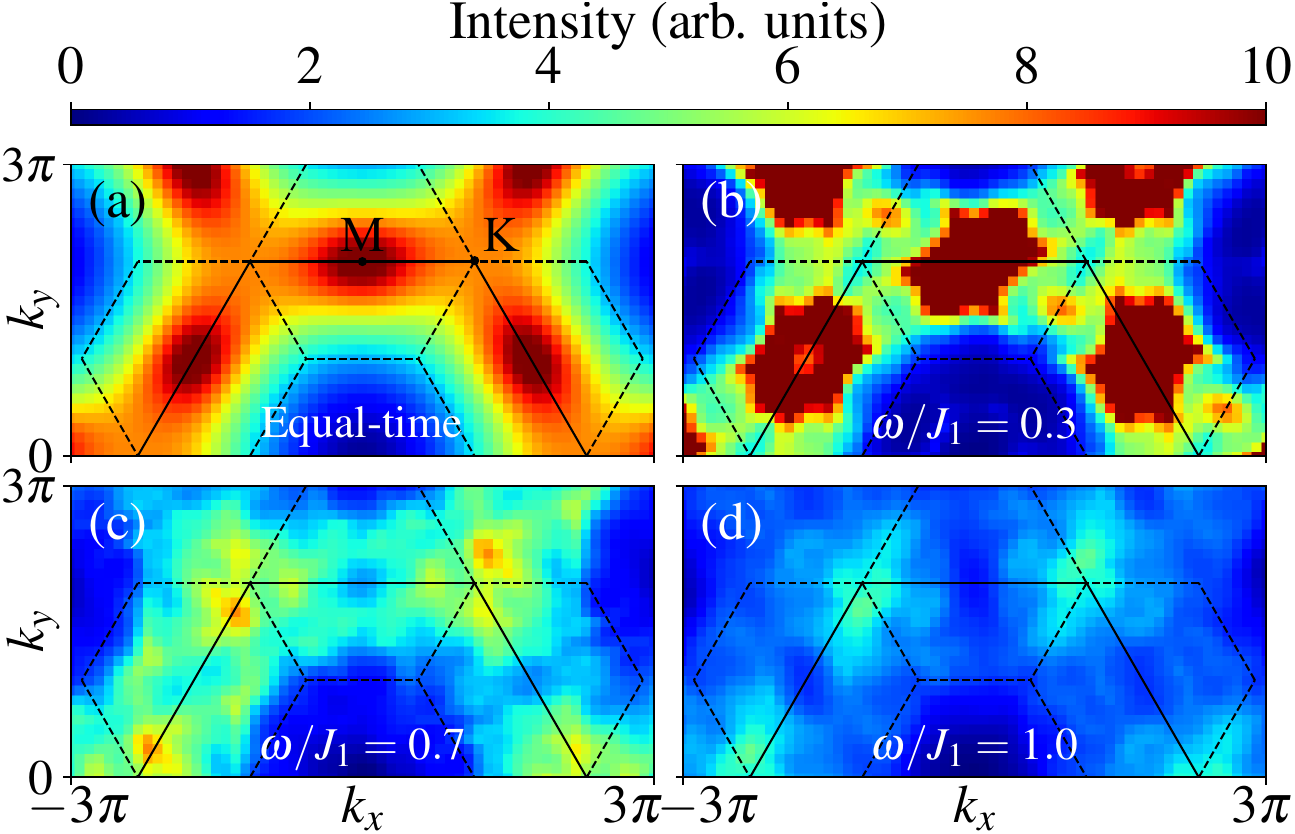}
\caption{{\bf Spin structure factor of the KHAF at $J_2 = 0$.} (a) Equal-time structure factor shown over the entire Brillouin zone. (b)-(d) Dynamic structure factors shown for $\omega = 0.3J_1$ (b), $0.7J_1$ (c), and $1.0J_1$ (d).
\label{fig2}}
\end{figure}

Before proceeding to a quantitative analysis of the low-energy spectrum, in Fig.~\ref{fig2} we characterize the spectral function of the QSL throughout the Brillouin zone by showing equal-time and dynamic structure factors at constant energy for $J_2 = 0$. The equal-time structure factor [Fig.~\ref{fig2}(a)] has maximal weight at the M point, but strong diffuse intensity spreads along the entire Brillouin-zone boundary. This behavior is in sharp contrast to the ordered phases, where one finds a single dominant peak, either at M in the $q = 0$ phase or at K in the $\sqrt{3}$$\times$$\sqrt{3}$ phase; for completeness we show the equal-time structure factor around the high-symmetry path in Sec.~S1 of the SM \cite{sm} and for a wide range of $J_2$ values in Sec.~S6. 

Turning to the dynamic structure factor, at low energies the spectral weight covers an extended but well bounded region around M [Fig.~\ref{fig2}(b)]. At energies above the lowest branches, the weight becomes nearly uniform and continuous around the Brillouin-zone boundary [Fig.~\ref{fig2}(c)]. While this result is reminiscent of some experimental observations~\cite{Han2012,breidenbach2025}, moving to still higher energies [Fig.~\ref{fig2}(d)] concentrates spectral weight at K, which may be a signature useful for future measurements.

\begin{figure}[t]
\centering
\includegraphics[width=0.48\textwidth]{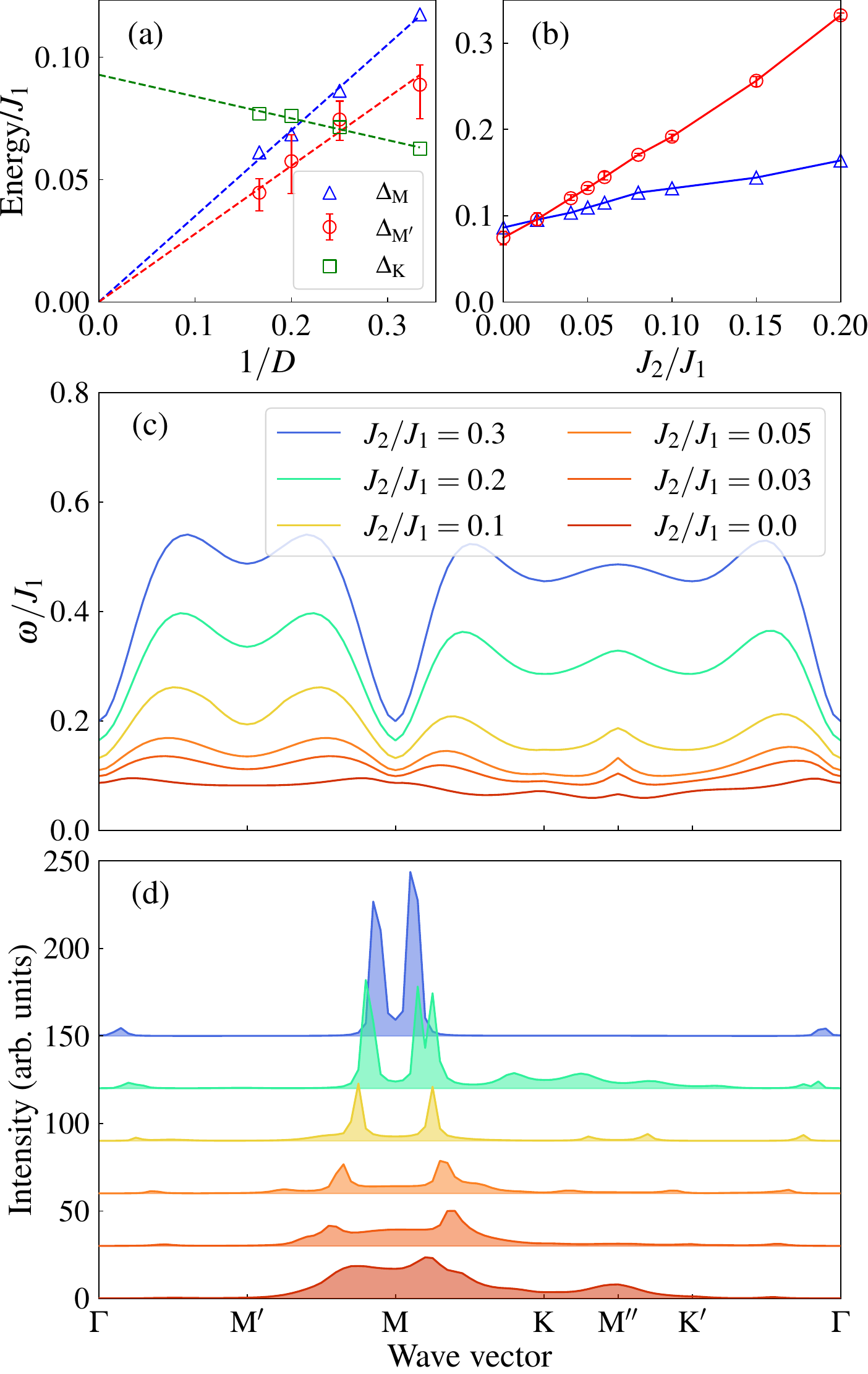}
\caption{{\bf Properties of the low-energy spectrum}. (a) Energy gaps at the M, M$^{\prime}$, and K points in the $J_2 = 0$ KHAF shown as functions of $1/D$. Error bars for $\Delta_{{\rm M}^{\prime}}$ represent the variation between M$^\prime$ points (the Ansatz does not preserve the $C_3$ symmetry making these equivalent). Dashed lines are guides to the eye. (b) Energy gaps at M and M$^{\prime}$ computed as functions of $J_2$ with $D = 4$. (c) Lowest excitation energy calculated around the high-symmetry path in the Brillouin zone for a range of $J_2$ values with $D = 4$. (d) Spin excitation spectra at fixed energy $\omega = 0.3J_1$ shown on the same path for the same $J_2$ and $D$ values.}
\label{fig3}
\end{figure}

The gapped or gapless nature of the QSL in the KHAF remains under intense debate. To determine its intrinsic nature from finite-$D$ spectral data, we first perform a bond-dimension analysis of the gap for the $J_2 = 0$ case, shown in Fig.~\ref{fig3}(a). The gaps at M and M$^{\prime}$ (with M$^{\prime\prime}$ symmetry-equivalent to M$^{\prime}$) decrease as $D$ is increased, following $\Delta \propto 1/D$ within the error bars, which indicates that the QSL is indeed gapless in the large-$D$ limit. By contrast, the energy at K increases with $D$, and in Fig.~\ref{fig3}(a) we apply a linear extrapolation to obtain an upper bound on the width of the lowest band.

As an alternative benchmarking approach, in Fig.~\ref{fig3}(b) we analyze the evolution of $\Delta_{\rm M}$ and $\Delta_{{\rm M}^{\prime}}$ as functions of $J_2$. At $J_2/J_1 > 0.05$, in the $q = 0$ phase, the system has a gapless Goldstone mode at M as SU(2) spin symmetry is broken (this mode has finite energy at M$^{\prime}$ and M$^{\prime\prime}$). Hence the entire blue ($\Delta_{\rm M}$) line in Fig.~\ref{fig3}(b) is a finite-$D$ gap. The fact that $\Delta_{\rm M}$ takes its smallest values for $J_2/J_1 < 0.05$ indicates that the QSL is indeed gapless, and we take this $\Delta_{\rm M}$ value as the criterion defining gap closure. Beyond Fig.~\ref{fig3}(a), the fact that the M$^{\prime}$-point gap falls below the finite-$D$ gap ($\Delta_{\rm M}$) near $J_2/J_1\approx 0.05$ in Fig.~\ref{fig3}(b) is another criterion for concluding that the M-, M$^{\prime}$-, and M$^{\prime\prime}$-point gaps all vanish simultaneously upon entering the QSL. The lowest excitation going to zero simultaneously at the M, M$^{\prime}$, and M$^{\prime\prime}$ points is a primary hallmark of the U(1) DSL \cite{Ran2007,Mei2015,Zhang2020} and of crystal momentum fractionalization \cite{Mei2015}. The equal-time structure factor of Fig.~\ref{fig2}(a) also matches projected-wave-function results obtained for the U(1) DSL~\cite{Mei2015,Zhang2020}, but is incompatible with alternatives including a gapped QSL and a spinon Fermi-surface QSL.

In Fig.~\ref{fig3}(c) we show the dispersion of the lowest excitation branch for a sequence of decreasing $J_2$ values. The band width is suppressed continuously until at $J_2 = 0$ we find an essentially flat band at the finite-$D$ gap. The emergence around $J_2/J_1 = 0.05$ of a cusp at M$^{\prime\prime}$ is a finite-$D$ effect, with M$^{\prime\prime}$ becoming the local minimum at higher $D$ (Fig.~S4 of the SM \cite{sm}). In Fig.~\ref{fig3}(d) we show that the constant-energy spectrum at $\omega = 0.3J_1$ has two sharp spin-wave peaks near M when the system is well in the $q = 0$ ordered phase. As $J_2$ is decreased, the peak intensities are suppressed systematically, while the spectral weight at the M point itself grows, until a broad feature centered at M emerges as the signature of the QSL.

\textit{Discussion---}The key to a meaningful analysis of QSL properties based on numerical data is achieving a quantitative definition of the terms ``gap'' and ``continuum.'' Previous studies of ordered magnetic states by the single-mode iPEPS Ansatz~\cite{Vanderstraeten2015, Vanderstraeten2019, Ponsioen2020, chi2022, chi2024, Xu2024} demonstrated that the finite-$D$ gap falls systematically with increasing $D$, restoring the gapless Goldstone mode. With this insight, our ability to scan $D$ over a narrow range and $J_2$ over a wide range allows us to establish a quantitative criterion for both the finite-$D$ gap and the gap $\Delta_{{\rm M}^\prime}$, and hence to deduce the gapless and U(1) Dirac nature of the QSL. 

Defining a continuum is a more delicate issue. Tensor-network methods are based on gapped systems and always return a spectrum of integer-spin excitations. The claim that this spectrum is only an effective description of a gapless QSL, which has deconfined $\Delta S = 1/2$ excitations (spinons), is a two-part argument. First, the spectral functions in Figs.~\ref{fig1}(b)-\ref{fig1}(e) show unbroken series of densely packed lines (a spacing below $0.02J_1$, less still at $D = 5$ in Fig.~S3 of the SM \cite{sm}) at all energies above the lowest three branches. There is no intensity hierarchy of the type expected from two- and higher-magnon continua \cite{chi2022}. Thus these have a clear physical interpretation as the edge of a continuum marking the onset of spinon deconfinement above a finite energy threshold that is both remarkably low [$0.2J_1$ at $J_2/J_1 = - 0.02$, Fig.~S7(j) of the SM \cite{sm}] and remarkably insensitive to the wave vector (presumably as a consequence of the very flat bands appearing in the KHAF problem). Second, these continua must extend to zero energy at the gapless points. Although we cannot observe this, because the entire lowest band lies at or below our finite-$D$ gap [approximately $0.08 J_1$; Fig.~\ref{fig1}(d)], we have deduced that this band is massless at M, M$^{\prime}$, and M$^{\prime\prime}$, while retaining a robust dispersion to an upper band edge close to the finite-$D$ gap. Gap closure at these fractions of the Brillouin-zone dimensions cannot be a property of a real $\Delta S = 1$ excitation, and hence we deduce that the low-energy spectrum must have developed into a continuum above the lowest band, as expected for a U(1) DSL.

The most distinctive features of our QSL spectra [Figs.~\ref{fig1}(c) and ~\ref{fig1}(d)] are (i) strong spectral weight at the finite-$D$ gap and (ii) the continuum at all wave vectors above a very low energy threshold. Property (i) is not shared by any fully gapped QSL, or by a Kitaev-type $\mathbb{Z}_2$ QSL \cite{Kitaev2003}, where a single Majorana mode is gapless but the spin response is gapped. However, it is a property of the U(1) DSL, whose low-energy physics is governed by four flavors of two-component Dirac fermion coupled to a U(1) gauge field \cite{Ran2007}. Property (ii) alone does not help to classify the QSL phase, but does offer a plausibility argument for why wave-vector-resolved measurements may perceive a full gap. The ongoing debate between gapped and gapless QSLs in the KHAF was shaped by early DMRG studies finding a full gap, and we comment on this in the End Matter. A projective symmetry-group (PSG) analysis of mean-field states in the KHAF found only one gapped $\mathbb{Z}_2$ state, classified as $\mathbb{Z}_2[0,\pi]\beta$, whose energy is competitive with the U(1) Dirac QSL (which is its parent state) \cite{Lu2011}. However, detailed studies found very few differences between the physical properties of these two states \cite{Iqbal2015, Zhang2020}, including a gapless spin response for the $\mathbb{Z}_2[0,\pi]\beta$ state, raising the possibility that the PSG mean-field classification may not be appropriate for the KHAF problem. Thus one expects that a gapless QSL in the KHAF can only be the U(1) DSL \cite{Ran2007,Hermele2008,Song2019}, and this is confirmed by our result for the M, M$^{\prime}$, and M$^{\prime\prime}$ gaps.

We remark in closing that, by the nature of the tensor-network Ansatz, our results are highly indicative rather than definitive. Our limitations lie in the ``finite-size effect'' of the bond dimension, $D$, which we can vary over only a small range (that is nevertheless sufficient to identify the lowest band and to reveal the continuum). We also applied a level-broadening factor, $\eta$, for visualization purposes, and verified that it has no effect on the overall spectral-weight distribution. Our advantages for deducing the intrinsic underlying physics lie in the systematic coverage of energy, wave vector, and $J_2$ that gives a maximum of certainty to our statements. As the key example, varying $J_2$ allowed us to benchmark the finite-$D$ gap and hence to make a definite statement that the QSL in the KHAF is gapless.

In summary, we have developed state-of-the-art tensor-network methods to investigate the spin excitation spectrum of the kagome Heisenberg model, using the next-neighbor coupling to control the evolution from the $q = 0$ and $\sqrt{3}$$\times$$\sqrt{3}$ magnetically ordered phases into the intermediate QSL regime. In the ordered phases, we compute accurate magnon (and multimagnon) branches lying below a near-uniform continuum, which drives strong quantum renormalization far below the predictions of spin-wave theory. In the QSL phase, we quantify the spectral-weight distribution across the Brillouin zone, finding a broad low-energy feature centered at the M point, a uniform intermediate-energy continuum, and high-energy intensity largely around K. By analyzing the evolution of the low-energy spectrum with $D$ and $J_2$, we identify vanishing gaps at M, M$^\prime$, and M$^{\prime\prime}$ in the QSL phase, indicating a gapless U(1) Dirac spin liquid. Beyond advancing the understanding of kagome QSLs and offering guidance to future experiments on candidate KHAF materials, our results open a wealth of possibilities for the exploration of exotic quantum states through their spectral functions.

\textit{Acknowledgments---} We thank Junyi Ji, Chun Zhang, Hong-Chen Jiang, and Wei Zhu for helpful discussions. This work was supported by the Strategic Priority Research Program of Chinese Academy of Sciences (Grant No.~XDB0500202), the National Key Research and Development Project of China (Grants No.~2024YFA1408604, No.~2021ZD0301800, and No.~2022YFA1403900), the National Natural Science Foundation of China (Grants No.~12488201, No.~12322403, and No.~12347107), and the Youth Innovation Promotion Association CAS (Grant No.~2021004). The numerical calculations in this work were carried out on the ORISE Supercomputer of the Chinese Academy of Sciences. 

\textit{Data availability---} The data that support the findings of
this article are openly available \cite{Hu2026data}.

\onecolumngrid

\vskip 10mm

\centerline{\large{\bf End Matter}}

\vskip 6mm

\twocolumngrid

\textit{Experiment---}Several $S = 1/2$ antiferromagnets with kagome geometry have been synthesized, including ZnCu$_3$(OH)$_6$Cl$_2$ (herbertsmithite)~\cite{shores2005}, Cu$_3$Zn(OH)$_6$FBr (Zn-barlowite)~\cite{feng2017}, ZnCu$_3$(OH)$_6$SO$_4$ (Zn-brochantite)~\cite{li2014}, and YCu$_3$(OH)$_{6.5}$Br$_{2.5}$ (YCOB)~\cite{Liu2022,zeng2022}. INS measurements on some of these materials~\cite{Han2012,nilsen2013,Zeng2024,breidenbach2025} have found broad excitation continua, which are a generic hallmark of fractionalized excitations. However, an interpretation is complicated by the prevalence of ionic site disorder in most of these systems, which can obscure or mimic the signatures of fractionalization. Further, Cu$^{2+}$ ions in a triangular geometry have unavoidable Dzyaloshinskii–Moriya (DM) interactions~\cite{zorko2008}, as well as possible further-neighbor Heisenberg interactions~\cite{jeschke2013}. Both terms act to stabilize magnetically ordered states, most notably the coplanar $q = 0$ and $\sqrt{3}$$\times$$\sqrt{3}$ ordering patterns~\cite{Rousochatzakis2009,Liao2017}, and to modify the spectrum accessible to INS. 

Experimental efforts over the past decade have focused on eliminating site disorder~\cite{zeng2022} and reducing the DM term~\cite{feng2017}. A recent study of Zn-barlowite favored the interpretation as a gapped QSL \cite{breidenbach2025} applied to early studies of herbertsmithite~\cite{Han2012}. By contrast, further recent work on a YCOB derivative with counterion site disorder reported the observation of Dirac cones, and hence of a gapless QSL, albeit with the cones appearing at the K$^\prime$ points \cite{Zeng2024}. Given the persisting discrepancies between the paradigm model and the spin Hamiltonians of these materials, we do not attempt to use our spectral functions for modeling purposes, but contribute instead by offering a deeper understanding of the intrinsic spin dynamics of the $J_1$-$J_2$ KHAF in both its ordered and QSL phases as a guide for future materials discovery and INS experiment. 

\textit{DMRG comparison---}Early evidence in favor of a gapped QSL in the KHAF with only nearest-neighbor interactions came from DMRG results obtained for relatively narrow cylinders \cite{Jiang2008, Yan2011, Stefan2012, Jiang2012, Nishimoto2013, gong2014}. More recent DMRG studies are divided among the scenarios of the fully gapped $\mathbb{Z}_2$ QSL \cite{mei2017}, the gapped CSL \cite{sun2024}, and a gapless (Dirac-cone) spinon spectrum \cite{he2017,Zhu2019}. Because DMRG, like tensor-network approaches, is based on matrix-product states, both methods share the weakness noted in the main text that a gapped spectrum is an intrinsic feature and the challenge is to deduce the underlying physics from this. 

The essential qualitative property of our tensor-network method is that we work on the infinite system, thereby capturing the full two-dimensional reciprocal space with no bias. To the extent that the physics of the U(1) DSL involves optimizing the kinetic energy of nearly-free spinonic quasiparticles, it is eminently possible that the restriction to only few wave vectors around the DMRG cylinder may constitute a strong bias. We draw attention once again to the very different QSL phase boundaries in $J_2$ obtained in our work \cite{Liao2017} and by finite-system methods~\cite{kolley2015, Zhu2019, jiang2026}, which indicate that the two approaches are studying a quite different phase competition. (This may not come as a total surprise, given the extremely delicate nature of the KHAF problem). We assert that any results obtained for the infinite system have an intrinsic advantage and justify this statement from the accuracy of the spectral functions we obtained at all energies in the $q = 0$ ordered phase near the continuous $J_2$-driven transition to the QSL.

\setcounter{figure}{0}
\renewcommand{\thefigure}{S\arabic{figure}}
\setcounter{section}{0}
\renewcommand{\thesection}{S\arabic{section}}
\setcounter{equation}{0}
\renewcommand{\theequation}{S\arabic{equation}}
\setcounter{table}{0}
\renewcommand{\thetable}{S\arabic{table}}

\onecolumngrid

\vskip 10mm

\noindent
{\bf{\large{Supplemental Material to accompany the manuscript}}}

\vskip 4mm

\noindent
{\bf{\large{Dynamical Spectral Function of the Kagome Quantum Spin Liquid}}}

\vskip 6mm

\noindent
Jiahang Hu, Runze Chi, Yibin Guo, B. Normand, Hai-Jun Liao, and T. Xiang

\vskip 10mm

\twocolumngrid

\section{Tensor-Network Methods}
\label{sec1}

\subsection{Ground State}

To calculate the dynamical structure factor, the initial step is to derive an accurate Ansatz to represent the ground state, $|0 \rangle$, and for this we employ an infinite projected entangled-pair state (iPEPS). For convenience in our subsequent calculations, we transform the kagome lattice into a square lattice by combining three sites on a triangle into a single effective site,
\begin{equation}
\label{eq1}
|0\rangle = |\Psi(A)\rangle = 
\begin{array}{l}
\begin{tikzpicture}[every node/.style={scale=1},scale=0.45]
      \kagomelattice{1.0}

      \triangular{-0.5}{0.5*1.7321}{1.04}
      \triangular{1.5}{0.5*1.7321}{1.04}
      \triangular{3.5}{0.5*1.7321}{1.04}
      \triangular{-1.5}{-0.5*1.7321}{1.04}
      \triangular{0.5}{-0.5*1.7321}{1.04}
      \triangular{2.5}{-0.5*1.7321}{1.04}
      \triangular{-0.5}{-1.5*1.7321}{1.04}
      \triangular{1.5}{-1.5*1.7321}{1.04}
      \triangular{3.5}{-1.5*1.7321}{1.04}

      \draw[gray,thick] (-2.2,0.5*1.7321)--(4.2,0.5*1.7321);
      \draw[gray,thick] (-2.2,-0.5*1.7321)--(4.2,-0.5*1.7321);
      \draw[gray,thick] (-2.2,-1.5*1.7321)--(4.2,-1.5*1.7321);

      \blackline{-1.0-0.1667}{-0.2887}{2.067}{2.3}
      \blackline{1.0-0.1667}{-0.2887}{3.3}{2.3}
      \blackline{3.0-0.1667}{-0.2887}{3.3}{2.3}
      \blackline{5.0-0.1667}{-0.2887}{3.3}{-1.267}

      \Tensor{-0.5}{0.5*1.7321}{0.15}{0.2}
      \Tensor{1.5}{0.5*1.7321}{0.15}{0.2}
      \Tensor{3.5}{0.5*1.7321}{0.15}{0.2}
      \Tensor{-1.5}{-0.5*1.7321}{0.15}{0.2}
      \Tensor{0.5}{-0.5*1.7321}{0.15}{0.2}
      \Tensor{2.5}{-0.5*1.7321}{0.15}{0.2}
      \Tensor{-0.5}{-1.5*1.7321}{0.15}{0.2}
      \Tensor{1.5}{-1.5*1.7321}{0.15}{0.2}
      \Tensor{3.5}{-1.5*1.7321}{0.15}{0.2}

     \end{tikzpicture}
    \end{array}
    \rightarrow
    \begin{array}{l}
    \begin{tikzpicture}[baseline,every node/.style={scale=1},scale=0.4]
    \draw[step=1 cm,gray] (-2.4,-2.4) grid (2.4,2.4);
    \Tensor{-1}{0}{0.15}{0.2}
    \Tensor{-1}{1}{0.15}{0.2}
    \Tensor{-1}{2}{0.15}{0.2}
    \Tensor{-1}{-1}{0.15}{0.2}
    \Tensor{-1}{-2}{0.15}{0.2}
    \Tensor{-2}{0}{0.15}{0.2}
    \Tensor{-2}{1}{0.15}{0.2}
    \Tensor{-2}{2}{0.15}{0.2}
    \Tensor{-2}{-1}{0.15}{0.2}
    \Tensor{-2}{-2}{0.15}{0.2}
    \CenterA{0}{0}{0.15}{0.2}
    \Tensor{0}{1}{0.15}{0.2}
    \Tensor{0}{2}{0.15}{0.2}
    \Tensor{0}{-1}{0.15}{0.2}
    \Tensor{0}{-2}{0.15}{0.2}
    \Tensor{1}{0}{0.15}{0.2}
    \Tensor{1}{1}{0.15}{0.2}
    \Tensor{1}{2}{0.15}{0.2}
    \Tensor{1}{-1}{0.15}{0.2}
    \Tensor{1}{-2}{0.15}{0.2}
    \Tensor{2}{0}{0.15}{0.2}
    \Tensor{2}{1}{0.15}{0.2}
    \Tensor{2}{2}{0.15}{0.2}
    \Tensor{2}{-1}{0.15}{0.2}
    \Tensor{2}{-2}{0.15}{0.2}
\end{tikzpicture}.
\end{array}
\end{equation}
%
Here one translationally invariant local tensor $A$ is used in representing both the QSL and the $q = 0$ magnetically ordered phase. $A$ is a rank-$5$ tensor with physical bond dimension $d = 8$ and virtual bond dimension $D$, where $D$ controls the maximum entanglement entropy of the tensor-network state and thus determines the accuracy of method. For the $\sqrt{3}$$\times$$\sqrt{3}$ ordered phase, three tensors ($A_1$, $A_2$, $A_3$) are required to represent this order in a 9-site unit cell, represented as
\begin{equation}
\label{eq2}
|0\rangle = 
    \begin{tikzpicture}[baseline,every node/.style={scale=1},scale=0.4]
    \draw[step=1 cm,gray] (-2.4,-2.4) grid (2.4,2.4);
    \CenterAc{-1}{0}{0.15}{0.2}
    \Tensora{-1}{1}{0.15}{0.2}
    \Tensorb{-1}{2}{0.15}{0.2}
    \Tensorb{-1}{-1}{0.15}{0.2}
    \Tensora{-1}{-2}{0.15}{0.2}
    \Tensora{-2}{0}{0.15}{0.2}
    \Tensorb{-2}{1}{0.15}{0.2}
    \Tensorc{-2}{2}{0.15}{0.2}
    \Tensorc{-2}{-1}{0.15}{0.2}
    \Tensorb{-2}{-2}{0.15}{0.2}
    \CenterAb{0}{0}{0.15}{0.2}
    \Tensorc{0}{1}{0.15}{0.2}
    \Tensora{0}{2}{0.15}{0.2}
    \Tensora{0}{-1}{0.15}{0.2}
    \Tensorc{0}{-2}{0.15}{0.2}
    \CenterAa{1}{0}{0.15}{0.2}
    \Tensorb{1}{1}{0.15}{0.2}
    \Tensorc{1}{2}{0.15}{0.2}
    \Tensorc{1}{-1}{0.15}{0.2}
    \Tensorb{1}{-2}{0.15}{0.2}
    \Tensorc{2}{0}{0.15}{0.2}
    \Tensora{2}{1}{0.15}{0.2}
    \Tensorb{2}{2}{0.15}{0.2}
    \Tensorb{2}{-1}{0.15}{0.2}
    \Tensora{2}{-2}{0.15}{0.2}
\end{tikzpicture}.
\end{equation}
For brevity, we focus our discussion below on the case of a single tensor $A$, while the generalization to multiple sublattices is straightforward.

For a given local tensor $A$, the expectation value of a local physical observable, such as the energy, is calculated by contracting a double-layer tensor network using the corner-transfer-matrix renormalization-group (CTMRG) method~\cite{corboz2014}. In this approach, the environment of each tensor is approximated by corner-transfer and edge-transfer matrices, and the convergence of the algorithm depends on their boundary bond dimension, $\chi$. In calculations of the excited states, $\chi = 50$ is sufficient to ensure convergence for tensor bond dimensions $D = 3$ and 4, while $\chi = 100$ is used for $D = 5$ and $\chi = 120$ for $D = 6$. To optimize $A$, we perform a variational minimization of the ground-state energy using the automatic differentiation method (AD)~\cite{Liao2019}. This approach, which optimizes the entanglement captured by the local basis, has been shown to be more accurate and effective compared to alternative algorithms, such as the simple-update (SU) method used in early work \cite{Liao2017}.

\begin{figure}[t]
\centering
\includegraphics[width=0.96\columnwidth]{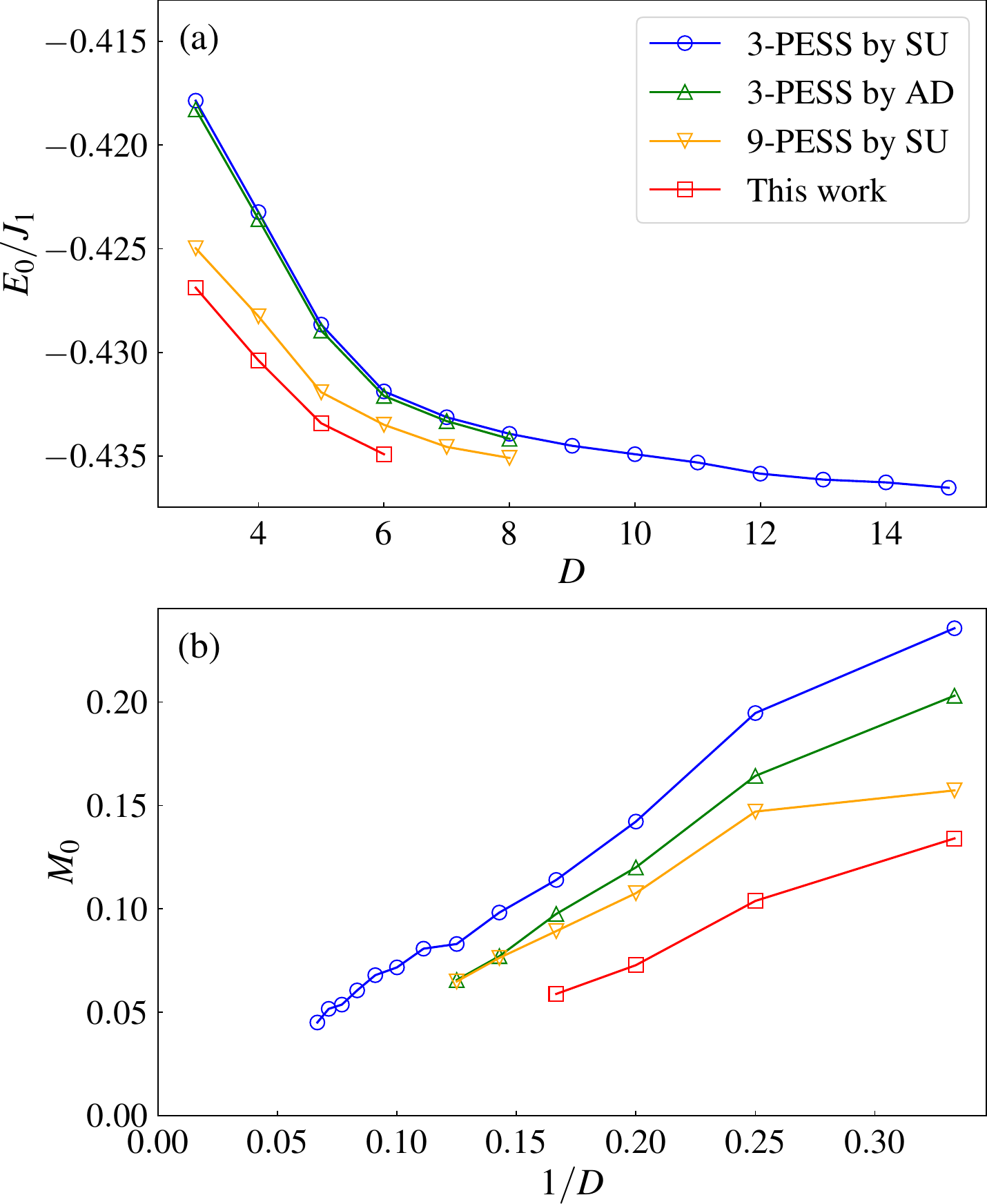}
\caption{Comparison of the ground-state energy, $E_0$, and the staggered magnetization, $M_0$, of the $J_2 = 0$ KHAF computed using 3-PESS, 9-PESS, and iPEPS bases and optimized by simple-update (SU) and automatic differentiation (AD) methods. 
\label{Gs_EM}}
\end{figure}

\begin{figure}[t]
\centering
\includegraphics[width=0.8\columnwidth]{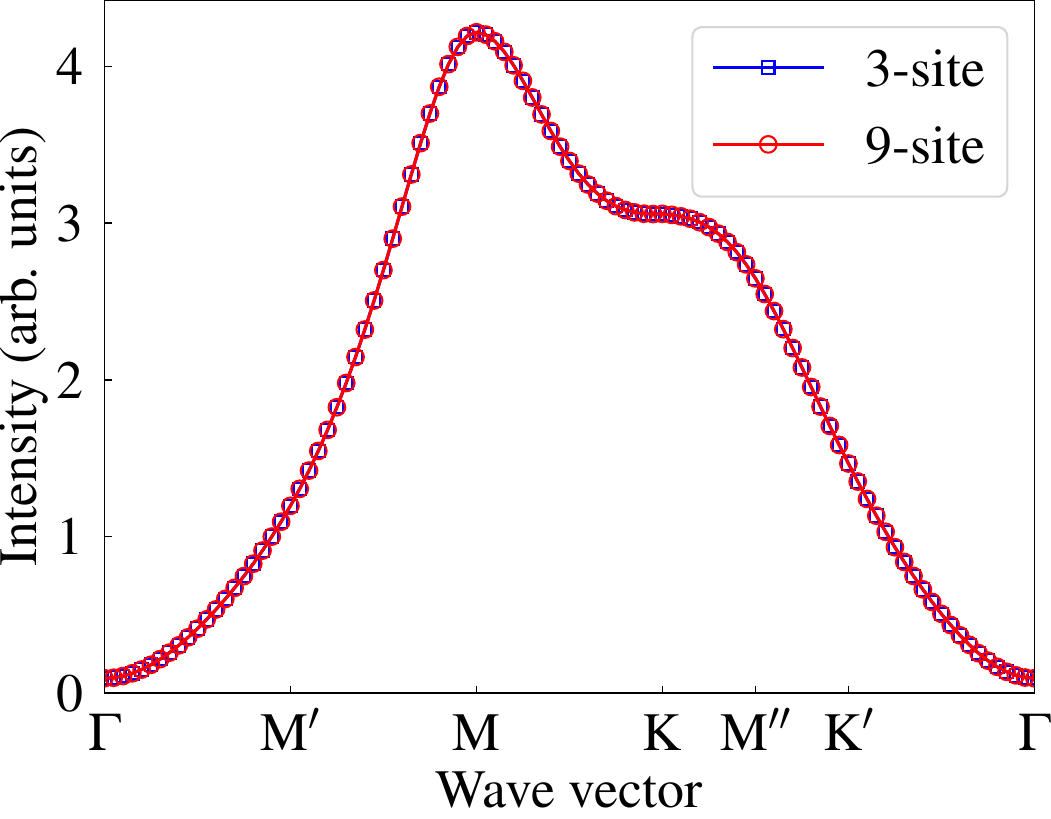}
\caption{Comparison of the equal-time structure factor for the $J_2 = 0$ KHAF calculated with $D = 4$ using a 3-site unit cell (one tensor) and a 9-site unit cell (three tensors). Both constructions yield identical structure factors, indicating that they represent the same state.
\label{StaticUnitCell}}
\end{figure}

To compare our present iPEPS approach with the projected entangled simplex-state (PESS) constructions used in our investigation of the ground-state properties of the KHAF \cite{Liao2017}, we specify that a PESS is composed of two types of tensor, namely the simplex tensor ($S$, without physical indices) and the projection tensor ($A$, with physical indices). Both 3-site and 9-site PESS can be mapped to an iPEPS by contracting all the connected simplex tensors and one projection tensor into a single rank-5 $A$ tensor. As an example, one simplex tensor and one projection tensor (circled) in a 9-PESS are contracted to form a single iPEPS $A$ tensor in the representation
\begin{equation}
\label{eq1}
\begin{array}{l}
\begin{tikzpicture}[every node/.style={scale=1},scale=0.45]
      
      \clip (-2.2,-0.2887-3.3*0.5*1.7321) rectangle (4.2,-0.2887+2.3*0.5*1.7321);
      \kagomelattice{1.0}

      \simplex{-1.5}{0.5*1.7321-1.7321/3}{1.04}
      \simplex{0.5}{0.5*1.7321-1.7321/3}{1.04}
      \simplex{2.5}{0.5*1.7321-1.7321/3}{1.04}
      \simplexS{-0.5}{-0.5*1.7321-1.7321/3}{1.04}
      \simplex{1.5}{-0.5*1.7321-1.7321/3}{1.04}
      \simplex{3.5}{-0.5*1.7321-1.7321/3}{1.04}

      \projector{-0.5}{0.5*1.7321}{1.04}
      \projector{1.5}{0.5*1.7321}{1.04}
      \projector{3.5}{0.5*1.7321}{1.04}
      \projector{-1.5}{-0.5*1.7321}{1.04}
      \projectorA{0.5}{-0.5*1.7321}{1.04}
      \projector{2.5}{-0.5*1.7321}{1.04}
      \projector{-0.5}{-1.5*1.7321}{1.04}
      \projector{1.5}{-1.5*1.7321}{1.04}
      \projector{3.5}{-1.5*1.7321}{1.04}

      \draw[dashed, myorange] (0.5-0.2,-0.5*1.7321) circle (1.1);

     \end{tikzpicture}
    \end{array}
    \rightarrow
\begin{array}{l}
\begin{tikzpicture}[every node/.style={scale=1},scale=0.45]
      \kagomelattice{1.0}

      \triangular{-0.5}{0.5*1.7321}{1.04}
      \triangular{1.5}{0.5*1.7321}{1.04}
      \triangular{3.5}{0.5*1.7321}{1.04}
      \triangular{-1.5}{-0.5*1.7321}{1.04}
      \triangular{0.5}{-0.5*1.7321}{1.04}
      \triangular{2.5}{-0.5*1.7321}{1.04}
      \triangular{-0.5}{-1.5*1.7321}{1.04}
      \triangular{1.5}{-1.5*1.7321}{1.04}
      \triangular{3.5}{-1.5*1.7321}{1.04}

      \draw[mygreen,thick] (-2.2,0.5*1.7321)--(4.2,0.5*1.7321);
      \draw[mygreen,thick] (-2.2,-0.5*1.7321)--(4.2,-0.5*1.7321);
      \draw[mygreen,thick] (-2.2,-1.5*1.7321)--(4.2,-1.5*1.7321);

      \blackline{-1.0-0.1667}{-0.2887}{2.067}{2.3}
      \blackline{1.0-0.1667}{-0.2887}{3.3}{2.3}
      \blackline{3.0-0.1667}{-0.2887}{3.3}{2.3}
      \blackline{5.0-0.1667}{-0.2887}{3.3}{-1.267}

      \Tensor{-0.5}{0.5*1.7321}{0.15}{0.2}
      \Tensor{1.5}{0.5*1.7321}{0.15}{0.2}
      \Tensor{3.5}{0.5*1.7321}{0.15}{0.2}
      \Tensor{-1.5}{-0.5*1.7321}{0.15}{0.2}
      \TensorA{0.5}{-0.5*1.7321}{0.15}{0.2}
      \Tensor{2.5}{-0.5*1.7321}{0.15}{0.2}
      \Tensor{-0.5}{-1.5*1.7321}{0.15}{0.2}
      \Tensor{1.5}{-1.5*1.7321}{0.15}{0.2}
      \Tensor{3.5}{-1.5*1.7321}{0.15}{0.2}

      \draw[dashed, myorange] (0.5,-0.5*1.7321) circle (1.1);

     \end{tikzpicture}.
    \end{array}
\end{equation}
In this sense the PESS constructions can be viewed as approximations of iPEPS.

To benchmark the PESS and iPEPS methods, Fig.~\ref{Gs_EM} compares the results obtained for the ground-state energy and magnetization obtained using the 3-site PESS (3-PESS) and the 9-PESS in our previous work~\cite{Liao2017} with the iPEPS used in our present study, also comparing optimization by the SU and AD methods. It is clear that, for the 3-PESS with the same bond dimension $D$, the AD method achieves a lower energy and magnetization than the SU method. Further, the ground-state energy and magnetization of the iPEPS optimized by AD are lower than those of the 3-PESS and 9-PESS optimized by either method, indicating that it is a superior variational state. Physically, the fact that the iPEPS wavefunction is formed by contracting the simplex and physical tensors of the 9-PESS means that it captures not only the bipartite entanglement of the original iPEPS wave function but also multipartite entanglement. Importantly, the convergence behavior of the two approaches with $D$ does not show significant differences, with both appearing to converge algebraically in $1/D$, and hence these results do not alter the conclusions of the previous work~\cite{Liao2017}.

Whereas the $\sqrt{3}$$\times$$\sqrt{3}$ ordered phase can only be represented by a 9-site unit cell, the QSL and the $q = 0$ ordered phase can be represented using both 3-site and 9-site cells. We have verified that both constructions yield identical ground-state energies and magnetizations, and in Fig.~\ref{StaticUnitCell} we show that they also yield the same equal-time structure factor. Thus we employ the 3-site unit cell in all our studies of these phases to save computational resources. Finally, we remark that mapping the kagome geometry to the square lattice entails a loss of $C_3$ symmetry in exchange for a very significant gain in computational convenience at the CTMRG step. Where the loss of symmetry is a small penalty that is readily benchmarked, the efficiency of the resulting iPEPS construction is what allowed us to access sufficiently high $D$ values in our calculations that we could identify the true nature of the KHAF spectrum.

\subsection{Excited States}

To calculate the spin excitation spectrum, we utilize the single-mode iPEPS Ansatz, which has been applied successfully to study a number of paradigm models in quantum magnetism~\cite{Vanderstraeten2015, Vanderstraeten2019, Ponsioen2020, Ponsioen2022, chi2022, chi2024, wang2024}. In this Ansatz, an excited state $|\Phi_{\boldsymbol k}(B)\rangle$ with momentum $\boldsymbol{k}$ is represented as a linear superposition of states $|\Phi_{\boldsymbol r}(B)\rangle$, which is constructed by substituting the local tensor $A$ at position $\boldsymbol r$ in $|\Psi(A)\rangle$ with a single new tensor $B$ to obtain
\begin{align}
|\Phi_{\boldsymbol k}(B)\rangle &= \sum_{\boldsymbol{r}} e^{i\boldsymbol{k}\cdot \boldsymbol r} |\Phi_{\boldsymbol r}(B)\rangle \\
&= \sum_{\boldsymbol{r}} e^{i\boldsymbol{k}\cdot \boldsymbol r}
\begin{array}{l}
\begin{tikzpicture}[every node/.style={scale=1},scale=0.5]
    \draw[step=1 cm,gray] (-2.4,-2.4) grid (2.4,2.4);
    \Tensor{-1}{0}{0.15}{0.2}
    \Tensor{-1}{1}{0.15}{0.2}
    \Tensor{-1}{2}{0.15}{0.2}
    \Tensor{-1}{-1}{0.15}{0.2}
    \Tensor{-1}{-2}{0.15}{0.2}
    \Tensor{-2}{0}{0.15}{0.2}
    \Tensor{-2}{1}{0.15}{0.2}
    \Tensor{-2}{2}{0.15}{0.2}
    \Tensor{-2}{-1}{0.15}{0.2}
    \Tensor{-2}{-2}{0.15}{0.2}
    \CenterB{0}{0}{0.15}{0.2}
    \Tensor{0}{1}{0.15}{0.2}
    \Tensor{0}{2}{0.15}{0.2}
    \Tensor{0}{-1}{0.15}{0.2}
    \Tensor{0}{-2}{0.15}{0.2}
    \Tensor{1}{0}{0.15}{0.2}
    \Tensor{1}{1}{0.15}{0.2}
    \Tensor{1}{2}{0.15}{0.2}
    \Tensor{1}{-1}{0.15}{0.2}
    \Tensor{1}{-2}{0.15}{0.2}
    \Tensor{2}{0}{0.15}{0.2}
    \Tensor{2}{1}{0.15}{0.2}
    \Tensor{2}{2}{0.15}{0.2}
    \Tensor{2}{-1}{0.15}{0.2}
    \Tensor{2}{-2}{0.15}{0.2}
\end{tikzpicture}
\end{array}.
\end{align}

Because all excited states are orthogonal to the ground state, $|\Phi_{\boldsymbol r}(B)\rangle$ is confined to the tangent space of the ground state, namely,
\begin{align}
  \langle \Phi_{\boldsymbol r}(B)|\Psi(A)\rangle = 0,
\end{align}
and this condition determines a set of basis vectors orthogonal to the ground state, $|\Phi_{\boldsymbol k}(\tilde B_m)\rangle$. In this basis, one calculates the effective Hamiltonian matrix
\begin{align}
  H^\mathrm{eff}_{\boldsymbol{k},mn} = \langle\Phi_{\boldsymbol{k}} (\tilde B_m) |H| \Phi_{\boldsymbol{k}} (\tilde B_n) \rangle
\label{Heff_matrix}  
\end{align} 
and the norm matrix 
\begin{align} 
  N^\mathrm{eff}_{\boldsymbol{k},mn} = \langle\Phi_{\boldsymbol{k}} (\tilde B_m)| \Phi_{\boldsymbol{k}} (\tilde B_n) \rangle.
\label{Neff_matrix}
\end{align} 
Here we employ AD~\cite{Liao2019,chi2022,Ponsioen2022} to speed up the most time-consuming calculations of these matrices. 
With these we are in a position to solve the generalized eigen-equation 
\begin{align} 
  \sum_n H^\mathrm{eff}_{\boldsymbol{k},mn} v_{np} = \sum_n E_p N^\mathrm{eff}_{\boldsymbol{k},mn} v_{np}
\end{align}
to obtain the excited energies $E_m$ and the corresponding excited states $|\Phi_{\boldsymbol k}(B_m)\rangle$, where $B_m = \sum_n \tilde B_n v_{nm}$. Finally, we calculate the spectral function using the expression
\begin{align}
\label{eq:delta}
& S^{\alpha \alpha}(\boldsymbol{k},\omega) = \sum_{m} |\langle \Phi_{\boldsymbol k}(B_m) | S^{\alpha}_{\boldsymbol{k}}  |  \Psi(A) \rangle|^2 \delta(\omega - E_m + E_0),
\end{align}
where we apply a Lorentzian broadening $\eta$ in evaluating the $\delta$-function. 

To calculate the static structure factor
\begin{align}
  & S^{\alpha \alpha}(\boldsymbol{k}) = \langle \Psi(A) | S^{\alpha}_{-\boldsymbol{k}} S^{\alpha}_{\boldsymbol{k}}  | \Psi(A) \rangle,
\end{align}
we treat $S^{\alpha}_{\boldsymbol{k}} | \Psi(A) \rangle$ as an excited state $|\Phi_{\boldsymbol k}(B)\rangle$, where $B = S_r^{\alpha}A$, and then compute the overlap $\langle\Phi_{\boldsymbol{k}} (B)| \Phi_{\boldsymbol{k}} (B) \rangle$ in the same way as the norm matrix.

\begin{figure}[t]
\centering
\includegraphics[width=0.96\columnwidth]{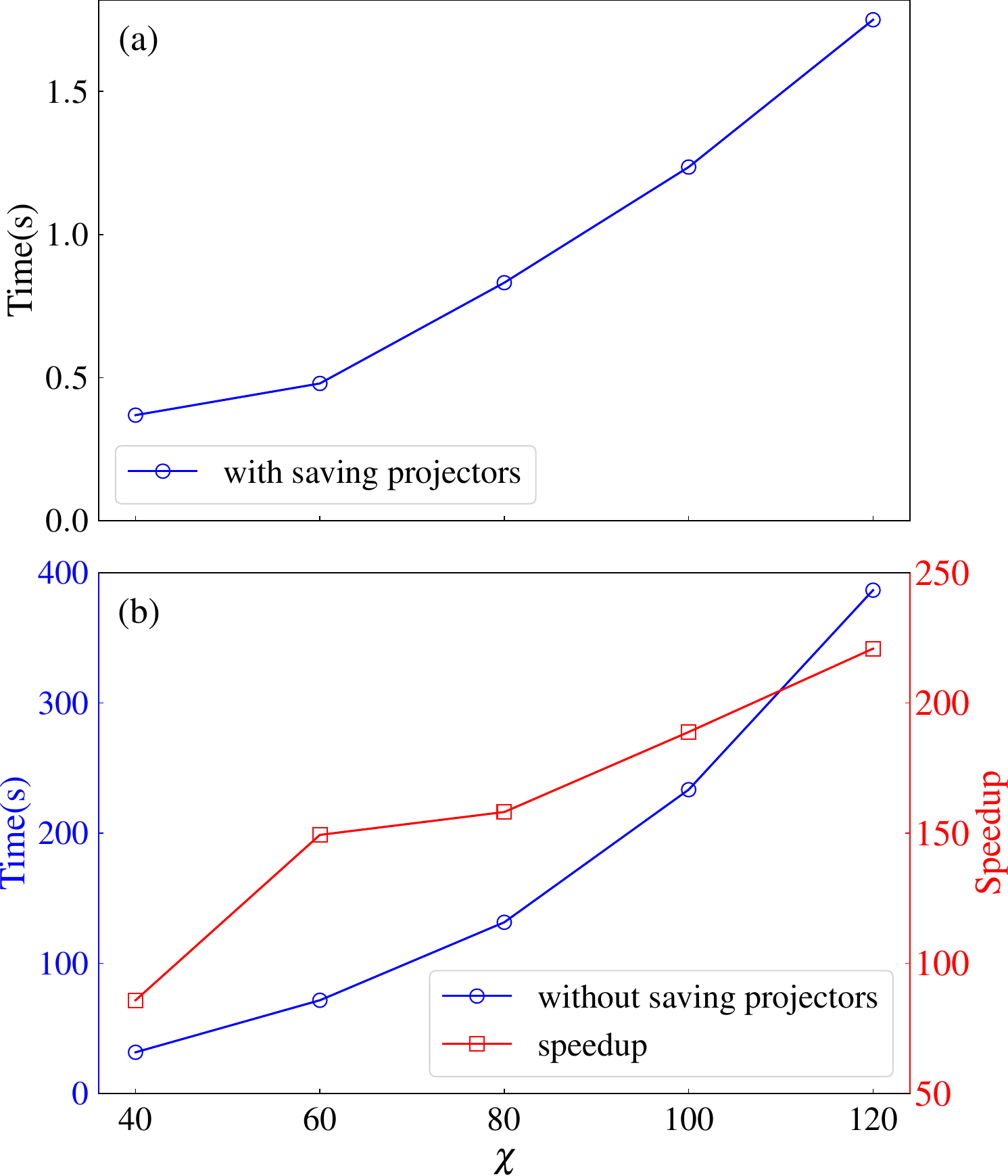}
\caption{Comparison of the computation times required for 20 CTMRG steps in the calculation of the excitation spectrum for the $D = 4$ iPEPS on an NVIDIA A100 GPU, with and without employing the method of saving projectors. Results are shown as a function of the bond dimension, $\chi$, of CTMRG. (a) Computation time $T_1$ achieved by saving projectors. (b) Computation time $T_0$ without saving projectors and the speed-up factor, defined as $T_0/T_1$.
\label{CTMRG_time}}
\end{figure}

\section{Development of Improved Tensor-Network Methods for Computing Excitation Spectra}
\label{sec2}

Calculating the full spectrum of excitations for a QSL in the infinite system is an extremely challenging problem, which is why very few comparable results may be found in the literature to date. These problems are manifest as computational limitations, meaning that they are quantitative rather than qualitative in nature. To overcome this bottleneck, we have made three specific types of improvement to the tensor-network method for calculating excitation spectra in order to enhance computational efficiency and reduce memory overhead.

\begin{figure*}[t]
  \centering
  \includegraphics[width=0.98\textwidth]{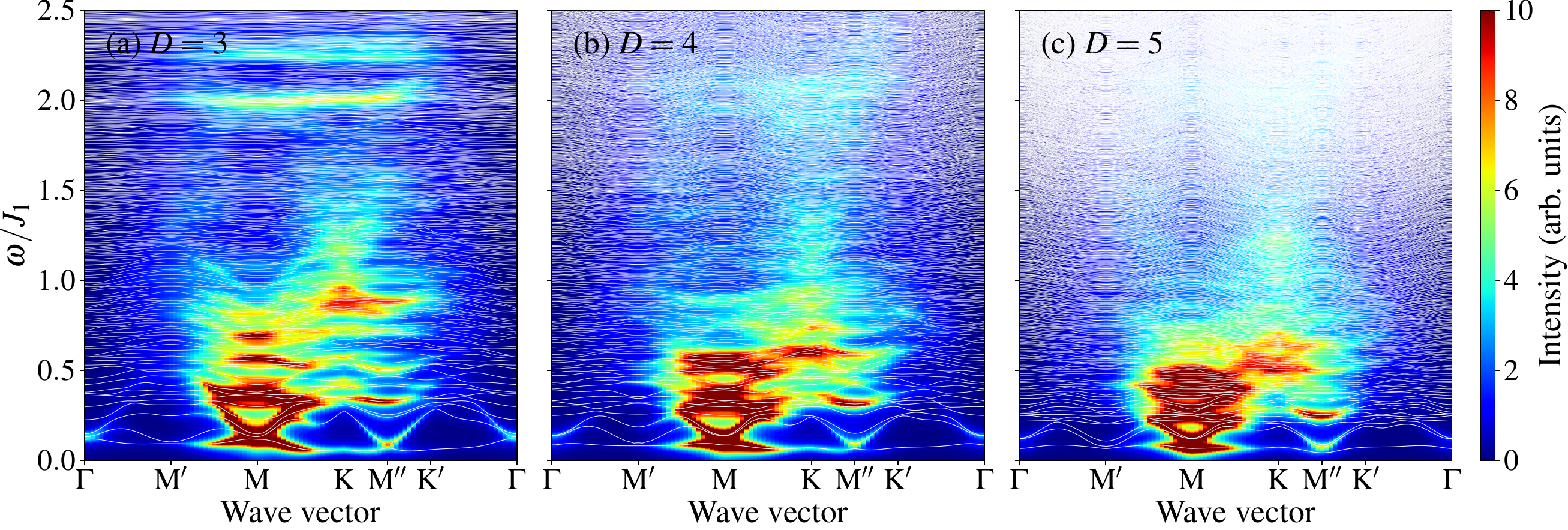}
  \caption{Spin excitation spectra of the $J_2 = 0$ KHAF computed with $\eta = 0.02$ for $D = 3$ (a), 4 (b), and 5 (c).
\label{SpectraVsD}}
\end{figure*}

First, during the CTMRG process for calculating the spectrum, we avoided recalculating the projectors at each step \cite{chi2022,Ponsioen2022,tu2024}. This is the single most time-consuming operation due to its reliance on singular value decomposition (SVD). We observe that these projectors depend only on the ground-state iPEPS and are independent of the $B$-tensors representing the excited states. As a consequence, they need be computed just once during the CTMRG process for the ground state, and can then be saved and reused directly in subsequent calculations of the excitation spectrum. This eliminates the need for repeated SVD operations, which are not only computationally expensive but also challenging to parallelize on a modern processor such as a GPU. The key improvement is practical, in that this reduces the calculations to matrix multiplications, which are highly parallelizable and hence can achieve significant acceleration on a GPU, meaning by factors of several hundred. This optimization therefore reduces the overall computation time quite dramatically. Figure~\ref{CTMRG_time} shows the time required to execute 20 CTMRG steps in calculating the excitation spectrum for the $D = 4$ iPEPS on an NVIDIA A100 GPU, with and without the saving of projectors. It is evident that this improvement achieves a speed-up that approaches a factor of 200, with the acceleration becoming more significant as the CTMRG bond dimension ($\chi$) is increased. 

Second, when using AD to compute the effective Hamiltonian matrices $H^\mathrm{eff}_{\boldsymbol{k},mn}$ and the norm matrices $N^\mathrm{eff}_{\boldsymbol{k},mn}$, it has been conventional to store all intermediate variables throughout the CTMRG iteration process. This results in a very high memory overhead, even with the checkpointing technique~\cite{Liao2019}. To reduce memory usage, we employ the fixed-point CTMRG method~\cite{Liao2019}, which stores only the intermediate variables of a single CTMRG step, thereby reducing the memory overhead by a factor of $N$, where $N$ is the number of CTMRG iterations. This procedure enables us to perform excitation spectrum calculations on a GPU with only tens of gigabytes of memory. 

Third, to obtain the fixed-point corner- and edge-transfer matrices in the fixed-point CTMRG method, it is necessary to eliminate the phase arbitrariness of the unitary matrices $U$ and $V$ in the SVD operation at each CTMRG step. In general, $U$ and $V$ can be multiplied by any diagonal phase matrix, $\Lambda = \mathrm{diag}\,(e^{i\alpha_1},\cdots, e^{i\alpha_n})$, while maintaining a constant $USV^{\dagger}$. Here we eliminate the phase arbitrariness by recording the location $x_j$ and the phase of the element $U_{x_j,j}$ with the largest absolute value for each column $j$ of $U$ during each SVD step. In the next SVD operation, $U^{\prime}S^{\prime}V^{\prime\dagger}$, we apply a matrix $\Lambda^{\prime}$ to $U^{\prime}$ and $V^{\prime}$, ensuring that $U^{\prime}_{x_j,j}$ and $U_{x_j,j}$ have the same phase. We then update the location $x^{\prime}_j$ and the phase of the element with the largest absolute value for each column in the new $U^{\prime}$, and repeat this process until all SVD operations are completed. We find that this method of eliminating phase arbitrariness is more stable than the approach proposed in previous work~\cite{Ponsioen2022}, which simply keeps the sign of the element with the largest absolute value in each column vector of $U$ positive at each SVD operation, and hence enhances the efficiency of our calculations.

\begin{figure}[t]
  \centering
  \includegraphics[width=0.96\columnwidth]{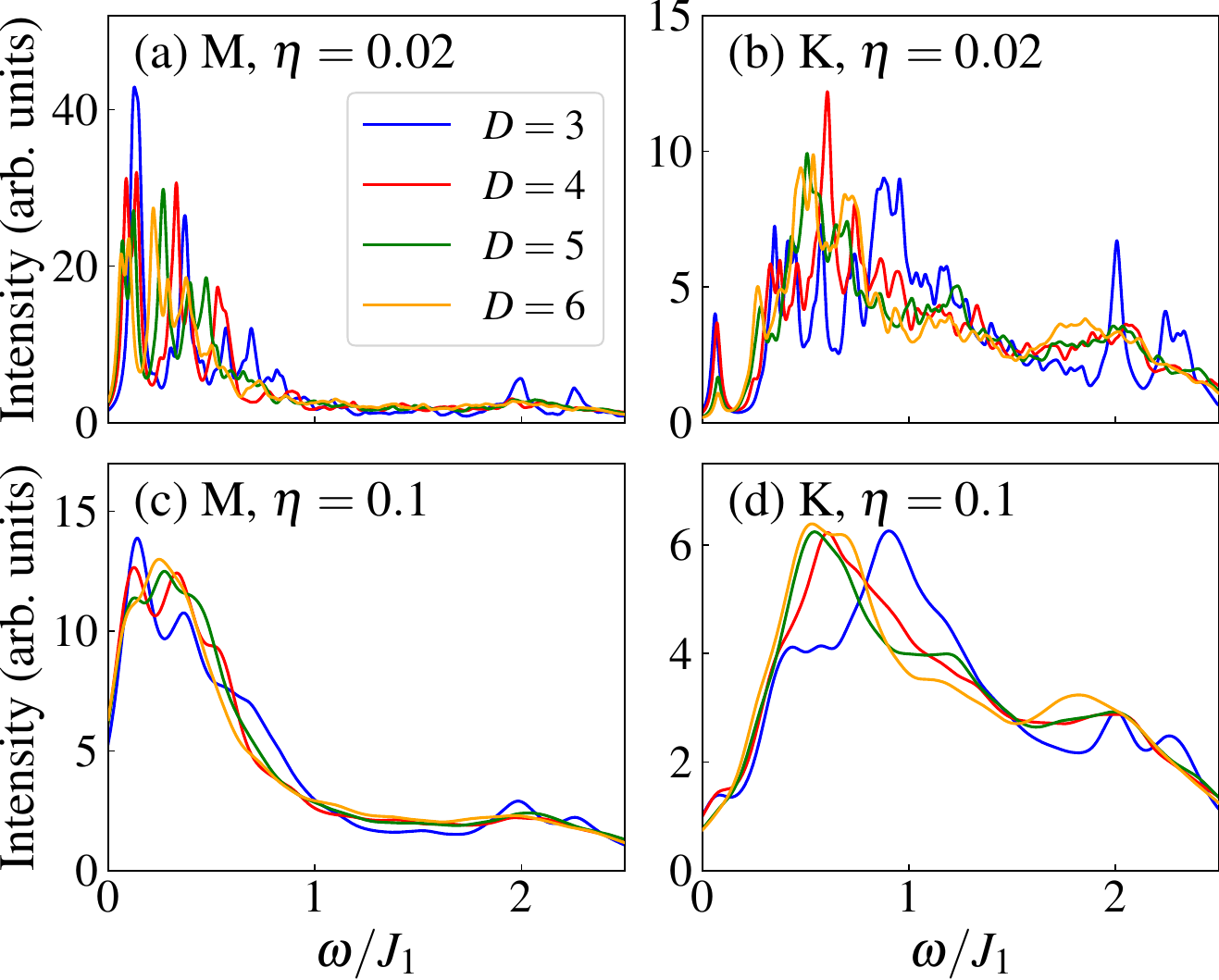}
  \caption{Spin excitation spectra of the $J_2 = 0$ KHAF computed with $D = 3$, 4, 5 and 6. Results with $\eta = 0.02$ (a,b) and $\eta = 0.1$ (c,d) are shown at the M point (a,c) and at the K point (b,d).}
\label{KpointSpectraVsD}
\end{figure}

\section{Evolution of QSL Spectrum with bond dimension}
\label{sec3}

The iPEPS method performs calculations directly for the infinite system, thereby avoiding any effects of a finite system size. Instead, the truncation parameter is the finite bond dimension, $D$, and the most important aspect of our study is to establish that $D$ is sufficiently large to capture the physics of the spectral function. Particularly in the QSL phase, it is already known~\cite{Liao2017} that the ground-state magnetization vanishes only in the limit $D \rightarrow \infty$. Excited-state calculations are more strongly limited by $D$ than ground-state computations: the latter scale with $O(D^{12})$, while the number of excited basis states scales as $dD^{4}$, leading to an overall computational cost of $O(D^{16})$. As a result, our spectral function calculations can access only rather limited values of $D$.

Figure \ref{SpectraVsD} shows the spectrum in the QSL phase for bond dimensions $D = 3$, 4, and 5. As noted in the main text, increasing $D$ makes the energy levels become progressively denser, which is a characteristic feature of a continuum spectrum. Beyond this effect, some nontrivial alterations are evident on passing from $D = 3$ to 4 in the band widths of certain low-lying excited states and in the intensities of higher-lying ones. In general, the changes in passing from $D = 4$ to 5 are limited to only minor alterations in level energies and intensities. However, we draw attention to the fact that the incommensurate band minima close to the M$^{\prime\prime}$ point visible at $D = 3$ and 4 [Figs.~\ref{SpectraVsD}(a,b)] and in Fig.~3(c) of the main text ($D = 4$) clearly converge to a single minimum at M$^{\prime\prime}$ when $D = 5$ [Fig.~\ref{SpectraVsD}(c)].

For a more quantitative comparison, in Fig.~\ref{KpointSpectraVsD} we show constant-wave-vector excitation spectra at the M and K points for the $D = 3$, 4, 5, and 6. Here we have also shown our results for two different choices of $\eta$, whose effect we discuss in detail in Sec.~\ref{sec4}. Small $\eta$ values reveal peaks at every white line in Fig.~\ref{SpectraVsD} whereas larger $\eta$ values give a smoother response function where the levels merge into broader peaks. Again we observe some clear shifts in peak positions and weights between $D = 3$ and 4, at both small and large $\eta$, whereas these effects are far more limited from $D = 4$ to 6. In particular, the overall distribution of spectral weight remains essentially the same for $D \ge 4$, which is clear to see at larger $\eta$. This means that calculations even with $D = 4$ already capture the weight distribution rather well, despite this bond dimension being relatively small for typical ground-state calculations. This result highlights the power of the single-mode iPEPS Ansatz in reproducing the qualitative and quantitative properties of the excitation spectrum.

\begin{figure}[t]
  \centering
  \includegraphics[width=\columnwidth]{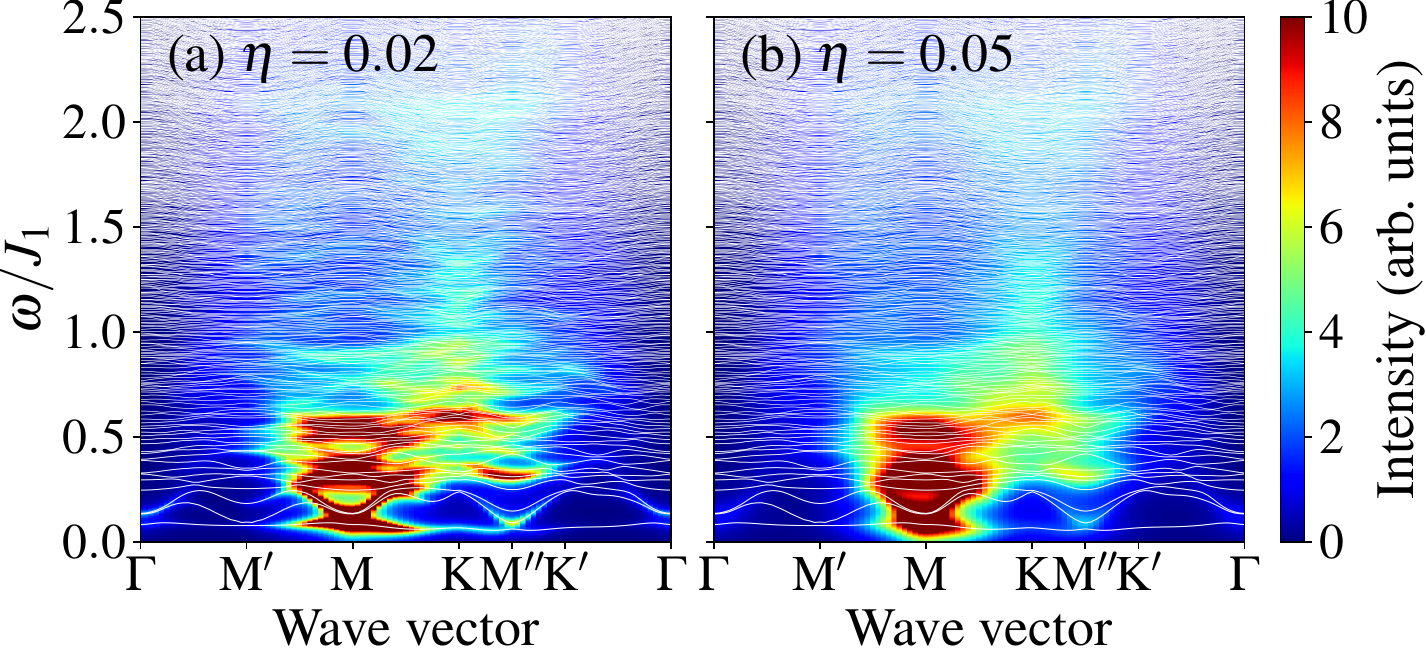}
  \caption{Spin excitation spectra of the $J_2 = 0$ KHAF computed with $D = 4$ to compare broadening $\eta = 0.02$ (a) with $\eta = 0.05$ (b).
\label{SpectraVsEta}}
\end{figure}

\section{Evolution of QSL Spectrum with $\eta$}
\label{sec4}

\begin{figure}[t]
  \centering
  \includegraphics[width=\columnwidth]{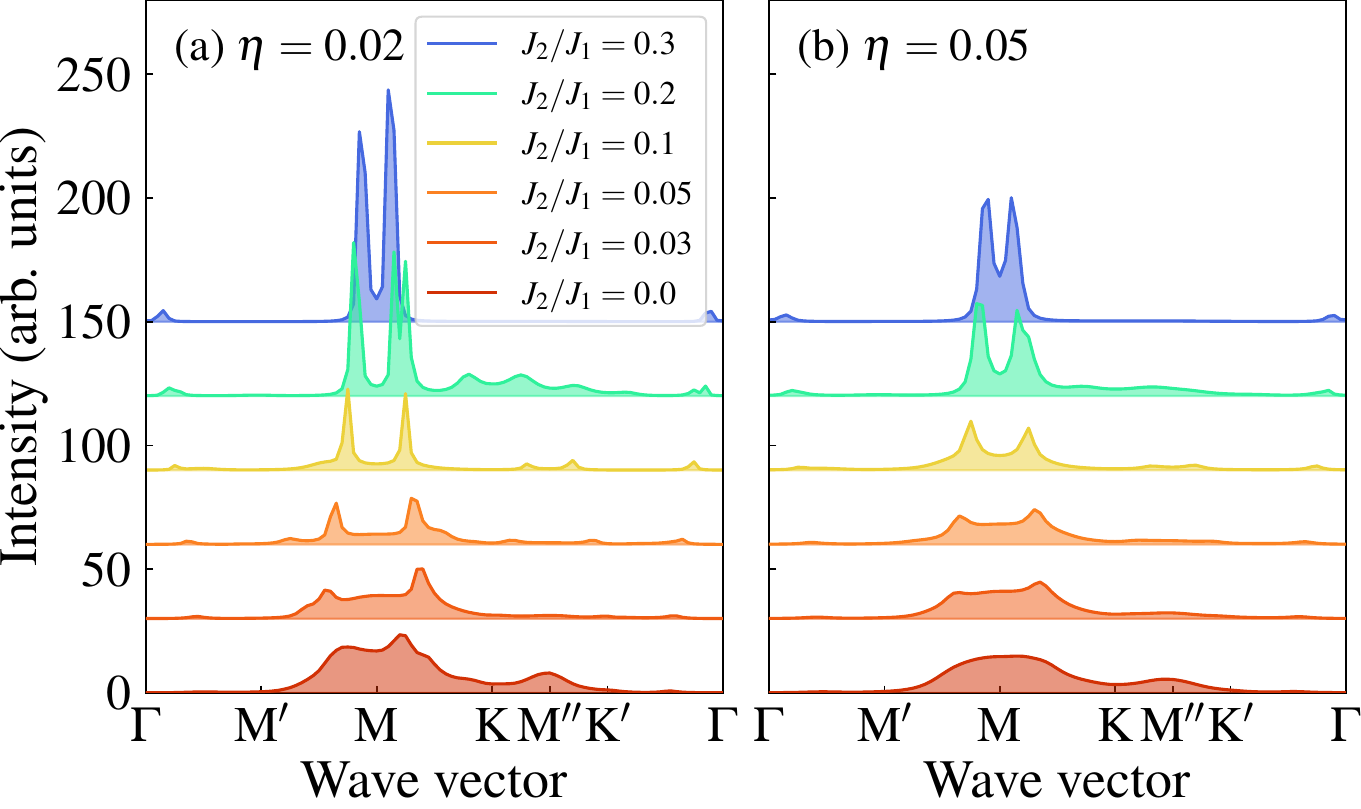}
  \caption{Spin excitation spectra shown at fixed energy $\omega = 0.3J_1$ for different values of $J_2$, computed with $D = 4$ for broadenings $\eta = 0.02$ (a) and $\eta = 0.05$ (b).
\label{LowEnergyVsEta}}
\end{figure}

The spectral function we compute consists of a sequence of levels (poles) that are continuous in wave vector but discrete in energy, each energetic $\delta$-function [Eq.~\eqref{eq:delta}] being associated with a spectral weight. To display our results we applied a Lorentzian broadening, $\eta$, which can be considered as analogous to the instrumental resolution function in an experimental measurement. Figure \ref{KpointSpectraVsD} showed the effects of $\eta$ in a constant-wave-vector spectrum, Fig.~\ref{SpectraVsEta} shows its effect in the complete spectral function, and Fig.~\ref{LowEnergyVsEta} shows how it affects a constant-energy spectrum. We stress again that $\eta$ is applied in a post-processing step and has no influence on the tensor-network calculation of the spectrum, only on the appearance of the results. Most importantly, $\eta$ does not shift the peak position of the Lorentzian and hence it cannot have the effect of closing an energy gap. The effect of $\eta$ in smoothing the spectral function will be valuable for future comparison with experiment,  but does not affect the overall distribution of spectral weight, which can be considered as independent of $\eta$ (but weakly dependent on $D$, as shown in Sec.~\ref{sec3}). 

\begin{figure*}[t]
\centering
\includegraphics[width=0.98\textwidth]{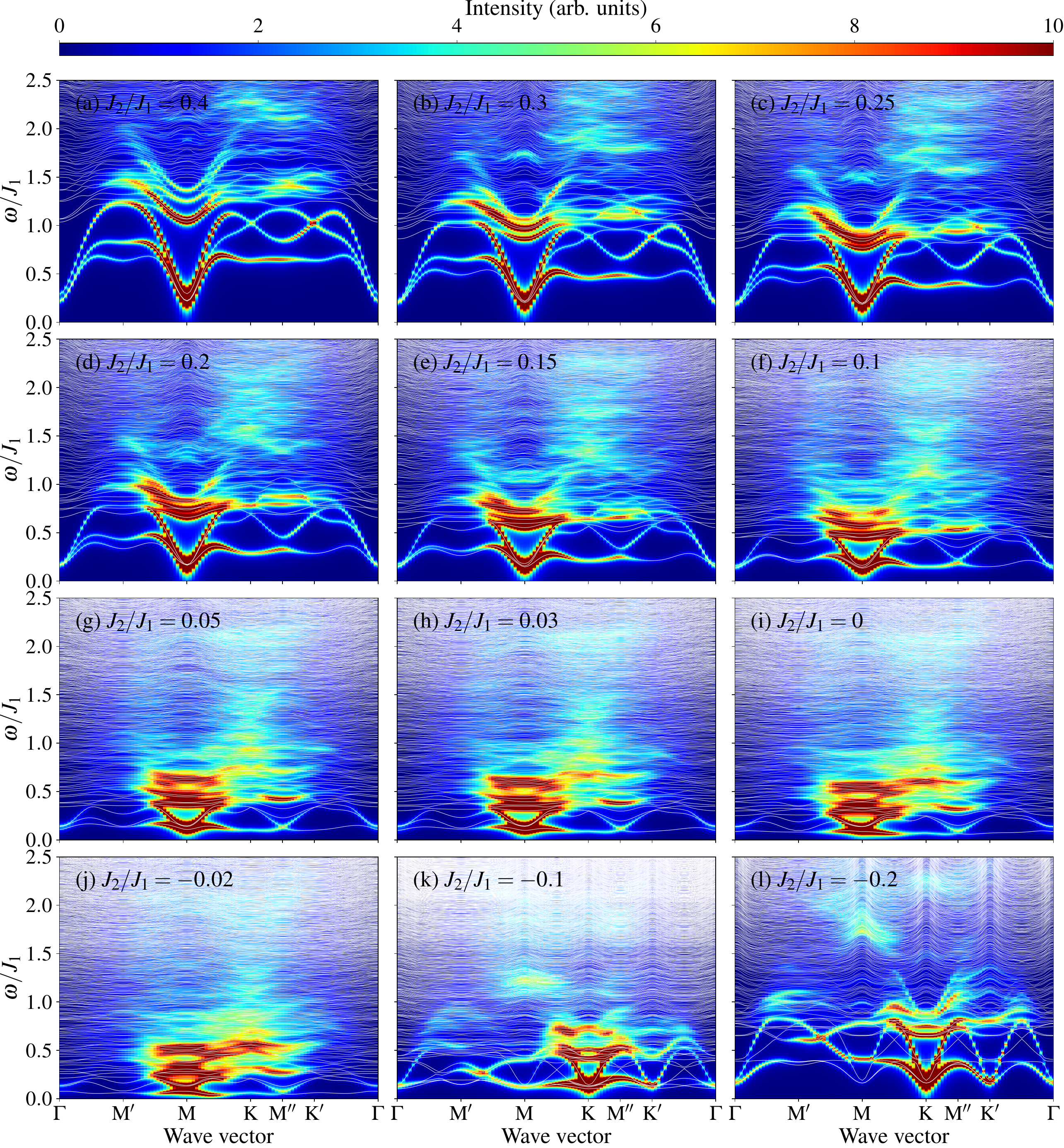}
\caption{Full spin excitation spectra for the $J_1$-$J_2$ KHAF, computed with $D = 4$ and $\eta = 0.02$ to compare a wide range of $J_2$ values.
\label{FullJ2Spectra}}
\end{figure*}

\section{Evolution of Excitation Spectra with $J_2$}
\label{sec5}

In Fig.~1 of the main text we showed full spin excitation spectra for the $J_1$-$J_2$ KHAF at six selected values of $J_2$. In Fig.~\ref{FullJ2Spectra} we show the spectra for 12 $J_2$ values in order to provide a more systematic interpolation of spectral features across the phase diagram. In the main text we focused primarily on describing the low-energy behavior of the spin-excitation spectrum, and the sequence of panels in Fig.~\ref{FullJ2Spectra} reinforces this discussion. 

Here we remark in addition that the high-energy sector also provides valuable information. In the $q = 0$ phase realized at positive $J_2$, the Goldstone mode is located at the M point while the high-energy continuum is centered at the K point. At large $J_2$ we find not only the conventional magnon branches but also distinct higher-lying modes with a clear two-magnon interpretation, some of which have been discussed as longitudinal or ``Higgs'' modes. As $J_2$ is decreased, the low-energy weight at M remains essentially fixed whereas the continuum shifts gradually downward in energy, spreading to all wave vectors and subsuming the formerly distinct excited levels. Thus, unlike the square lattice \cite{DallaPiazza2015,Shaik2025}, deconfinement and consequent continuum formation in the KHAF appear to be a largely wave-vector-independent consequence of exceeding a threshold energy that is driven towards zero by decreasing $J_2$. 

This trend continues into the QSL phase at $J_2/J_1 < 0.05$, producing a nearly uniform continuum at all energies above the finite-$D$ gap. When $J_2$ becomes sufficiently negative that the system enters the $\sqrt{3}$$\times$$\sqrt{3}$ phase, the Goldstone mode moves from M to K and the high-energy continuum relocates from K to M. These trends suggest that one interpretation for the emergence of the nearly uniform continuum in the QSL is that it arises from a competition of spin correlations related to the $q = 0$ and $\sqrt{3}$$\times$$\sqrt{3}$ ordered states. Within the QSL phase, decreasing $J_2$ from $0.03J_1$ to $-0.02J_1$ [Figs.~\ref{FullJ2Spectra}(h-j)] pushes the continuum at K to lower energies, whereas the M-point weight remains largely unchanged. This rather low energy scale for the onset of the uniform continuum coincides with the value that has been interpreted as a spin gap in some experimental measurements~\cite{Han2012,breidenbach2025}. 

\begin{figure*}[t]
\centering
\includegraphics[width=0.98\textwidth]{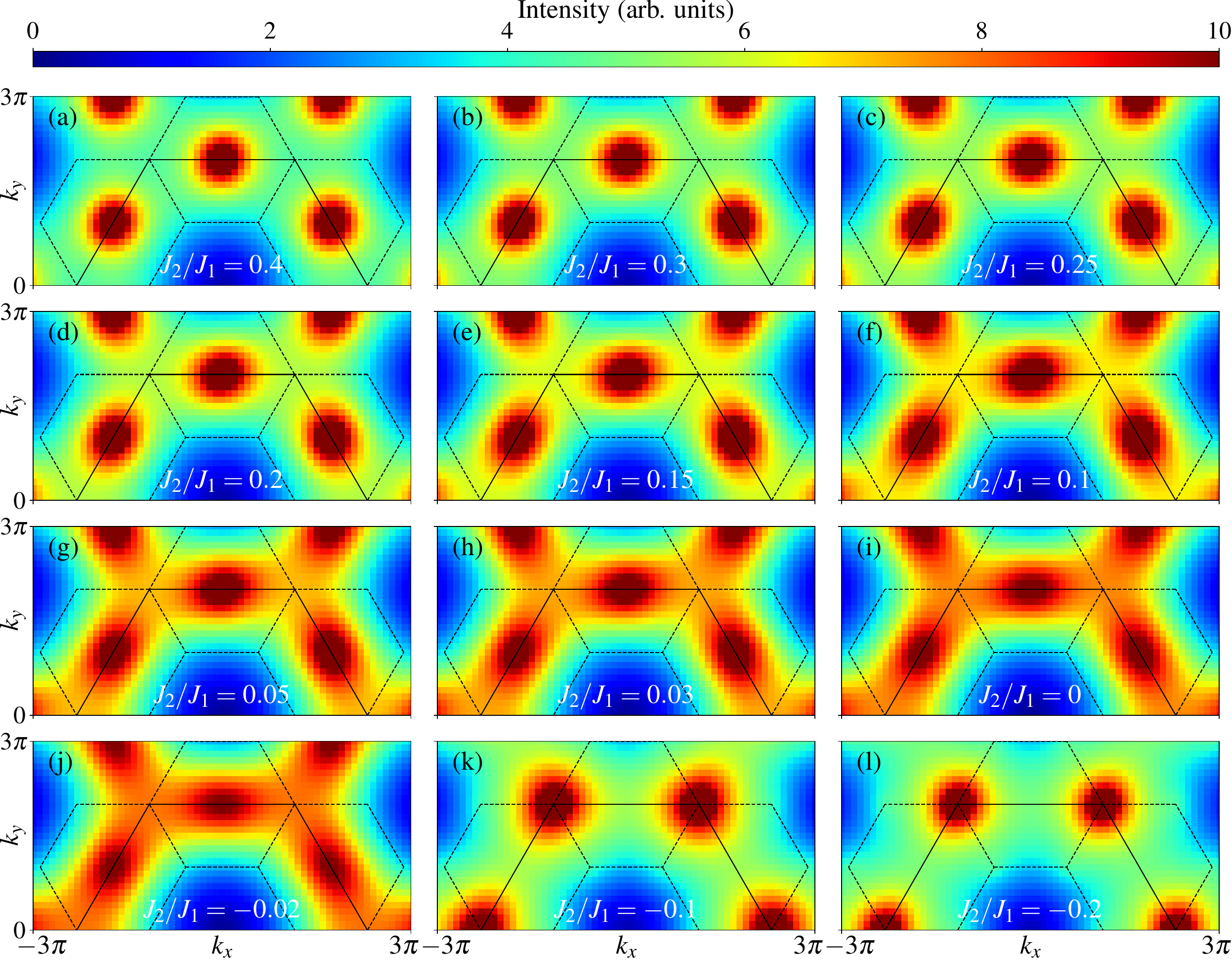}
\caption{Static spin structure factors for the $J_1$-$J_2$ KHAF computed with different values of $J_2$.
\label{J2BZStatic}}
\end{figure*}

\section{Evolution of Static Spin Structure Factor with $J_2$}
\label{sec6}

In Fig.~2(a) of the main text we showed the static spin structure factor for the $J_1$-$J_2$ KHAF at $J_2 = 0$. For further systematic 
insight into the nature of the QSL phase, in Fig.~\ref{J2BZStatic} we show the static structure factor over a wide range of $J_2$ values. 
Well in the $q = 0$ ordered phase [Figs.~\ref{J2BZStatic}(a-c)], the static structure factor has a sharp peak at the M point, consistent with our results for the spectral function in Fig.~1 of the main text. As $J_2$ decreases, these peaks become increasingly broad and spread toward the K point [Figs.~\ref{J2BZStatic}(d,e)]. Within the QSL phase, some spectral weight has already accumulated at the K point, but the strongest weight remains at the M point [Figs.~\ref{J2BZStatic}(f-i)]. Only when the system enters the $\sqrt3$$\times$$\sqrt3$ phase does the peak shift discontinuously to the K point, marking the transition to a different ordered state.

We remark for completeness that the dynamical spectral functions shown in Figs.~2(b-d) of the main text break the $C_3$ symmetry of the system. This effect, which is not visible in the static structure factor, is a consequence of the fact that the iPEPS Ansatz does not preserve this symmetry, which is broken at the step where we combine the kagome lattice into a square lattice [Eq.~\eqref{eq1}]. We overlook this minor quantitative discrepancy in exchange for the benefits it affords in the scope of our spectral calculations. 

\section{Linear Spin-Wave Theory}
\label{sec7}

\begin{figure}[t]
  \centering
  \includegraphics[width=0.45\textwidth]{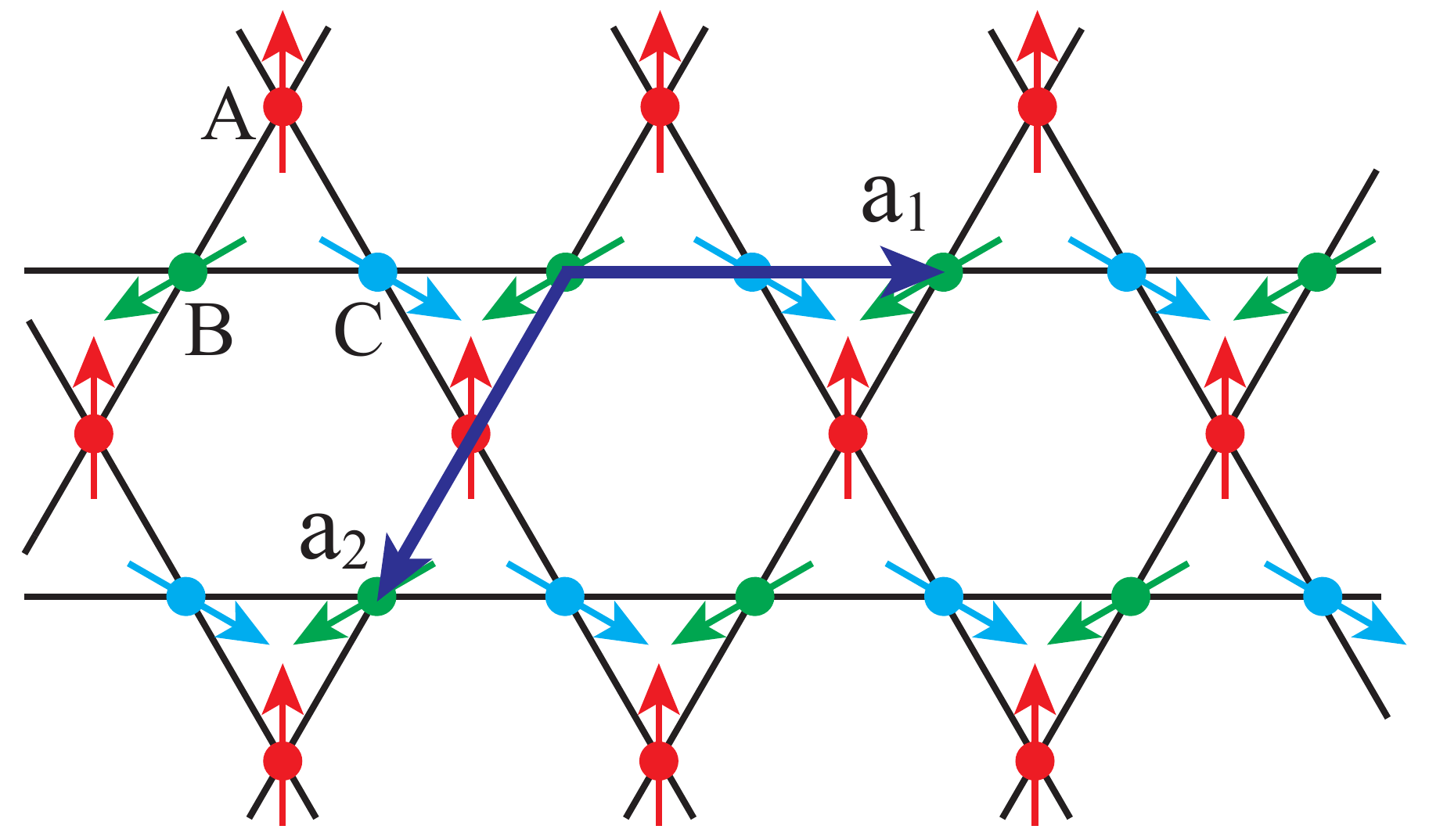}
  \caption{Schematic representation of the spin configuration in the $q = 0$ magnetically ordered state.
\label{kagome_q0}}
\end{figure}

\begin{figure*}[t]
\centering
\includegraphics[width=0.98\textwidth]{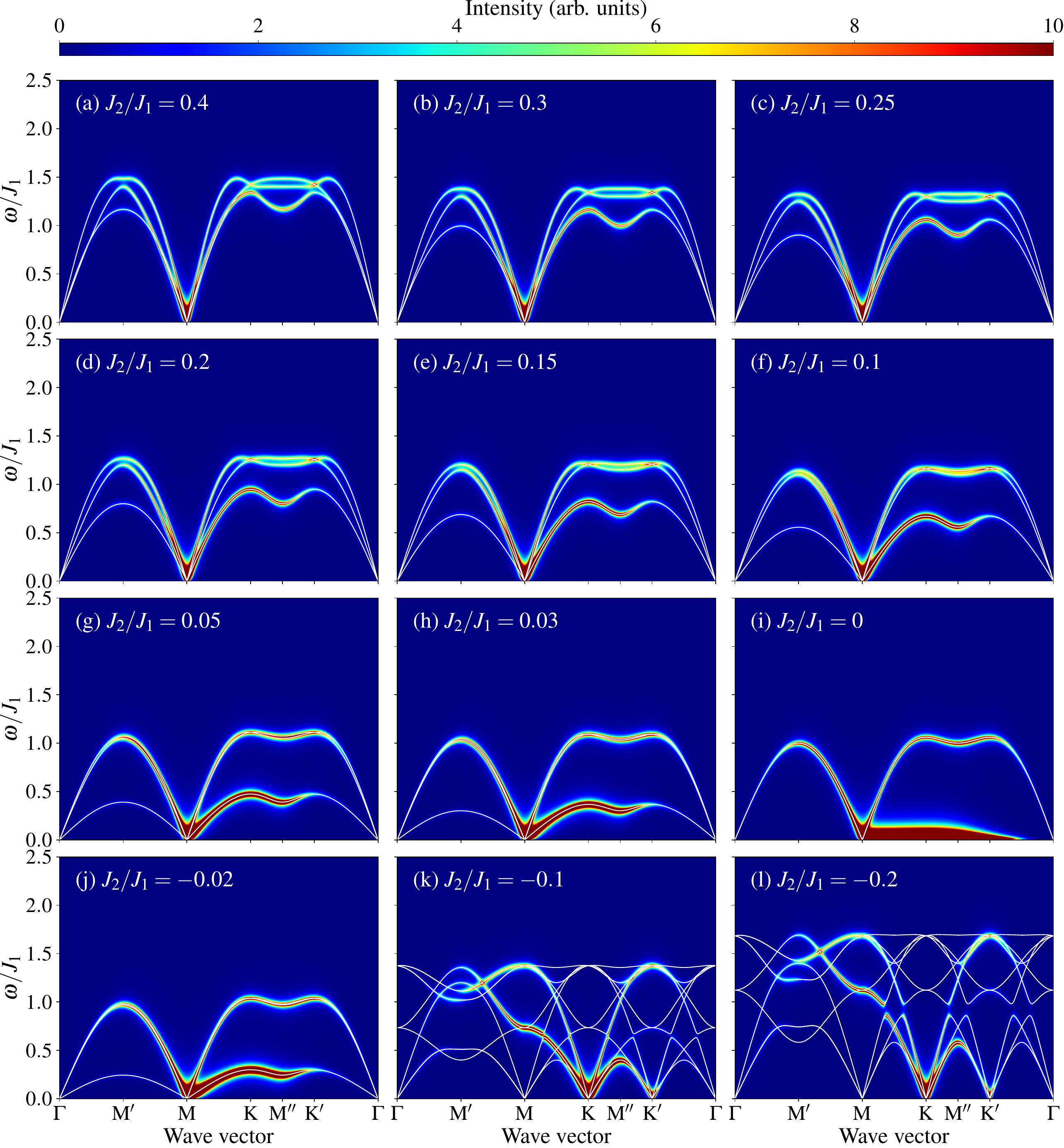}
\caption{Spectral functions obtained from linear spin-wave theory for the $J_1$-$J_2$ KHAF. White lines denote the spin-wave dispersion relations and intensities were obtained using $\eta = 0.02$.
\label{LSW}}
\end{figure*}

In the main text we discuss how our calculated spectra differ from the results of linear spin-wave theory (LSWT), which can always be taken as a baseline comparison for the spectra of ordered magnetic phases. Here we document our LSWT calculations for the case of the $q = 0$ magnetically ordered phase. The Hamiltonian of the $J_1$-$J_2$ KHAF can be expressed in the form
\begin{equation}
  \label{eq:origin-Hamt}
 H = J_1 \sum_{\langle i j\rangle} \vec{S}_{i} \cdot \vec{S}_{j} + J_2 \sum_{\langle\langle i j\rangle\rangle} \vec{S}_{i} \cdot \vec{S}_{j} = {\textstyle \frac12} \sum_{i\alpha,j\beta} \mathbf{S}^{T}_{i\alpha} \,J_{i\alpha,j\beta} \,\mathbf{S}_{j\beta},
\end{equation}
where $\vec{S}_{i\alpha}$ is the spin operator in $i$th unit cell on sublattice index $\alpha$. This spin operator is expressed in terms of Holstein-Primakoff bosons as
\begin{equation}
\vec{S}_{i\alpha} \! = \! \sqrt{S \! - \! {\textstyle \frac12} n_{i\alpha}} e_{\alpha}^{-} a_{i\alpha} \! + \! e_{\alpha}^{+} a^{\dagger}_{i\alpha}  \sqrt{S \! - \! {\textstyle \frac12} n_{i\alpha}} \! + \! e_{\alpha}^{z} (S - n_{i\alpha}),
\label{eq:HP-transformation}
\end{equation}
where $e_{\alpha}^{z}$ is the classical ordering direction of the spin on sublattice $\alpha$, $e_{\alpha}^{\pm} = (e_{\alpha}^{x} \pm i \, e_{\alpha}^{y})/\sqrt{2}$, and $e_{\alpha}^{x,y}$ are basis vectors orthogonal to $e_{\alpha}^{z}$. Explicitly, we describe the $q = 0$ magnetically ordered state shown in Fig.~\ref{kagome_q0} by 
\begin{align}
&e_{\alpha}^{x} = [\cos \theta_{\alpha} \cos \phi_{\alpha}, \cos \theta_{\alpha} \sin \phi_{\alpha}, - \sin \theta_{\alpha}]^{T}, \nonumber \\
&e_{\alpha}^{y} = [-\sin \phi_{\alpha}, \cos \phi_{\alpha}, 0]^{T}, \\
&e_{\alpha}^{z} = [\sin \theta_{\alpha} \cos \phi_{\alpha}, \sin \theta_{\alpha} \sin \phi_{\alpha}, \cos \theta_{\alpha}]^{T}, \nonumber
\label{eq:rotation_basis}
\end{align}
i.e.~the spins form a coplanar $120^{\circ}$ structure in the $xz$ plane. With this convention we have $(\theta_A, \theta_B, \theta_C) = (0,2\pi/3,2\pi/3)$ and $(\phi_A, \phi_B, \phi_C) = (0,\pi,0)$. 

Substituting Eq.~(\ref{eq:HP-transformation}) into the Hamiltonian of Eq.~(\ref{eq:origin-Hamt}) and expanding up to quadratic order yields 
\begin{equation}
H \approx E_{cl} + H_1 + H_2
\end{equation}
with
\newpage
\begin{align}
E_{cl} & = {\textstyle \frac12} N_c\,S(S+1) \sum_{\alpha\beta,\delta}  J^{z,z}_{\delta, \alpha\beta}, 
\\
H_1 & = {\textstyle \frac12} S\sqrt{S} \sum_{i\alpha}\Big\{
\sum_{\beta,\delta}\Big(J^{-,z}_{\delta, \alpha\beta} + J^{z,-}_{-\delta, \beta\alpha} \Big) a_{i\alpha} 
\nonumber \\ 
& \quad\qquad\qquad + \sum_{\beta,\delta}\Big(J^{+,z}_{\delta, \alpha\beta}  + J^{z,+}_{-\delta, \beta\alpha} \Big) a^{\dagger}_{i\alpha}\Big\}, 
\\
H_2 &= {\textstyle \frac12} S \sum_{i\alpha,j\beta} \Big\{ \Big( 
  J^{-,+}_{i\alpha,j\beta}\, a_{i\alpha} a_{j\beta}^{\dagger} 
+ J^{-,-}_{i\alpha,j\beta}\, a_{i\alpha} a_{j\beta} + h.c. \Big) \nonumber\\
& \quad\qquad\quad\, - J^{z,z}_{i\alpha,j\beta}\, \Big( a_{j\beta} a_{j\beta}^{\dagger} + a_{i\alpha}^{\dagger} a_{i\alpha}\Big)\Big\},
\end{align}
where $J^{\mu,\nu}_{i\alpha,j\beta} \equiv (e_{\alpha}^{\mu})^T J_{i\alpha,j\beta} \, e_{\beta}^{\nu}$ and the linear term, $H_1$, vanishes identically for the classical $120^{\circ}$ spin structure [Fig.~\ref{kagome_q0}].

After Fourier transformation we obtain the quadratic term
\begin{align}
H_2 &= {\textstyle \frac12} S \sum_{k,\alpha\beta} \Big[ A_{k}^{\alpha \beta} a_{k\alpha}^{\dagger} a_{k\beta} + A_{-k}^{\beta\alpha} a_{-k\alpha} a_{-k\beta}^{\dagger} \nonumber \\
&+( B_{k}^{\alpha \beta} a_{k\alpha}^{\dagger} a_{-k\beta}^{\dagger} + h.c.  ) \Big] = {\textstyle \frac12} S \sum_{k} \Psi^{\dagger}_k M_k \Psi_k,
\label{eq:H2}
\end{align}
where 
\begin{align}
&\Psi_k = (a_{kA},a_{kB},a_{kC},a^{\dagger}_{-kA},a^{\dagger}_{-kB},a^{\dagger}_{-kC})^T, \nonumber\\
& M_k = \begin{bmatrix}
A_k & B_{k} \\
B^{\dagger}_{k} & A^T_{-k} 
\end{bmatrix}, \nonumber\\
&A_k = 2(J_1 + J_2)I_3 - B_k/3, \nonumber\\
& B_k = \frac{1}{2}\begin{bmatrix}
0      & b_{12}  & b_{13} \\
b_{21} & 0       & b_{23} \\
b_{31} & b_{32}  & 0      \\
\end{bmatrix}, 
\quad I_3 = \begin{bmatrix}
1 & 0 & 0 \\
0 & 1 & 0 \\
0 & 0 & 1 \\
\end{bmatrix}, \nonumber \\
& b_{12} = b_{21} = \,3[J_1 \cos q_2 + J_2 \cos(2q_1+q_2)], \nonumber \\
& b_{13} = b_{31} = -3[J_1 \cos (q_1 + q_2) + J_2 \cos(q_1-q_2)],\nonumber \\
& b_{23} = b_{32} = \,3[J_1 \cos q_1 + J_2 \cos(q_1+2q_2)], \nonumber \\
& q_1 = {\textstyle \frac12} \boldsymbol{k} \cdot {\bf a}_1, \quad {\rm and} \quad q_2 = {\textstyle \frac12} \boldsymbol{k} \cdot {\bf a}_2. \nonumber
\end{align}
We then diagonalize $H_2$ by an appropriate Bogoliubov transformation~\cite{Colpa1978} to obtain the excitation spectrum of LSWT in the form
\begin{align}
& H_2 = {\textstyle \frac12} S \sum_{k} \Psi^{\dagger}_k M_k \Psi_k = {\textstyle \frac12} S \sum_{k} \Phi^{\dagger}_k D_k \Phi_k  , \\
& \Psi_k = U_k \Phi_k, \,\, \Phi_k = (\alpha_{kA}, \alpha_{kB}, \alpha_{kC}, \alpha^{\dagger}_{-kA}, \alpha^{\dagger}_{-kB},\alpha^{\dagger}_{-kC})^T, \nonumber \\
& D_k = U_k^{\dagger} M_K U_k = \mathrm{diag}(\epsilon_{1,k},\epsilon_{2,k},\epsilon_{3,k},\epsilon_{1,k},\epsilon_{2,k},\epsilon_{3,k}), \nonumber \\
& U_k \sigma_6 U_k^{\dagger} = U_k^{\dagger} \sigma_6 U_k = \sigma_6,  \,\, \sigma_6 = \mathrm{diag}(1,1,1,-1,-1,-1). \nonumber
\end{align}

For the case $J_2 = 0$, the three spin-wave branches can be expressed analytically as~\cite{Harris1992}
\begin{align}
& \epsilon_{1,k} = 0, \nonumber \\
& \epsilon_{2,k} = \epsilon_{3,k} = J_1 S \sqrt{2[\sin^2 q_1  + \sin^2 q_2 + \sin^2 (q_1 + q_2)]}. \nonumber 
\end{align}
When $J_2 \neq 0$, we obtain the eigenvalues numerically for all three branches and all $\boldsymbol{k}$ values. However, at the point K $\equiv 2(b_1 + b_2)/3$, the quadratic Hamiltonian of LSWT is 
\begin{align}
& H_2 = {\textstyle \frac12} S \sum_{k} \Psi^{\dagger}_k M_k \Psi_k, \quad M_k = \begin{bmatrix}
A_k & B_{k} \\
B^{\dagger}_{k} & A^T_{-k} 
\end{bmatrix}, \\
&A_k = 2(J_1 + J_2)I_3 - B_k/3, \nonumber\\
& B_k = -\frac{3(J_1-2J_2)}{4}\begin{bmatrix}
 0 &  1  & -1 \\
 1 &  0  &  1 \\
-1 &  1  &  0      \\
\end{bmatrix}, 
\quad I_3 = \begin{bmatrix}
1 & 0 & 0 \\
0 & 1 & 0 \\
0 & 0 & 1 \\
\end{bmatrix}, \nonumber
\end{align}
and the three eigenvalues have the analytical expressions 
\begin{align}
& \epsilon_{1,k} = 3S\sqrt{2J_1 J_2}, \nonumber \\
& \epsilon_{2,k} = \epsilon_{3,k} = {\textstyle \frac{1}{\sqrt{2}}} 3S \sqrt{J_1(J_1+2J_2)}. 
\end{align}
From these we read that the second and third spin-wave branches are always degenerate at the K point for all cases with $J_1 > 0$ and $J_2 > 0$. 

Figure~\ref{LSW} shows our results for the spectral function of the $J_1$-$J_2$ KHAF within LSWT. In the $q = 0$ phase we observe a universal band-crossing of the second and third branches at the K and K$^{\prime}$ points. As $J_2$ decreases, the lowest-energy branch softens systematically, ultimately becoming a zero-energy flat band at $J_2 = 0$, where the second and third branches also become completely degenerate. The spectrum changes abruptly on entering the $\sqrt{3}$$\times$$\sqrt{3}$ phase, where the lowest branches also have zero energy at the K and K$^\prime$ points, which are those with finite spectral intensity. Other than the absence of the QSL phase, the primary difference between the LSWT spectrum and the true spectral function obtained from our tensor-network calculations (Fig.~\ref{FullJ2Spectra} and Fig.~1 of the main text) concerns the energy scale of the lowest excitation branches, whose suppression towards zero on entering the QSL regime indicates the strength of renormalization effects arising from the dominant quantum fluctuation effects in the KHAF. 


\begin{thebibliography}{70}%
  \makeatletter
  \providecommand \@ifxundefined [1]{%
   \@ifx{#1\undefined}
  }%
  \providecommand \@ifnum [1]{%
   \ifnum #1\expandafter \@firstoftwo
   \else \expandafter \@secondoftwo
   \fi
  }%
  \providecommand \@ifx [1]{%
   \ifx #1\expandafter \@firstoftwo
   \else \expandafter \@secondoftwo
   \fi
  }%
  \providecommand \natexlab [1]{#1}%
  \providecommand \enquote  [1]{``#1''}%
  \providecommand \bibnamefont  [1]{#1}%
  \providecommand \bibfnamefont [1]{#1}%
  \providecommand \citenamefont [1]{#1}%
  \providecommand \href@noop [0]{\@secondoftwo}%
  \providecommand \href [0]{\begingroup \@sanitize@url \@href}%
  \providecommand \@href[1]{\@@startlink{#1}\@@href}%
  \providecommand \@@href[1]{\endgroup#1\@@endlink}%
  \providecommand \@sanitize@url [0]{\catcode `\\12\catcode `\$12\catcode `\&12\catcode `\#12\catcode `\^12\catcode `\_12\catcode `\%12\relax}%
  \providecommand \@@startlink[1]{}%
  \providecommand \@@endlink[0]{}%
  \providecommand \url  [0]{\begingroup\@sanitize@url \@url }%
  \providecommand \@url [1]{\endgroup\@href {#1}{\urlprefix }}%
  \providecommand \urlprefix  [0]{URL }%
  \providecommand \Eprint [0]{\href }%
  \providecommand \doibase [0]{https://doi.org/}%
  \providecommand \selectlanguage [0]{\@gobble}%
  \providecommand \bibinfo  [0]{\@secondoftwo}%
  \providecommand \bibfield  [0]{\@secondoftwo}%
  \providecommand \translation [1]{[#1]}%
  \providecommand \BibitemOpen [0]{}%
  \providecommand \bibitemStop [0]{}%
  \providecommand \bibitemNoStop [0]{.\EOS\space}%
  \providecommand \EOS [0]{\spacefactor3000\relax}%
  \providecommand \BibitemShut  [1]{\csname bibitem#1\endcsname}%
  \let\auto@bib@innerbib\@empty
  \bibitem [{\citenamefont {Anderson}(1973)}]{Anderson1973}%
    \BibitemOpen
    \bibfield  {author} {\bibinfo {author} {\bibfnamefont {P.}~\bibnamefont {Anderson}},\ }\bibfield  {title} {\bibinfo {title} {{Resonating valence bonds: A new kind of insulator?}},\ }\href {https://doi.org/10.1016/0025-5408(73)90167-0} {\bibfield  {journal} {\bibinfo  {journal} {Mater. Res. Bull.}\ }\textbf {\bibinfo {volume} {8}},\ \bibinfo {pages} {153} (\bibinfo {year} {1973})}\BibitemShut {NoStop}%
  \bibitem [{\citenamefont {Savary}\ and\ \citenamefont {Balents}(2016)}]{Savary2016}%
    \BibitemOpen
    \bibfield  {author} {\bibinfo {author} {\bibfnamefont {L.}~\bibnamefont {Savary}}\ and\ \bibinfo {author} {\bibfnamefont {L.}~\bibnamefont {Balents}},\ }\bibfield  {title} {\bibinfo {title} {{Quantum Spin Liquids: A Review}},\ }\href {https://doi.org/10.1088/0034-4885/80/1/016502} {\bibfield  {journal} {\bibinfo  {journal} {Rep. Prog. Phys.}\ }\textbf {\bibinfo {volume} {80}},\ \bibinfo {pages} {016502} (\bibinfo {year} {2016})}\BibitemShut {NoStop}%
  \bibitem [{\citenamefont {Zhou}\ \emph {et~al.}(2017)\citenamefont {Zhou}, \citenamefont {Kanoda},\ and\ \citenamefont {Ng}}]{Zhou2017}%
    \BibitemOpen
    \bibfield  {author} {\bibinfo {author} {\bibfnamefont {Y.}~\bibnamefont {Zhou}}, \bibinfo {author} {\bibfnamefont {K.}~\bibnamefont {Kanoda}},\ and\ \bibinfo {author} {\bibfnamefont {T.-K.}\ \bibnamefont {Ng}},\ }\bibfield  {title} {\bibinfo {title} {Quantum spin liquid states},\ }\href {https://doi.org/10.1103/RevModPhys.89.025003} {\bibfield  {journal} {\bibinfo  {journal} {Rev. Mod. Phys.}\ }\textbf {\bibinfo {volume} {89}},\ \bibinfo {pages} {025003} (\bibinfo {year} {2017})}\BibitemShut {NoStop}%
  \bibitem [{\citenamefont {Sindzingre}\ and\ \citenamefont {Lhuillier}(2009)}]{sindzingre2009}%
    \BibitemOpen
    \bibfield  {author} {\bibinfo {author} {\bibfnamefont {P.}~\bibnamefont {Sindzingre}}\ and\ \bibinfo {author} {\bibfnamefont {C.}~\bibnamefont {Lhuillier}},\ }\bibfield  {title} {\bibinfo {title} {Low-energy excitations of the kagomé antiferromagnet and the spin-gap issue},\ }\href {https://doi.org/10.1209/0295-5075/88/27009} {\bibfield  {journal} {\bibinfo  {journal} {Europhys. Lett.}\ }\textbf {\bibinfo {volume} {88}},\ \bibinfo {pages} {27009} (\bibinfo {year} {2009})}\BibitemShut {NoStop}%
  \bibitem [{\citenamefont {Sachdev}(1992)}]{Sachdev1992}%
    \BibitemOpen
    \bibfield  {author} {\bibinfo {author} {\bibfnamefont {S.}~\bibnamefont {Sachdev}},\ }\bibfield  {title} {\bibinfo {title} {{Kagom\'e- and triangular-lattice Heisenberg antiferromagnets: Ordering from quantum fluctuations and quantum-disordered ground states with unconfined bosonic spinons}},\ }\href {https://doi.org/10.1103/PhysRevB.45.12377} {\bibfield  {journal} {\bibinfo  {journal} {Phys. Rev. B}\ }\textbf {\bibinfo {volume} {45}},\ \bibinfo {pages} {12377} (\bibinfo {year} {1992})}\BibitemShut {NoStop}%
  \bibitem [{\citenamefont {Yan}\ \emph {et~al.}(2011)\citenamefont {Yan}, \citenamefont {Huse},\ and\ \citenamefont {White}}]{Yan2011}%
    \BibitemOpen
    \bibfield  {author} {\bibinfo {author} {\bibfnamefont {S.}~\bibnamefont {Yan}}, \bibinfo {author} {\bibfnamefont {D.~A.}\ \bibnamefont {Huse}},\ and\ \bibinfo {author} {\bibfnamefont {S.~R.}\ \bibnamefont {White}},\ }\bibfield  {title} {\bibinfo {title} {{Spin-Liquid Ground State of the $S = 1/2$ Kagome Heisenberg Antiferromagnet}},\ }\href {https://doi.org/10.1126/science.1201080} {\bibfield  {journal} {\bibinfo  {journal} {Science}\ }\textbf {\bibinfo {volume} {332}},\ \bibinfo {pages} {1173} (\bibinfo {year} {2011})}\BibitemShut {NoStop}%
  \bibitem [{\citenamefont {He}\ \emph {et~al.}(2017)\citenamefont {He}, \citenamefont {Zaletel}, \citenamefont {Oshikawa},\ and\ \citenamefont {Pollmann}}]{he2017}%
    \BibitemOpen
    \bibfield  {author} {\bibinfo {author} {\bibfnamefont {Y.-C.}\ \bibnamefont {He}}, \bibinfo {author} {\bibfnamefont {M.~P.}\ \bibnamefont {Zaletel}}, \bibinfo {author} {\bibfnamefont {M.}~\bibnamefont {Oshikawa}},\ and\ \bibinfo {author} {\bibfnamefont {F.}~\bibnamefont {Pollmann}},\ }\bibfield  {title} {\bibinfo {title} {{Signatures of {{Dirac}} Cones in a {{DMRG}} Study of the Kagome {{Heisenberg}} Model}},\ }\href {https://doi.org/10.1103/PhysRevX.7.031020} {\bibfield  {journal} {\bibinfo  {journal} {Phys. Rev. X}\ }\textbf {\bibinfo {volume} {7}},\ \bibinfo {pages} {031020} (\bibinfo {year} {2017})}\BibitemShut {NoStop}%
  \bibitem [{\citenamefont {L\"auchli}\ \emph {et~al.}(2019)\citenamefont {L\"auchli}, \citenamefont {Sudan},\ and\ \citenamefont {Moessner}}]{Lauchli2019}%
    \BibitemOpen
    \bibfield  {author} {\bibinfo {author} {\bibfnamefont {A.~M.}\ \bibnamefont {L\"auchli}}, \bibinfo {author} {\bibfnamefont {J.}~\bibnamefont {Sudan}},\ and\ \bibinfo {author} {\bibfnamefont {R.}~\bibnamefont {Moessner}},\ }\bibfield  {title} {\bibinfo {title} {${{S}}=\frac{1}{2}$ kagome {{Heisenberg}} antiferromagnet revisited},\ }\href {https://doi.org/10.1103/PhysRevB.100.155142} {\bibfield  {journal} {\bibinfo  {journal} {Phys. Rev. B}\ }\textbf {\bibinfo {volume} {100}},\ \bibinfo {pages} {155142} (\bibinfo {year} {2019})}\BibitemShut {NoStop}%
  \bibitem [{\citenamefont {Gong}\ \emph {et~al.}(2014)\citenamefont {Gong}, \citenamefont {Zhu},\ and\ \citenamefont {Sheng}}]{gong2014}%
    \BibitemOpen
    \bibfield  {author} {\bibinfo {author} {\bibfnamefont {S.-s.}\ \bibnamefont {Gong}}, \bibinfo {author} {\bibfnamefont {W.}~\bibnamefont {Zhu}},\ and\ \bibinfo {author} {\bibfnamefont {D.~N.}\ \bibnamefont {Sheng}},\ }\bibfield  {title} {\bibinfo {title} {Emergent chiral spin liquid: Fractional quantum {{Hall}} effect in a kagome {{Heisenberg}} model},\ }\href {https://doi.org/10.1038/srep06317} {\bibfield  {journal} {\bibinfo  {journal} {Sci. Rep.}\ }\textbf {\bibinfo {volume} {4}},\ \bibinfo {pages} {6317} (\bibinfo {year} {2014})}\BibitemShut {NoStop}%
  \bibitem [{\citenamefont {Suttner}\ \emph {et~al.}(2014)\citenamefont {Suttner}, \citenamefont {Platt}, \citenamefont {Reuther},\ and\ \citenamefont {Thomale}}]{suttner2014}%
    \BibitemOpen
    \bibfield  {author} {\bibinfo {author} {\bibfnamefont {R.}~\bibnamefont {Suttner}}, \bibinfo {author} {\bibfnamefont {C.}~\bibnamefont {Platt}}, \bibinfo {author} {\bibfnamefont {J.}~\bibnamefont {Reuther}},\ and\ \bibinfo {author} {\bibfnamefont {R.}~\bibnamefont {Thomale}},\ }\bibfield  {title} {\bibinfo {title} {Renormalization group analysis of competing quantum phases in the {$J_1$-$J_2$ Heisenberg} model on the kagome lattice},\ }\href {https://doi.org/10.1103/PhysRevB.89.020408} {\bibfield  {journal} {\bibinfo  {journal} {Phys. Rev. B}\ }\textbf {\bibinfo {volume} {89}},\ \bibinfo {pages} {020408(R)} (\bibinfo {year} {2014})}\BibitemShut {NoStop}%
  \bibitem [{\citenamefont {Iqbal}\ \emph {et~al.}(2015)\citenamefont {Iqbal}, \citenamefont {Poilblanc},\ and\ \citenamefont {Becca}}]{Iqbal2015}%
    \BibitemOpen
    \bibfield  {author} {\bibinfo {author} {\bibfnamefont {Y.}~\bibnamefont {Iqbal}}, \bibinfo {author} {\bibfnamefont {D.}~\bibnamefont {Poilblanc}},\ and\ \bibinfo {author} {\bibfnamefont {F.}~\bibnamefont {Becca}},\ }\bibfield  {title} {\bibinfo {title} {Spin-$\frac{1}{2}$ {Heisenberg} {$J_{1}$-$J_{2}$} antiferromagnet on the kagome lattice},\ }\href {https://doi.org/10.1103/PhysRevB.91.020402} {\bibfield  {journal} {\bibinfo  {journal} {Phys. Rev. B}\ }\textbf {\bibinfo {volume} {91}},\ \bibinfo {pages} {020402(R)} (\bibinfo {year} {2015})}\BibitemShut {NoStop}%
  \bibitem [{\citenamefont {Kolley}\ \emph {et~al.}(2015)\citenamefont {Kolley}, \citenamefont {Depenbrock}, \citenamefont {McCulloch}, \citenamefont {Schollwöck},\ and\ \citenamefont {Alba}}]{kolley2015}%
    \BibitemOpen
    \bibfield  {author} {\bibinfo {author} {\bibfnamefont {F.}~\bibnamefont {Kolley}}, \bibinfo {author} {\bibfnamefont {S.}~\bibnamefont {Depenbrock}}, \bibinfo {author} {\bibfnamefont {I.~P.}\ \bibnamefont {McCulloch}}, \bibinfo {author} {\bibfnamefont {U.}~\bibnamefont {Schollwöck}},\ and\ \bibinfo {author} {\bibfnamefont {V.}~\bibnamefont {Alba}},\ }\bibfield  {title} {\bibinfo {title} {Phase diagram of the {$J_1$-$J_2$ Heisenberg} model on the kagome lattice},\ }\href {https://doi.org/10.1103/PhysRevB.91.104418} {\bibfield  {journal} {\bibinfo  {journal} {Phys. Rev. B}\ }\textbf {\bibinfo {volume} {91}},\ \bibinfo {pages} {104418} (\bibinfo {year} {2015})}\BibitemShut {NoStop}%
  \bibitem [{\citenamefont {Liao}\ \emph {et~al.}(2017)\citenamefont {Liao}, \citenamefont {Xie}, \citenamefont {Chen}, \citenamefont {Liu}, \citenamefont {Xie}, \citenamefont {Huang}, \citenamefont {Normand},\ and\ \citenamefont {Xiang}}]{Liao2017}%
    \BibitemOpen
    \bibfield  {author} {\bibinfo {author} {\bibfnamefont {H.~J.}\ \bibnamefont {Liao}}, \bibinfo {author} {\bibfnamefont {Z.~Y.}\ \bibnamefont {Xie}}, \bibinfo {author} {\bibfnamefont {J.}~\bibnamefont {Chen}}, \bibinfo {author} {\bibfnamefont {Z.~Y.}\ \bibnamefont {Liu}}, \bibinfo {author} {\bibfnamefont {H.~D.}\ \bibnamefont {Xie}}, \bibinfo {author} {\bibfnamefont {R.~Z.}\ \bibnamefont {Huang}}, \bibinfo {author} {\bibfnamefont {B.}~\bibnamefont {Normand}},\ and\ \bibinfo {author} {\bibfnamefont {T.}~\bibnamefont {Xiang}},\ }\bibfield  {title} {\bibinfo {title} {{Gapless Spin-Liquid Ground State in the ${{S}}=1/2$ Kagome Antiferromagnet}},\ }\href {https://doi.org/10.1103/PhysRevLett.118.137202} {\bibfield  {journal} {\bibinfo  {journal} {Phys. Rev. Lett.}\ }\textbf {\bibinfo {volume} {118}},\ \bibinfo {pages} {137202} (\bibinfo {year} {2017})}\BibitemShut {NoStop}%
  \bibitem [{\citenamefont {Zhu}\ \emph {et~al.}(2019)\citenamefont {Zhu}, \citenamefont {Gong},\ and\ \citenamefont {Sheng}}]{Zhu2019}%
    \BibitemOpen
    \bibfield  {author} {\bibinfo {author} {\bibfnamefont {W.}~\bibnamefont {Zhu}}, \bibinfo {author} {\bibfnamefont {S.-s.}\ \bibnamefont {Gong}},\ and\ \bibinfo {author} {\bibfnamefont {D.~N.}\ \bibnamefont {Sheng}},\ }\bibfield  {title} {\bibinfo {title} {{Identifying spinon excitations from dynamic structure factor of spin-1/2 Heisenberg antiferromagnet on the kagome lattice}},\ }\href {https://doi.org/10.1073/pnas.1807840116} {\bibfield  {journal} {\bibinfo  {journal} {Proc. Natl. Acad. Sci. U.S.A.}\ }\textbf {\bibinfo {volume} {116}},\ \bibinfo {pages} {5437} (\bibinfo {year} {2019})}\BibitemShut {NoStop}%
  \bibitem [{\citenamefont {Nakano}\ and\ \citenamefont {Sakai}(2011)}]{nakano2011}%
    \BibitemOpen
    \bibfield  {author} {\bibinfo {author} {\bibfnamefont {H.}~\bibnamefont {Nakano}}\ and\ \bibinfo {author} {\bibfnamefont {T.}~\bibnamefont {Sakai}},\ }\bibfield  {title} {\bibinfo {title} {Numerical-{{Diagonalization Study}} of {{Spin Gap Issue}} of the {{Kagome Lattice Heisenberg Antiferromagnet}}},\ }\href {https://doi.org/10.1143/JPSJ.80.053704} {\bibfield  {journal} {\bibinfo  {journal} {J. Phys. Soc. Jpn.}\ }\textbf {\bibinfo {volume} {80}},\ \bibinfo {pages} {053704} (\bibinfo {year} {2011})}\BibitemShut {NoStop}%
  \bibitem [{\citenamefont {L{\"a}uchli}\ \emph {et~al.}(2011)\citenamefont {L{\"a}uchli}, \citenamefont {Sudan},\ and\ \citenamefont {S{\o}rensen}}]{Lauchli2011}%
    \BibitemOpen
    \bibfield  {author} {\bibinfo {author} {\bibfnamefont {A.~M.}\ \bibnamefont {L{\"a}uchli}}, \bibinfo {author} {\bibfnamefont {J.}~\bibnamefont {Sudan}},\ and\ \bibinfo {author} {\bibfnamefont {E.~S.}\ \bibnamefont {S{\o}rensen}},\ }\bibfield  {title} {\bibinfo {title} {Ground-state energy and spin gap of spin-$\frac{1}{2}$ {{Kagom}}\'e-{{Heisenberg}} antiferromagnetic clusters: {{Large-scale}} exact diagonalization results},\ }\href {https://doi.org/10.1103/PhysRevB.83.212401} {\bibfield  {journal} {\bibinfo  {journal} {Phys. Rev. B}\ }\textbf {\bibinfo {volume} {83}},\ \bibinfo {pages} {212401} (\bibinfo {year} {2011})}\BibitemShut {NoStop}%
  \bibitem [{\citenamefont {Prelov\ifmmode~\check{s}\else \v{s}\fi{}ek}\ \emph {et~al.}(2021)\citenamefont {Prelov\ifmmode~\check{s}\else \v{s}\fi{}ek}, \citenamefont {Gomil\ifmmode~\check{s}\else \v{s}\fi{}ek}, \citenamefont {Arh},\ and\ \citenamefont {Zorko}}]{Prelovsek2021}%
    \BibitemOpen
    \bibfield  {author} {\bibinfo {author} {\bibfnamefont {P.}~\bibnamefont {Prelov\ifmmode~\check{s}\else \v{s}\fi{}ek}}, \bibinfo {author} {\bibfnamefont {M.}~\bibnamefont {Gomil\ifmmode~\check{s}\else \v{s}\fi{}ek}}, \bibinfo {author} {\bibfnamefont {T.}~\bibnamefont {Arh}},\ and\ \bibinfo {author} {\bibfnamefont {A.}~\bibnamefont {Zorko}},\ }\bibfield  {title} {\bibinfo {title} {Dynamical spin correlations of the kagome antiferromagnet},\ }\href {https://doi.org/10.1103/PhysRevB.103.014431} {\bibfield  {journal} {\bibinfo  {journal} {Phys. Rev. B}\ }\textbf {\bibinfo {volume} {103}},\ \bibinfo {pages} {014431} (\bibinfo {year} {2021})}\BibitemShut {NoStop}%
  \bibitem [{\citenamefont {Jiang}\ \emph {et~al.}(2008)\citenamefont {Jiang}, \citenamefont {Weng},\ and\ \citenamefont {Sheng}}]{Jiang2008}%
    \BibitemOpen
    \bibfield  {author} {\bibinfo {author} {\bibfnamefont {H.~C.}\ \bibnamefont {Jiang}}, \bibinfo {author} {\bibfnamefont {Z.~Y.}\ \bibnamefont {Weng}},\ and\ \bibinfo {author} {\bibfnamefont {D.~N.}\ \bibnamefont {Sheng}},\ }\bibfield  {title} {\bibinfo {title} {{Density Matrix Renormalization Group Numerical Study of the Kagome Antiferromagnet}},\ }\href {https://doi.org/10.1103/PhysRevLett.101.117203} {\bibfield  {journal} {\bibinfo  {journal} {Phys. Rev. Lett.}\ }\textbf {\bibinfo {volume} {101}},\ \bibinfo {pages} {117203} (\bibinfo {year} {2008})}\BibitemShut {NoStop}%
  \bibitem [{\citenamefont {Depenbrock}\ \emph {et~al.}(2012)\citenamefont {Depenbrock}, \citenamefont {McCulloch},\ and\ \citenamefont {Schollw\"ock}}]{Stefan2012}%
    \BibitemOpen
    \bibfield  {author} {\bibinfo {author} {\bibfnamefont {S.}~\bibnamefont {Depenbrock}}, \bibinfo {author} {\bibfnamefont {I.~P.}\ \bibnamefont {McCulloch}},\ and\ \bibinfo {author} {\bibfnamefont {U.}~\bibnamefont {Schollw\"ock}},\ }\bibfield  {title} {\bibinfo {title} {{Nature of the Spin-Liquid Ground State of the {{$S=1/2$}} Heisenberg Model on the Kagome Lattice}},\ }\href {https://doi.org/10.1103/PhysRevLett.109.067201} {\bibfield  {journal} {\bibinfo  {journal} {Phys. Rev. Lett.}\ }\textbf {\bibinfo {volume} {109}},\ \bibinfo {pages} {067201} (\bibinfo {year} {2012})}\BibitemShut {NoStop}%
  \bibitem [{\citenamefont {Jiang}\ \emph {et~al.}(2012)\citenamefont {Jiang}, \citenamefont {Wang},\ and\ \citenamefont {Balents}}]{Jiang2012}%
    \BibitemOpen
    \bibfield  {author} {\bibinfo {author} {\bibfnamefont {H.-C.}\ \bibnamefont {Jiang}}, \bibinfo {author} {\bibfnamefont {Z.}~\bibnamefont {Wang}},\ and\ \bibinfo {author} {\bibfnamefont {L.}~\bibnamefont {Balents}},\ }\bibfield  {title} {\bibinfo {title} {Identifying topological order by entanglement entropy},\ }\href {https://doi.org/10.1038/nphys2465} {\bibfield  {journal} {\bibinfo  {journal} {Nat. Phys.}\ }\textbf {\bibinfo {volume} {8}},\ \bibinfo {pages} {902} (\bibinfo {year} {2012})}\BibitemShut {NoStop}%
  \bibitem [{\citenamefont {Nishimoto}\ \emph {et~al.}(2013)\citenamefont {Nishimoto}, \citenamefont {Shibata},\ and\ \citenamefont {Hotta}}]{Nishimoto2013}%
    \BibitemOpen
    \bibfield  {author} {\bibinfo {author} {\bibfnamefont {S.}~\bibnamefont {Nishimoto}}, \bibinfo {author} {\bibfnamefont {N.}~\bibnamefont {Shibata}},\ and\ \bibinfo {author} {\bibfnamefont {C.}~\bibnamefont {Hotta}},\ }\bibfield  {title} {\bibinfo {title} {Controlling frustrated liquids and solids with an applied field in a kagome {Heisenberg} antiferromagnet},\ }\href {https://doi.org/10.1038/ncomms3287} {\bibfield  {journal} {\bibinfo  {journal} {Nat. Commun.}\ }\textbf {\bibinfo {volume} {4}},\ \bibinfo {pages} {2287} (\bibinfo {year} {2013})}\BibitemShut {NoStop}%
  \bibitem [{\citenamefont {Sun}\ \emph {et~al.}(2024)\citenamefont {Sun}, \citenamefont {Jin}, \citenamefont {Tu},\ and\ \citenamefont {Zhou}}]{sun2024}%
    \BibitemOpen
    \bibfield  {author} {\bibinfo {author} {\bibfnamefont {R.-Y.}\ \bibnamefont {Sun}}, \bibinfo {author} {\bibfnamefont {H.-K.}\ \bibnamefont {Jin}}, \bibinfo {author} {\bibfnamefont {H.-H.}\ \bibnamefont {Tu}},\ and\ \bibinfo {author} {\bibfnamefont {Y.}~\bibnamefont {Zhou}},\ }\bibfield  {title} {\bibinfo {title} {Possible chiral spin liquid state in the $s = 1/2$ kagome {{Heisenberg}} model},\ }\href {https://doi.org/10.1038/s41535-024-00627-5} {\bibfield  {journal} {\bibinfo  {journal} {npj Quantum Mater.}\ }\textbf {\bibinfo {volume} {9}},\ \bibinfo {pages} {1} (\bibinfo {year} {2024})}\BibitemShut {NoStop}%
  \bibitem [{\citenamefont {Ran}\ \emph {et~al.}(2007)\citenamefont {Ran}, \citenamefont {Hermele}, \citenamefont {Lee},\ and\ \citenamefont {Wen}}]{Ran2007}%
    \BibitemOpen
    \bibfield  {author} {\bibinfo {author} {\bibfnamefont {Y.}~\bibnamefont {Ran}}, \bibinfo {author} {\bibfnamefont {M.}~\bibnamefont {Hermele}}, \bibinfo {author} {\bibfnamefont {P.~A.}\ \bibnamefont {Lee}},\ and\ \bibinfo {author} {\bibfnamefont {X.-G.}\ \bibnamefont {Wen}},\ }\bibfield  {title} {\bibinfo {title} {{Projected-Wave-Function Study of the Spin-1/2 Heisenberg Model on the Kagom\'e Lattice}},\ }\href {https://doi.org/10.1103/PhysRevLett.98.117205} {\bibfield  {journal} {\bibinfo  {journal} {Phys. Rev. Lett.}\ }\textbf {\bibinfo {volume} {98}},\ \bibinfo {pages} {117205} (\bibinfo {year} {2007})}\BibitemShut {NoStop}%
  \bibitem [{\citenamefont {Iqbal}\ \emph {et~al.}(2013)\citenamefont {Iqbal}, \citenamefont {Becca}, \citenamefont {Sorella},\ and\ \citenamefont {Poilblanc}}]{Iqbal2013}%
    \BibitemOpen
    \bibfield  {author} {\bibinfo {author} {\bibfnamefont {Y.}~\bibnamefont {Iqbal}}, \bibinfo {author} {\bibfnamefont {F.}~\bibnamefont {Becca}}, \bibinfo {author} {\bibfnamefont {S.}~\bibnamefont {Sorella}},\ and\ \bibinfo {author} {\bibfnamefont {D.}~\bibnamefont {Poilblanc}},\ }\bibfield  {title} {\bibinfo {title} {Gapless spin-liquid phase in the kagome spin-$\frac{1}{2}$ {Heisenberg} antiferromagnet},\ }\href {https://doi.org/10.1103/PhysRevB.87.060405} {\bibfield  {journal} {\bibinfo  {journal} {Phys. Rev. B}\ }\textbf {\bibinfo {volume} {87}},\ \bibinfo {pages} {060405(R)} (\bibinfo {year} {2013})}\BibitemShut {NoStop}%
  \bibitem [{\citenamefont {Iqbal}\ \emph {et~al.}(2014)\citenamefont {Iqbal}, \citenamefont {Poilblanc},\ and\ \citenamefont {Becca}}]{iqbal2014}%
    \BibitemOpen
    \bibfield  {author} {\bibinfo {author} {\bibfnamefont {Y.}~\bibnamefont {Iqbal}}, \bibinfo {author} {\bibfnamefont {D.}~\bibnamefont {Poilblanc}},\ and\ \bibinfo {author} {\bibfnamefont {F.}~\bibnamefont {Becca}},\ }\bibfield  {title} {\bibinfo {title} {Vanishing spin gap in a competing spin-liquid phase in the kagome {{Heisenberg}} antiferromagnet},\ }\href {https://doi.org/10.1103/PhysRevB.89.020407} {\bibfield  {journal} {\bibinfo  {journal} {Phys. Rev. B}\ }\textbf {\bibinfo {volume} {89}},\ \bibinfo {pages} {020407(R)} (\bibinfo {year} {2014})}\BibitemShut {NoStop}%
  \bibitem [{\citenamefont {Mei}\ and\ \citenamefont {Wen}(2015)}]{Mei2015}%
    \BibitemOpen
    \bibfield  {author} {\bibinfo {author} {\bibfnamefont {J.-W.}\ \bibnamefont {Mei}}\ and\ \bibinfo {author} {\bibfnamefont {X.-G.}\ \bibnamefont {Wen}},\ }\href {https://doi.org/10.48550/arXiv.1507.03007} {\bibinfo {title} {Fractionalized spin-wave continuum in spin liquid states on the kagome lattice}},\ \Eprint {https://arxiv.org/abs/1507.03007} {arXiv:1507.03007} \BibitemShut {NoStop}%
  \bibitem [{\citenamefont {Zhang}\ and\ \citenamefont {Li}(2020)}]{Zhang2020}%
    \BibitemOpen
    \bibfield  {author} {\bibinfo {author} {\bibfnamefont {C.}~\bibnamefont {Zhang}}\ and\ \bibinfo {author} {\bibfnamefont {T.}~\bibnamefont {Li}},\ }\bibfield  {title} {\bibinfo {title} {{Variational study of the ground state and spin dynamics of the spin-$\frac{1}{2}$ kagome antiferromagnetic Heisenberg model and its implication for herbertsmithite ${\mathrm{ZnCu}}_{3}{(\mathrm{OH})}_{6}{\mathrm{Cl}}_{2}$}},\ }\href {https://doi.org/10.1103/PhysRevB.102.195106} {\bibfield  {journal} {\bibinfo  {journal} {Phys. Rev. B}\ }\textbf {\bibinfo {volume} {102}},\ \bibinfo {pages} {195106} (\bibinfo {year} {2020})}\BibitemShut {NoStop}%
  \bibitem [{\citenamefont {Ferrari}\ \emph {et~al.}(2021)\citenamefont {Ferrari}, \citenamefont {Parola},\ and\ \citenamefont {Becca}}]{Ferrari2021}%
    \BibitemOpen
    \bibfield  {author} {\bibinfo {author} {\bibfnamefont {F.}~\bibnamefont {Ferrari}}, \bibinfo {author} {\bibfnamefont {A.}~\bibnamefont {Parola}},\ and\ \bibinfo {author} {\bibfnamefont {F.}~\bibnamefont {Becca}},\ }\bibfield  {title} {\bibinfo {title} {Gapless spin liquids in disguise},\ }\href {https://doi.org/10.1103/PhysRevB.103.195140} {\bibfield  {journal} {\bibinfo  {journal} {Phys. Rev. B}\ }\textbf {\bibinfo {volume} {103}},\ \bibinfo {pages} {195140} (\bibinfo {year} {2021})}\BibitemShut {NoStop}%
  \bibitem [{\citenamefont {Jiang}\ \emph {et~al.}(2019)\citenamefont {Jiang}, \citenamefont {Kim}, \citenamefont {Han},\ and\ \citenamefont {Ran}}]{jiang2019}%
    \BibitemOpen
    \bibfield  {author} {\bibinfo {author} {\bibfnamefont {S.}~\bibnamefont {Jiang}}, \bibinfo {author} {\bibfnamefont {P.}~\bibnamefont {Kim}}, \bibinfo {author} {\bibfnamefont {J.~H.}\ \bibnamefont {Han}},\ and\ \bibinfo {author} {\bibfnamefont {Y.}~\bibnamefont {Ran}},\ }\bibfield  {title} {\bibinfo {title} {Competing spin liquid phases in the $s = \frac{1}{2}$ {{Heisenberg}} model on the kagome lattice},\ }\href {https://doi.org/10.21468/SciPostPhys.7.1.006} {\bibfield  {journal} {\bibinfo  {journal} {SciPost Phys.}\ }\textbf {\bibinfo {volume} {7}},\ \bibinfo {pages} {006} (\bibinfo {year} {2019})}\BibitemShut {NoStop}%
  \bibitem [{\citenamefont {Jahromi}\ \emph {et~al.}(2020)\citenamefont {Jahromi}, \citenamefont {Orus}, \citenamefont {Poilblanc},\ and\ \citenamefont {Mila}}]{jahromi2020}%
    \BibitemOpen
    \bibfield  {author} {\bibinfo {author} {\bibfnamefont {S.}~\bibnamefont {Jahromi}}, \bibinfo {author} {\bibfnamefont {R.}~\bibnamefont {Orus}}, \bibinfo {author} {\bibfnamefont {D.}~\bibnamefont {Poilblanc}},\ and\ \bibinfo {author} {\bibfnamefont {F.}~\bibnamefont {Mila}},\ }\bibfield  {title} {\bibinfo {title} {Spin-$\frac{1}{2}$ kagome {{Heisenberg}} antiferromagnet with strong breathing anisotropy},\ }\href {https://doi.org/10.21468/SciPostPhys.9.6.092} {\bibfield  {journal} {\bibinfo  {journal} {SciPost Phys.}\ }\textbf {\bibinfo {volume} {9}},\ \bibinfo {pages} {092} (\bibinfo {year} {2020})}\BibitemShut {NoStop}%
  \bibitem [{\citenamefont {Mei}\ \emph {et~al.}(2017)\citenamefont {Mei}, \citenamefont {Chen}, \citenamefont {He},\ and\ \citenamefont {Wen}}]{mei2017}%
    \BibitemOpen
    \bibfield  {author} {\bibinfo {author} {\bibfnamefont {J.-W.}\ \bibnamefont {Mei}}, \bibinfo {author} {\bibfnamefont {J.-Y.}\ \bibnamefont {Chen}}, \bibinfo {author} {\bibfnamefont {H.}~\bibnamefont {He}},\ and\ \bibinfo {author} {\bibfnamefont {X.-G.}\ \bibnamefont {Wen}},\ }\bibfield  {title} {\bibinfo {title} {Gapped spin liquid with $\mathbb{{Z}}_{2}$ topological order for the kagome {{Heisenberg}} model},\ }\href {https://doi.org/10.1103/PhysRevB.95.235107} {\bibfield  {journal} {\bibinfo  {journal} {Phys. Rev. B}\ }\textbf {\bibinfo {volume} {95}},\ \bibinfo {pages} {235107} (\bibinfo {year} {2017})}\BibitemShut {NoStop}%
  \bibitem [{\citenamefont {L\"auchli}\ and\ \citenamefont {Lhuillier}(2009)}]{laeuchli2009}%
    \BibitemOpen
    \bibfield  {author} {\bibinfo {author} {\bibfnamefont {A.}~\bibnamefont {L\"auchli}}\ and\ \bibinfo {author} {\bibfnamefont {C.}~\bibnamefont {Lhuillier}},\ }\href {https://doi.org/10.48550/arXiv.0901.1065} {\bibinfo {title} {Dynamical {{Correlations}} of the {{Kagome}} $s = 1/2$ {{Heisenberg Quantum Antiferromagnet}}}},\ \Eprint {https://arxiv.org/abs/0901.1065} {arXiv:0901.1065} \BibitemShut {NoStop}%
  \bibitem [{\citenamefont {Endo}\ \emph {et~al.}(2018)\citenamefont {Endo}, \citenamefont {Hotta},\ and\ \citenamefont {Shimizu}}]{endo2018}%
    \BibitemOpen
    \bibfield  {author} {\bibinfo {author} {\bibfnamefont {H.}~\bibnamefont {Endo}}, \bibinfo {author} {\bibfnamefont {C.}~\bibnamefont {Hotta}},\ and\ \bibinfo {author} {\bibfnamefont {A.}~\bibnamefont {Shimizu}},\ }\bibfield  {title} {\bibinfo {title} {From {{Linear}} to {{Nonlinear Responses}} of {{Thermal Pure Quantum States}}},\ }\href {https://doi.org/10.1103/PhysRevLett.121.220601} {\bibfield  {journal} {\bibinfo  {journal} {Phys. Rev. Lett.}\ }\textbf {\bibinfo {volume} {121}},\ \bibinfo {pages} {220601} (\bibinfo {year} {2018})}\BibitemShut {NoStop}%
  \bibitem [{\citenamefont {Jiang}\ \emph {et~al.}(2026)\citenamefont {Jiang}, \citenamefont {Campello}, \citenamefont {He}, \citenamefont {Wen}, \citenamefont {Pajerowski}, \citenamefont {Lee},\ and\ \citenamefont {Jiang}}]{jiang2026}%
    \BibitemOpen
    \bibfield  {author} {\bibinfo {author} {\bibfnamefont {S.}~\bibnamefont {Jiang}}, \bibinfo {author} {\bibfnamefont {A.~C.}\ \bibnamefont {Campello}}, \bibinfo {author} {\bibfnamefont {W.}~\bibnamefont {He}}, \bibinfo {author} {\bibfnamefont {J.}~\bibnamefont {Wen}}, \bibinfo {author} {\bibfnamefont {D.~M.}\ \bibnamefont {Pajerowski}}, \bibinfo {author} {\bibfnamefont {Y.~S.}\ \bibnamefont {Lee}},\ and\ \bibinfo {author} {\bibfnamefont {H.-C.}\ \bibnamefont {Jiang}},\ }\bibfield  {title} {\bibinfo {title} {Quantifying the phase diagram and {{Hamiltonian}} of {{S}} = 1/2 kagome antiferromagnets: Bridging theory and experiment},\ }\href {https://doi.org/10.1038/s41524-026-01959-5} {\bibfield  {journal} {\bibinfo  {journal} {npj Comput. Mater.}\ }\textbf {\bibinfo {volume} {12}},\ \bibinfo {pages} {91} (\bibinfo {year} {2026})}\BibitemShut {NoStop}%
  \bibitem [{\citenamefont {\"Ostlund}\ and\ \citenamefont {Rommer}(1995)}]{Ostlund1995}%
    \BibitemOpen
    \bibfield  {author} {\bibinfo {author} {\bibfnamefont {S.}~\bibnamefont {\"Ostlund}}\ and\ \bibinfo {author} {\bibfnamefont {S.}~\bibnamefont {Rommer}},\ }\bibfield  {title} {\bibinfo {title} {Thermodynamic limit of density matrix renormalization},\ }\href {https://doi.org/10.1103/PhysRevLett.75.3537} {\bibfield  {journal} {\bibinfo  {journal} {Phys. Rev. Lett.}\ }\textbf {\bibinfo {volume} {75}},\ \bibinfo {pages} {3537} (\bibinfo {year} {1995})}\BibitemShut {NoStop}%
  \bibitem [{\citenamefont {Haegeman}\ \emph {et~al.}(2012)\citenamefont {Haegeman}, \citenamefont {Pirvu}, \citenamefont {Weir}, \citenamefont {Cirac}, \citenamefont {Osborne}, \citenamefont {Verschelde},\ and\ \citenamefont {Verstraete}}]{Haegeman2012}%
    \BibitemOpen
    \bibfield  {author} {\bibinfo {author} {\bibfnamefont {J.}~\bibnamefont {Haegeman}}, \bibinfo {author} {\bibfnamefont {B.}~\bibnamefont {Pirvu}}, \bibinfo {author} {\bibfnamefont {D.~J.}\ \bibnamefont {Weir}}, \bibinfo {author} {\bibfnamefont {J.~I.}\ \bibnamefont {Cirac}}, \bibinfo {author} {\bibfnamefont {T.~J.}\ \bibnamefont {Osborne}}, \bibinfo {author} {\bibfnamefont {H.}~\bibnamefont {Verschelde}},\ and\ \bibinfo {author} {\bibfnamefont {F.}~\bibnamefont {Verstraete}},\ }\bibfield  {title} {\bibinfo {title} {Variational matrix product ansatz for dispersion relations},\ }\href {https://doi.org/10.1103/PhysRevB.85.100408} {\bibfield  {journal} {\bibinfo  {journal} {Phys. Rev. B}\ }\textbf {\bibinfo {volume} {85}},\ \bibinfo {pages} {100408(R)} (\bibinfo {year} {2012})}\BibitemShut {NoStop}%
  \bibitem [{\citenamefont {Vanderstraeten}\ \emph {et~al.}(2015)\citenamefont {Vanderstraeten}, \citenamefont {Mari\"en}, \citenamefont {Verstraete},\ and\ \citenamefont {Haegeman}}]{Vanderstraeten2015}%
    \BibitemOpen
    \bibfield  {author} {\bibinfo {author} {\bibfnamefont {L.}~\bibnamefont {Vanderstraeten}}, \bibinfo {author} {\bibfnamefont {M.}~\bibnamefont {Mari\"en}}, \bibinfo {author} {\bibfnamefont {F.}~\bibnamefont {Verstraete}},\ and\ \bibinfo {author} {\bibfnamefont {J.}~\bibnamefont {Haegeman}},\ }\bibfield  {title} {\bibinfo {title} {{Excitations and the tangent space of projected entangled-pair states}},\ }\href {https://doi.org/10.1103/physrevb.92.201111} {\bibfield  {journal} {\bibinfo  {journal} {Phys. Rev. B}\ }\textbf {\bibinfo {volume} {92}},\ \bibinfo {pages} {201111(R)} (\bibinfo {year} {2015})}\BibitemShut {NoStop}%
  \bibitem [{sm()}]{sm}%
    \BibitemOpen
    \href@noop {} {}\bibinfo {howpublished} {See Supplemental Material, which includes Refs. ~\cite{corboz2014,wang2024,DallaPiazza2015,Shaik2025,Colpa1978,Harris1992}, for detailed discussions of (i) the single-mode Ansatz for computing excitation spectra, (ii) the technical developments we implemented to improve these calculations, (iii) a benchmarking of the evolution of the spectrum with the bond dimension $D$, (iv) an illustration of the effects of $\eta$, (v) the evolution of the spectrum with $J_2$, (vi) the evolution of the static structure factor with $J_2$, and (vii) the way we performed our linear spin-wave calculations for comparison with the tensor-network spectral function}\BibitemShut {NoStop}%
  \bibitem [{\citenamefont {Corboz}\ \emph {et~al.}(2014)\citenamefont {Corboz}, \citenamefont {Rice},\ and\ \citenamefont {Troyer}}]{corboz2014}%
    \BibitemOpen
    \bibfield  {author} {\bibinfo {author} {\bibfnamefont {P.}~\bibnamefont {Corboz}}, \bibinfo {author} {\bibfnamefont {T.~M.}\ \bibnamefont {Rice}},\ and\ \bibinfo {author} {\bibfnamefont {M.}~\bibnamefont {Troyer}},\ }\bibfield  {title} {\bibinfo {title} {Competing {{States}} in the $t$-${{J}}$ {{Model}}: {{Uniform}} $d$-{{Wave State}} versus {{Stripe State}}},\ }\href {https://doi.org/10.1103/PhysRevLett.113.046402} {\bibfield  {journal} {\bibinfo  {journal} {Phys. Rev. Lett.}\ }\textbf {\bibinfo {volume} {113}},\ \bibinfo {pages} {046402} (\bibinfo {year} {2014})}\BibitemShut {NoStop}%
  \bibitem [{\citenamefont {Wang}\ \emph {et~al.}(2025)\citenamefont {Wang}, \citenamefont {Feng}, \citenamefont {Zhu}, \citenamefont {Chi}, \citenamefont {Liao}, \citenamefont {Trivedi},\ and\ \citenamefont {Xiang}}]{wang2024}%
    \BibitemOpen
    \bibfield  {author} {\bibinfo {author} {\bibfnamefont {K.}~\bibnamefont {Wang}}, \bibinfo {author} {\bibfnamefont {S.}~\bibnamefont {Feng}}, \bibinfo {author} {\bibfnamefont {P.}~\bibnamefont {Zhu}}, \bibinfo {author} {\bibfnamefont {R.}~\bibnamefont {Chi}}, \bibinfo {author} {\bibfnamefont {H.-J.}\ \bibnamefont {Liao}}, \bibinfo {author} {\bibfnamefont {N.}~\bibnamefont {Trivedi}},\ and\ \bibinfo {author} {\bibfnamefont {T.}~\bibnamefont {Xiang}},\ }\bibfield  {title} {\bibinfo {title} {Fractionalization signatures in the dynamics of quantum spin liquids},\ }\href {https://doi.org/10.1103/PhysRevB.111.L100402} {\bibfield  {journal} {\bibinfo  {journal} {Phys. Rev. B}\ }\textbf {\bibinfo {volume} {111}},\ \bibinfo {pages} {L100402} (\bibinfo {year} {2025})}\BibitemShut {NoStop}%
  \bibitem [{\citenamefont {{Dalla Piazza}}\ \emph {et~al.}(2015)\citenamefont {{Dalla Piazza}}, \citenamefont {Mourigal}, \citenamefont {Christensen}, \citenamefont {Nilsen}, \citenamefont {Tregenna-Piggott}, \citenamefont {Perring}, \citenamefont {Enderle}, \citenamefont {McMorrow}, \citenamefont {Ivanov},\ and\ \citenamefont {Rønnow}}]{DallaPiazza2015}%
    \BibitemOpen
    \bibfield  {author} {\bibinfo {author} {\bibfnamefont {B.}~\bibnamefont {{Dalla Piazza}}}, \bibinfo {author} {\bibfnamefont {M.}~\bibnamefont {Mourigal}}, \bibinfo {author} {\bibfnamefont {N.~B.}\ \bibnamefont {Christensen}}, \bibinfo {author} {\bibfnamefont {G.~J.}\ \bibnamefont {Nilsen}}, \bibinfo {author} {\bibfnamefont {P.}~\bibnamefont {Tregenna-Piggott}}, \bibinfo {author} {\bibfnamefont {T.~G.}\ \bibnamefont {Perring}}, \bibinfo {author} {\bibfnamefont {M.}~\bibnamefont {Enderle}}, \bibinfo {author} {\bibfnamefont {D.~F.}\ \bibnamefont {McMorrow}}, \bibinfo {author} {\bibfnamefont {D.~A.}\ \bibnamefont {Ivanov}},\ and\ \bibinfo {author} {\bibfnamefont {H.~M.}\ \bibnamefont {Rønnow}},\ }\bibfield  {title} {\bibinfo {title} {Fractional excitations in the square-lattice quantum antiferromagnet},\ }\href {https://doi.org/10.1038/nphys3172} {\bibfield  {journal} {\bibinfo  {journal} {Nat. Phys.}\ }\textbf {\bibinfo {volume} {11}},\ \bibinfo {pages} {62} (\bibinfo {year} {2015})}\BibitemShut {NoStop}%
  \bibitem [{\citenamefont {Shaik}\ \emph {et~al.}(2025)\citenamefont {Shaik}, \citenamefont {Fogh}, \citenamefont {Piazza}, \citenamefont {Normand}, \citenamefont {Ivanov},\ and\ \citenamefont {Rønnow}}]{Shaik2025}%
    \BibitemOpen
    \bibfield  {author} {\bibinfo {author} {\bibfnamefont {N.~E.}\ \bibnamefont {Shaik}}, \bibinfo {author} {\bibfnamefont {E.}~\bibnamefont {Fogh}}, \bibinfo {author} {\bibfnamefont {B.~Dalla}\ \bibnamefont {Piazza}}, \bibinfo {author} {\bibfnamefont {B.}~\bibnamefont {Normand}}, \bibinfo {author} {\bibfnamefont {D.~A.}\ \bibnamefont {Ivanov}},\ and\ \bibinfo {author} {\bibfnamefont {H.~M.}\ \bibnamefont {Rønnow}},\ }\bibfield  {title} {\bibinfo {title} {Confined and deconfined spinon excitations in the rectangular-lattice quantum antiferromagnet},\ }\href {https://doi.org/10.1103/zntl-4622} {\bibfield  {journal} {\bibinfo  {journal} {Phys. Rev. B}\ }\textbf {\bibinfo {volume} {112}},\ \bibinfo {pages} {L020410} (\bibinfo {year} {2025})}\BibitemShut {NoStop}%
  \bibitem [{\citenamefont {Colpa}(1978)}]{Colpa1978}%
    \BibitemOpen
    \bibfield  {author} {\bibinfo {author} {\bibfnamefont {J.~H.~P.}\ \bibnamefont {Colpa}},\ }\bibfield  {title} {\bibinfo {title} {{Diagonalization of the quadratic boson Hamiltonian}},\ }\href {https://doi.org/https://doi.org/10.1016/0378-4371(78)90160-7} {\bibfield  {journal} {\bibinfo  {journal} {Physica (Amsterdam)}\ }\textbf {\bibinfo {volume} {93A}},\ \bibinfo {pages} {327} (\bibinfo {year} {1978})}\BibitemShut {NoStop}%
  \bibitem [{\citenamefont {Harris}\ \emph {et~al.}(1992)\citenamefont {Harris}, \citenamefont {Kallin},\ and\ \citenamefont {Berlinsky}}]{Harris1992}%
    \BibitemOpen
    \bibfield  {author} {\bibinfo {author} {\bibfnamefont {A.~B.}\ \bibnamefont {Harris}}, \bibinfo {author} {\bibfnamefont {C.}~\bibnamefont {Kallin}},\ and\ \bibinfo {author} {\bibfnamefont {A.~J.}\ \bibnamefont {Berlinsky}},\ }\bibfield  {title} {\bibinfo {title} {{Possible N\'eel orderings of the kagom\'e antiferromagnet}},\ }\href {https://doi.org/10.1103/PhysRevB.45.2899} {\bibfield  {journal} {\bibinfo  {journal} {Phys. Rev. B}\ }\textbf {\bibinfo {volume} {45}},\ \bibinfo {pages} {2899} (\bibinfo {year} {1992})}\BibitemShut {NoStop}%
  \bibitem [{\citenamefont {Liao}\ \emph {et~al.}(2019)\citenamefont {Liao}, \citenamefont {Liu}, \citenamefont {Wang},\ and\ \citenamefont {Xiang}}]{Liao2019}%
    \BibitemOpen
    \bibfield  {author} {\bibinfo {author} {\bibfnamefont {H.-J.}\ \bibnamefont {Liao}}, \bibinfo {author} {\bibfnamefont {J.-G.}\ \bibnamefont {Liu}}, \bibinfo {author} {\bibfnamefont {L.}~\bibnamefont {Wang}},\ and\ \bibinfo {author} {\bibfnamefont {T.}~\bibnamefont {Xiang}},\ }\bibfield  {title} {\bibinfo {title} {{Differentiable Programming Tensor Networks}},\ }\href {https://doi.org/10.1103/physrevx.9.031041} {\bibfield  {journal} {\bibinfo  {journal} {Phys. Rev. X}\ }\textbf {\bibinfo {volume} {9}},\ \bibinfo {pages} {031041} (\bibinfo {year} {2019})}\BibitemShut {NoStop}%
  \bibitem [{\citenamefont {Chi}\ \emph {et~al.}(2022)\citenamefont {Chi}, \citenamefont {Liu}, \citenamefont {Wan}, \citenamefont {Liao},\ and\ \citenamefont {Xiang}}]{chi2022}%
    \BibitemOpen
    \bibfield  {author} {\bibinfo {author} {\bibfnamefont {R.}~\bibnamefont {Chi}}, \bibinfo {author} {\bibfnamefont {Y.}~\bibnamefont {Liu}}, \bibinfo {author} {\bibfnamefont {Y.}~\bibnamefont {Wan}}, \bibinfo {author} {\bibfnamefont {H.-J.}\ \bibnamefont {Liao}},\ and\ \bibinfo {author} {\bibfnamefont {T.}~\bibnamefont {Xiang}},\ }\bibfield  {title} {\bibinfo {title} {{Spin Excitation Spectra of Anisotropic Spin-$1/2$ Triangular Lattice Heisenberg Antiferromagnets}},\ }\href {https://doi.org/10.1103/PhysRevLett.129.227201} {\bibfield  {journal} {\bibinfo  {journal} {Phys. Rev. Lett.}\ }\textbf {\bibinfo {volume} {129}},\ \bibinfo {pages} {227201} (\bibinfo {year} {2022})}\BibitemShut {NoStop}%
  \bibitem [{\citenamefont {Ponsioen}\ \emph {et~al.}(2022)\citenamefont {Ponsioen}, \citenamefont {Assaad},\ and\ \citenamefont {Corboz}}]{Ponsioen2022}%
    \BibitemOpen
    \bibfield  {author} {\bibinfo {author} {\bibfnamefont {B.}~\bibnamefont {Ponsioen}}, \bibinfo {author} {\bibfnamefont {F.~F.}\ \bibnamefont {Assaad}},\ and\ \bibinfo {author} {\bibfnamefont {P.}~\bibnamefont {Corboz}},\ }\bibfield  {title} {\bibinfo {title} {{Automatic differentiation applied to excitations with Projected Entangled Pair States}},\ }\href {https://doi.org/10.21468/SciPostPhys.12.1.006} {\bibfield  {journal} {\bibinfo  {journal} {SciPost Phys.}\ }\textbf {\bibinfo {volume} {12}},\ \bibinfo {pages} {6} (\bibinfo {year} {2022})}\BibitemShut {NoStop}%
  \bibitem [{\citenamefont {Tu}\ \emph {et~al.}(2024)\citenamefont {Tu}, \citenamefont {Vanderstraeten}, \citenamefont {Schuch}, \citenamefont {Lee}, \citenamefont {Kawashima},\ and\ \citenamefont {Chen}}]{tu2024}%
    \BibitemOpen
    \bibfield  {author} {\bibinfo {author} {\bibfnamefont {W.-L.}\ \bibnamefont {Tu}}, \bibinfo {author} {\bibfnamefont {L.}~\bibnamefont {Vanderstraeten}}, \bibinfo {author} {\bibfnamefont {N.}~\bibnamefont {Schuch}}, \bibinfo {author} {\bibfnamefont {H.-Y.}\ \bibnamefont {Lee}}, \bibinfo {author} {\bibfnamefont {N.}~\bibnamefont {Kawashima}},\ and\ \bibinfo {author} {\bibfnamefont {J.-Y.}\ \bibnamefont {Chen}},\ }\bibfield  {title} {\bibinfo {title} {Generating {{Function}} for {{Projected Entangled-Pair States}}},\ }\href {https://doi.org/10.1103/PRXQuantum.5.010335} {\bibfield  {journal} {\bibinfo  {journal} {PRX Quantum}\ }\textbf {\bibinfo {volume} {5}},\ \bibinfo {pages} {010335} (\bibinfo {year} {2024})}\BibitemShut {NoStop}%
  \bibitem [{\citenamefont {Fransson}\ \emph {et~al.}(2016)\citenamefont {Fransson}, \citenamefont {Black-Schaffer},\ and\ \citenamefont {Balatsky}}]{Fransson2016}%
    \BibitemOpen
    \bibfield  {author} {\bibinfo {author} {\bibfnamefont {J.}~\bibnamefont {Fransson}}, \bibinfo {author} {\bibfnamefont {A.~M.}\ \bibnamefont {Black-Schaffer}},\ and\ \bibinfo {author} {\bibfnamefont {A.~V.}\ \bibnamefont {Balatsky}},\ }\bibfield  {title} {\bibinfo {title} {{Magnon Dirac materials}},\ }\href {https://doi.org/10.1103/PhysRevB.94.075401} {\bibfield  {journal} {\bibinfo  {journal} {Phys. Rev. B}\ }\textbf {\bibinfo {volume} {94}},\ \bibinfo {pages} {075401} (\bibinfo {year} {2016})}\BibitemShut {NoStop}%
  \bibitem [{\citenamefont {Han}\ \emph {et~al.}(2012)\citenamefont {Han}, \citenamefont {Helton}, \citenamefont {Chu}, \citenamefont {Nocera}, \citenamefont {Rodriguez-Rivera}, \citenamefont {Broholm},\ and\ \citenamefont {Lee}}]{Han2012}%
    \BibitemOpen
    \bibfield  {author} {\bibinfo {author} {\bibfnamefont {T.-H.}\ \bibnamefont {Han}}, \bibinfo {author} {\bibfnamefont {J.~S.}\ \bibnamefont {Helton}}, \bibinfo {author} {\bibfnamefont {S.}~\bibnamefont {Chu}}, \bibinfo {author} {\bibfnamefont {D.~G.}\ \bibnamefont {Nocera}}, \bibinfo {author} {\bibfnamefont {J.~A.}\ \bibnamefont {Rodriguez-Rivera}}, \bibinfo {author} {\bibfnamefont {C.}~\bibnamefont {Broholm}},\ and\ \bibinfo {author} {\bibfnamefont {Y.~S.}\ \bibnamefont {Lee}},\ }\bibfield  {title} {\bibinfo {title} {Fractionalized excitations in the spin-liquid state of a kagome-lattice antiferromagnet},\ }\href {https://doi.org/10.1038/nature11659} {\bibfield  {journal} {\bibinfo  {journal} {Nature (London)}\ }\textbf {\bibinfo {volume} {492}},\ \bibinfo {pages} {406} (\bibinfo {year} {2012})}\BibitemShut {NoStop}%
  \bibitem [{\citenamefont {Breidenbach}\ \emph {et~al.}(2025)\citenamefont {Breidenbach}, \citenamefont {Campello}, \citenamefont {Wen}, \citenamefont {Jiang}, \citenamefont {Pajerowski}, \citenamefont {Smaha},\ and\ \citenamefont {Lee}}]{breidenbach2025}%
    \BibitemOpen
    \bibfield  {author} {\bibinfo {author} {\bibfnamefont {A.~T.}\ \bibnamefont {Breidenbach}}, \bibinfo {author} {\bibfnamefont {A.~C.}\ \bibnamefont {Campello}}, \bibinfo {author} {\bibfnamefont {J.}~\bibnamefont {Wen}}, \bibinfo {author} {\bibfnamefont {H.-C.}\ \bibnamefont {Jiang}}, \bibinfo {author} {\bibfnamefont {D.~M.}\ \bibnamefont {Pajerowski}}, \bibinfo {author} {\bibfnamefont {R.~W.}\ \bibnamefont {Smaha}},\ and\ \bibinfo {author} {\bibfnamefont {Y.~S.}\ \bibnamefont {Lee}},\ }\bibfield  {title} {\bibinfo {title} {Identifying universal spin excitations in candidate spin-1/2 kagome quantum spin liquid materials},\ }\href {https://doi.org/10.1038/s41567-025-03069-3} {\bibfield  {journal} {\bibinfo  {journal} {Nat. Phys.}\ }\textbf {\bibinfo {volume} {21}},\ \bibinfo {pages} {1957} (\bibinfo {year} {2025})}\BibitemShut {NoStop}%
  \bibitem [{\citenamefont {Vanderstraeten}\ \emph {et~al.}(2019)\citenamefont {Vanderstraeten}, \citenamefont {Haegeman},\ and\ \citenamefont {Verstraete}}]{Vanderstraeten2019}%
    \BibitemOpen
    \bibfield  {author} {\bibinfo {author} {\bibfnamefont {L.}~\bibnamefont {Vanderstraeten}}, \bibinfo {author} {\bibfnamefont {J.}~\bibnamefont {Haegeman}},\ and\ \bibinfo {author} {\bibfnamefont {F.}~\bibnamefont {Verstraete}},\ }\bibfield  {title} {\bibinfo {title} {{Simulating excitation spectra with projected entangled-pair states}},\ }\href {https://doi.org/10.1103/physrevb.99.165121} {\bibfield  {journal} {\bibinfo  {journal} {Phys. Rev. B}\ }\textbf {\bibinfo {volume} {99}},\ \bibinfo {pages} {165121} (\bibinfo {year} {2019})}\BibitemShut {NoStop}%
  \bibitem [{\citenamefont {Ponsioen}\ and\ \citenamefont {Corboz}(2020)}]{Ponsioen2020}%
    \BibitemOpen
    \bibfield  {author} {\bibinfo {author} {\bibfnamefont {B.}~\bibnamefont {Ponsioen}}\ and\ \bibinfo {author} {\bibfnamefont {P.}~\bibnamefont {Corboz}},\ }\bibfield  {title} {\bibinfo {title} {{Excitations with projected entangled pair states using the corner transfer matrix method}},\ }\href {https://doi.org/10.1103/physrevb.101.195109} {\bibfield  {journal} {\bibinfo  {journal} {Phys. Rev. B}\ }\textbf {\bibinfo {volume} {101}},\ \bibinfo {pages} {195109} (\bibinfo {year} {2020})}\BibitemShut {NoStop}%
  \bibitem [{\citenamefont {Chi}\ \emph {et~al.}(2024)\citenamefont {Chi}, \citenamefont {Hu}, \citenamefont {Liao},\ and\ \citenamefont {Xiang}}]{chi2024}%
    \BibitemOpen
    \bibfield  {author} {\bibinfo {author} {\bibfnamefont {R.}~\bibnamefont {Chi}}, \bibinfo {author} {\bibfnamefont {J.}~\bibnamefont {Hu}}, \bibinfo {author} {\bibfnamefont {H.-J.}\ \bibnamefont {Liao}},\ and\ \bibinfo {author} {\bibfnamefont {T.}~\bibnamefont {Xiang}},\ }\bibfield  {title} {\bibinfo {title} {Dynamical spectra of spin supersolid states in triangular antiferromagnets},\ }\href {https://doi.org/10.1103/PhysRevB.110.L180404} {\bibfield  {journal} {\bibinfo  {journal} {Phys. Rev. B}\ }\textbf {\bibinfo {volume} {110}},\ \bibinfo {pages} {L180404} (\bibinfo {year} {2024})}\BibitemShut {NoStop}%
  \bibitem [{\citenamefont {Xu}\ \emph {et~al.}(2025)\citenamefont {Xu}, \citenamefont {Hasik}, \citenamefont {Ponsioen},\ and\ \citenamefont {Nevidomskyy}}]{Xu2024}%
    \BibitemOpen
    \bibfield  {author} {\bibinfo {author} {\bibfnamefont {Y.}~\bibnamefont {Xu}}, \bibinfo {author} {\bibfnamefont {J.}~\bibnamefont {Hasik}}, \bibinfo {author} {\bibfnamefont {B.}~\bibnamefont {Ponsioen}},\ and\ \bibinfo {author} {\bibfnamefont {A.~H.}\ \bibnamefont {Nevidomskyy}},\ }\bibfield  {title} {\bibinfo {title} {Simulating spin dynamics of supersolid states in a quantum {Ising} magnet},\ }\href {https://doi.org/10.1103/PhysRevB.111.L060402} {\bibfield  {journal} {\bibinfo  {journal} {Phys. Rev. B}\ }\textbf {\bibinfo {volume} {111}},\ \bibinfo {pages} {L060402} (\bibinfo {year} {2025})}\BibitemShut {NoStop}%
  \bibitem [{\citenamefont {Kitaev}(2003)}]{Kitaev2003}%
    \BibitemOpen
    \bibfield  {author} {\bibinfo {author} {\bibfnamefont {A.}~\bibnamefont {Kitaev}},\ }\bibfield  {title} {\bibinfo {title} {Fault-tolerant quantum computation by anyons},\ }\href {https://doi.org/https://doi.org/10.1016/S0003-4916(02)00018-0} {\bibfield  {journal} {\bibinfo  {journal} {Ann. Phys. (Amsterdam)}\ }\textbf {\bibinfo {volume} {303}},\ \bibinfo {pages} {2} (\bibinfo {year} {2003})}\BibitemShut {NoStop}%
  \bibitem [{\citenamefont {Lu}\ \emph {et~al.}(2011)\citenamefont {Lu}, \citenamefont {Ran},\ and\ \citenamefont {Lee}}]{Lu2011}%
    \BibitemOpen
    \bibfield  {author} {\bibinfo {author} {\bibfnamefont {Y.-M.}\ \bibnamefont {Lu}}, \bibinfo {author} {\bibfnamefont {Y.}~\bibnamefont {Ran}},\ and\ \bibinfo {author} {\bibfnamefont {P.~A.}\ \bibnamefont {Lee}},\ }\bibfield  {title} {\bibinfo {title} {{$\mathbb{Z}_2$ spin liquids in the $S = 1/2$ Heisenberg model on the kagome lattice: A projective symmetry-group study of Schwinger fermion mean-field states}},\ }\href {https://doi.org/10.1103/PhysRevB.83.224413} {\bibfield  {journal} {\bibinfo  {journal} {Phys. Rev. B}\ }\textbf {\bibinfo {volume} {83}},\ \bibinfo {pages} {224413} (\bibinfo {year} {2011})}\BibitemShut {NoStop}%
  \bibitem [{\citenamefont {Hermele}\ \emph {et~al.}(2008)\citenamefont {Hermele}, \citenamefont {Ran}, \citenamefont {Lee},\ and\ \citenamefont {Wen}}]{Hermele2008}%
    \BibitemOpen
    \bibfield  {author} {\bibinfo {author} {\bibfnamefont {M.}~\bibnamefont {Hermele}}, \bibinfo {author} {\bibfnamefont {Y.}~\bibnamefont {Ran}}, \bibinfo {author} {\bibfnamefont {P.~A.}\ \bibnamefont {Lee}},\ and\ \bibinfo {author} {\bibfnamefont {X.-G.}\ \bibnamefont {Wen}},\ }\bibfield  {title} {\bibinfo {title} {Properties of an algebraic spin liquid on the kagome lattice},\ }\href {https://doi.org/10.1103/PhysRevB.77.224413} {\bibfield  {journal} {\bibinfo  {journal} {Phys. Rev. B}\ }\textbf {\bibinfo {volume} {77}},\ \bibinfo {pages} {224413} (\bibinfo {year} {2008})}\BibitemShut {NoStop}%
  \bibitem [{\citenamefont {Song}\ \emph {et~al.}(2019)\citenamefont {Song}, \citenamefont {Wang}, \citenamefont {Vishwanath},\ and\ \citenamefont {He}}]{Song2019}%
    \BibitemOpen
    \bibfield  {author} {\bibinfo {author} {\bibfnamefont {X.-Y.}\ \bibnamefont {Song}}, \bibinfo {author} {\bibfnamefont {C.}~\bibnamefont {Wang}}, \bibinfo {author} {\bibfnamefont {A.}~\bibnamefont {Vishwanath}},\ and\ \bibinfo {author} {\bibfnamefont {Y.-C.}\ \bibnamefont {He}},\ }\bibfield  {title} {\bibinfo {title} {Unifying description of competing orders in two-dimensional quantum magnets},\ }\href {https://doi.org/10.1038/s41467-019-11727-3} {\bibfield  {journal} {\bibinfo  {journal} {Nat. Commun.}\ }\textbf {\bibinfo {volume} {10}},\ \bibinfo {pages} {4254} (\bibinfo {year} {2019})}\BibitemShut {NoStop}%
  \bibitem [{\citenamefont {Hu}(2026)}]{Hu2026data}%
    \BibitemOpen
    \bibfield  {author} {\bibinfo {author} {\bibfnamefont {J.}~\bibnamefont {Hu}},\ }\bibfield  {title} {\bibinfo {title} {Dynamical spectral function of the Kagome quantum spin liquid},\ }\href {https://doi.org/10.5281/zenodo.21319354} {\bibinfo {howpublished} {Zenodo, 10.5281/zenodo.21319354}} (\bibinfo {year} {2026})\BibitemShut {NoStop}%
  \bibitem [{\citenamefont {Shores}\ \emph {et~al.}(2005)\citenamefont {Shores}, \citenamefont {Nytko}, \citenamefont {Bartlett},\ and\ \citenamefont {Nocera}}]{shores2005}%
    \BibitemOpen
    \bibfield  {author} {\bibinfo {author} {\bibfnamefont {M.~P.}\ \bibnamefont {Shores}}, \bibinfo {author} {\bibfnamefont {E.~A.}\ \bibnamefont {Nytko}}, \bibinfo {author} {\bibfnamefont {B.~M.}\ \bibnamefont {Bartlett}},\ and\ \bibinfo {author} {\bibfnamefont {D.~G.}\ \bibnamefont {Nocera}},\ }\bibfield  {title} {\bibinfo {title} {A {{Structurally Perfect}} $s = 1/2$ {{Kagomé Antiferromagnet}}},\ }\href {https://doi.org/10.1021/ja053891p} {\bibfield  {journal} {\bibinfo  {journal} {J. Am. Chem. Soc.}\ }\textbf {\bibinfo {volume} {127}},\ \bibinfo {pages} {13462} (\bibinfo {year} {2005})}\BibitemShut {NoStop}%
  \bibitem [{\citenamefont {Feng}\ \emph {et~al.}(2017)\citenamefont {Feng}, \citenamefont {Li}, \citenamefont {Meng}, \citenamefont {Yi}, \citenamefont {Wei}, \citenamefont {Zhang}, \citenamefont {Wang}, \citenamefont {Jiang}, \citenamefont {Liu}, \citenamefont {Li}, \citenamefont {Liu}, \citenamefont {Luo}, \citenamefont {Li}, \citenamefont {Zheng}, \citenamefont {Meng}, \citenamefont {Mei},\ and\ \citenamefont {Shi}}]{feng2017}%
    \BibitemOpen
    \bibfield  {author} {\bibinfo {author} {\bibfnamefont {Z.}~\bibnamefont {Feng}}, \bibinfo {author} {\bibfnamefont {Z.}~\bibnamefont {Li}}, \bibinfo {author} {\bibfnamefont {X.}~\bibnamefont {Meng}}, \bibinfo {author} {\bibfnamefont {W.}~\bibnamefont {Yi}}, \bibinfo {author} {\bibfnamefont {Y.}~\bibnamefont {Wei}}, \bibinfo {author} {\bibfnamefont {J.}~\bibnamefont {Zhang}}, \bibinfo {author} {\bibfnamefont {Y.-C.}\ \bibnamefont {Wang}}, \bibinfo {author} {\bibfnamefont {W.}~\bibnamefont {Jiang}}, \bibinfo {author} {\bibfnamefont {Z.}~\bibnamefont {Liu}}, \bibinfo {author} {\bibfnamefont {S.}~\bibnamefont {Li}}, \bibinfo {author} {\bibfnamefont {F.}~\bibnamefont {Liu}}, \bibinfo {author} {\bibfnamefont {J.}~\bibnamefont {Luo}}, \bibinfo {author} {\bibfnamefont {S.}~\bibnamefont {Li}}, \bibinfo {author} {\bibfnamefont {G.-q.}\ \bibnamefont {Zheng}}, \bibinfo {author} {\bibfnamefont {Z.~Y.}\ \bibnamefont {Meng}}, \bibinfo {author} {\bibfnamefont {J.-W.}\ \bibnamefont {Mei}},\ and\ \bibinfo {author} {\bibfnamefont {Y.}~\bibnamefont {Shi}},\ }\bibfield  {title} {\bibinfo {title} {Gapped spin-1/2 spinon excitations in a new kagome quantum spin liquid compound {{Cu}}$_{3}${{Zn}}({{OH}})$_{6}${{FBr}}},\ }\href {https://doi.org/10.1088/0256-307X/34/7/077502} {\bibfield  {journal} {\bibinfo  {journal} {Chin. Phys. Lett.}\ }\textbf {\bibinfo {volume} {34}},\ \bibinfo {pages} {077502} (\bibinfo {year} {2017})}\BibitemShut {NoStop}%
  \bibitem [{\citenamefont {Li}\ \emph {et~al.}(2014)\citenamefont {Li}, \citenamefont {Pan}, \citenamefont {Li}, \citenamefont {Tong}, \citenamefont {Ling}, \citenamefont {Yang}, \citenamefont {Wang}, \citenamefont {Chen}, \citenamefont {Wu},\ and\ \citenamefont {Zhang}}]{li2014}%
    \BibitemOpen
    \bibfield  {author} {\bibinfo {author} {\bibfnamefont {Y.}~\bibnamefont {Li}}, \bibinfo {author} {\bibfnamefont {B.}~\bibnamefont {Pan}}, \bibinfo {author} {\bibfnamefont {S.}~\bibnamefont {Li}}, \bibinfo {author} {\bibfnamefont {W.}~\bibnamefont {Tong}}, \bibinfo {author} {\bibfnamefont {L.}~\bibnamefont {Ling}}, \bibinfo {author} {\bibfnamefont {Z.}~\bibnamefont {Yang}}, \bibinfo {author} {\bibfnamefont {J.}~\bibnamefont {Wang}}, \bibinfo {author} {\bibfnamefont {Z.}~\bibnamefont {Chen}}, \bibinfo {author} {\bibfnamefont {Z.}~\bibnamefont {Wu}},\ and\ \bibinfo {author} {\bibfnamefont {Q.}~\bibnamefont {Zhang}},\ }\bibfield  {title} {\bibinfo {title} {Gapless quantum spin liquid in the $s = 1/2$ anisotropic kagome antiferromagnet {{ZnCu}}$_3$({{OH}})$_6${{SO}}$_4$},\ }\href {https://doi.org/10.1088/1367-2630/16/9/093011} {\bibfield  {journal} {\bibinfo  {journal} {New J. Phys.}\ }\textbf {\bibinfo {volume} {16}},\ \bibinfo {pages} {093011} (\bibinfo {year} {2014})}\BibitemShut {NoStop}%
  \bibitem [{\citenamefont {Liu}\ \emph {et~al.}(2022)\citenamefont {Liu}, \citenamefont {Yuan}, \citenamefont {Li}, \citenamefont {Li}, \citenamefont {Zhao}, \citenamefont {Liao},\ and\ \citenamefont {Li}}]{Liu2022}%
    \BibitemOpen
    \bibfield  {author} {\bibinfo {author} {\bibfnamefont {J.}~\bibnamefont {Liu}}, \bibinfo {author} {\bibfnamefont {L.}~\bibnamefont {Yuan}}, \bibinfo {author} {\bibfnamefont {X.}~\bibnamefont {Li}}, \bibinfo {author} {\bibfnamefont {B.}~\bibnamefont {Li}}, \bibinfo {author} {\bibfnamefont {K.}~\bibnamefont {Zhao}}, \bibinfo {author} {\bibfnamefont {H.}~\bibnamefont {Liao}},\ and\ \bibinfo {author} {\bibfnamefont {Y.}~\bibnamefont {Li}},\ }\bibfield  {title} {\bibinfo {title} {{Gapless spin liquid behavior in a kagome Heisenberg antiferromagnet with randomly distributed hexagons of alternate bonds}},\ }\href {https://doi.org/10.1103/PhysRevB.105.024418} {\bibfield  {journal} {\bibinfo  {journal} {Phys. Rev. B}\ }\textbf {\bibinfo {volume} {105}},\ \bibinfo {pages} {024418} (\bibinfo {year} {2022})}\BibitemShut {NoStop}%
  \bibitem [{\citenamefont {Zeng}\ \emph {et~al.}(2022)\citenamefont {Zeng}, \citenamefont {Ma}, \citenamefont {Wu}, \citenamefont {Li}, \citenamefont {Tao}, \citenamefont {Lu}, \citenamefont {Chen}, \citenamefont {Mi}, \citenamefont {Song}, \citenamefont {Cao}, \citenamefont {Che}, \citenamefont {Li}, \citenamefont {Li}, \citenamefont {Luo}, \citenamefont {Meng},\ and\ \citenamefont {Li}}]{zeng2022}%
    \BibitemOpen
    \bibfield  {author} {\bibinfo {author} {\bibfnamefont {Z.}~\bibnamefont {Zeng}}, \bibinfo {author} {\bibfnamefont {X.}~\bibnamefont {Ma}}, \bibinfo {author} {\bibfnamefont {S.}~\bibnamefont {Wu}}, \bibinfo {author} {\bibfnamefont {H.-F.}\ \bibnamefont {Li}}, \bibinfo {author} {\bibfnamefont {Z.}~\bibnamefont {Tao}}, \bibinfo {author} {\bibfnamefont {X.}~\bibnamefont {Lu}}, \bibinfo {author} {\bibfnamefont {X.-h.}\ \bibnamefont {Chen}}, \bibinfo {author} {\bibfnamefont {J.-X.}\ \bibnamefont {Mi}}, \bibinfo {author} {\bibfnamefont {S.-J.}\ \bibnamefont {Song}}, \bibinfo {author} {\bibfnamefont {G.-H.}\ \bibnamefont {Cao}}, \bibinfo {author} {\bibfnamefont {G.}~\bibnamefont {Che}}, \bibinfo {author} {\bibfnamefont {K.}~\bibnamefont {Li}}, \bibinfo {author} {\bibfnamefont {G.}~\bibnamefont {Li}}, \bibinfo {author} {\bibfnamefont {H.}~\bibnamefont {Luo}}, \bibinfo {author} {\bibfnamefont {Z.~Y.}\ \bibnamefont {Meng}},\ and\ \bibinfo {author} {\bibfnamefont {S.}~\bibnamefont {Li}},\ }\bibfield  {title} {\bibinfo {title} {Possible {{Dirac}} quantum spin liquid in the kagome quantum antiferromagnet {{YCu}}$_3$({{OH}})$_6${{Br}}$_2$[{{Br}}$_x$({{OH}})$_{1 - x}$]},\ }\href {https://doi.org/10.1103/PhysRevB.105.L121109} {\bibfield  {journal} {\bibinfo  {journal} {Phys. Rev. B}\ }\textbf {\bibinfo {volume} {105}},\ \bibinfo {pages} {L121109} (\bibinfo {year} {2022})}\BibitemShut {NoStop}%
  \bibitem [{\citenamefont {Nilsen}\ \emph {et~al.}(2013)\citenamefont {Nilsen}, \citenamefont {de.~Vries}, \citenamefont {Stewart}, \citenamefont {Harrison},\ and\ \citenamefont {Rønnow}}]{nilsen2013}%
    \BibitemOpen
    \bibfield  {author} {\bibinfo {author} {\bibfnamefont {G.~J.}\ \bibnamefont {Nilsen}}, \bibinfo {author} {\bibfnamefont {M.~A.}\ \bibnamefont {de.~Vries}}, \bibinfo {author} {\bibfnamefont {J.~R.}\ \bibnamefont {Stewart}}, \bibinfo {author} {\bibfnamefont {A.}~\bibnamefont {Harrison}},\ and\ \bibinfo {author} {\bibfnamefont {H.~M.}\ \bibnamefont {Rønnow}},\ }\bibfield  {title} {\bibinfo {title} {Low-energy spin dynamics of the $s = 1/2$ kagome system herbertsmithite},\ }\href {https://iopscience.iop.org/article/10.1088/0953-8984/25/10/106001} {\bibfield  {journal} {\bibinfo  {journal} {J. Phys. Condens. Matter}\ }\textbf {\bibinfo {volume} {25}},\ \bibinfo {pages} {106001} (\bibinfo {year} {2013})}\BibitemShut {NoStop}%
  \bibitem [{\citenamefont {Zeng}\ \emph {et~al.}(2024)\citenamefont {Zeng}, \citenamefont {Zhou}, \citenamefont {Zhou}, \citenamefont {Han}, \citenamefont {Chi}, \citenamefont {Li}, \citenamefont {Kofu}, \citenamefont {Nakajima}, \citenamefont {Wei}, \citenamefont {Zhang}, \citenamefont {Mazzone}, \citenamefont {Meng},\ and\ \citenamefont {Li}}]{Zeng2024}%
    \BibitemOpen
    \bibfield  {author} {\bibinfo {author} {\bibfnamefont {Z.}~\bibnamefont {Zeng}}, \bibinfo {author} {\bibfnamefont {C.}~\bibnamefont {Zhou}}, \bibinfo {author} {\bibfnamefont {H.}~\bibnamefont {Zhou}}, \bibinfo {author} {\bibfnamefont {L.}~\bibnamefont {Han}}, \bibinfo {author} {\bibfnamefont {R.}~\bibnamefont {Chi}}, \bibinfo {author} {\bibfnamefont {K.}~\bibnamefont {Li}}, \bibinfo {author} {\bibfnamefont {M.}~\bibnamefont {Kofu}}, \bibinfo {author} {\bibfnamefont {K.}~\bibnamefont {Nakajima}}, \bibinfo {author} {\bibfnamefont {Y.}~\bibnamefont {Wei}}, \bibinfo {author} {\bibfnamefont {W.}~\bibnamefont {Zhang}}, \bibinfo {author} {\bibfnamefont {D.~G.}\ \bibnamefont {Mazzone}}, \bibinfo {author} {\bibfnamefont {Z.~Y.}\ \bibnamefont {Meng}},\ and\ \bibinfo {author} {\bibfnamefont {S.}~\bibnamefont {Li}},\ }\bibfield  {title} {\bibinfo {title} {Spectral evidence for {{Dirac}} spinons in a kagome lattice antiferromagnet},\ }\href {https://doi.org/10.1038/s41567-024-02495-z} {\bibfield  {journal} {\bibinfo  {journal} {Nat. Phys.}\ }\textbf {\bibinfo {volume} {20}},\ \bibinfo {pages} {1097} (\bibinfo {year} {2024})}\BibitemShut {NoStop}%
  \bibitem [{\citenamefont {Zorko}\ \emph {et~al.}(2008)\citenamefont {Zorko}, \citenamefont {Nellutla}, \citenamefont {van~Tol}, \citenamefont {Brunel}, \citenamefont {Bert}, \citenamefont {Duc}, \citenamefont {Trombe}, \citenamefont {De~Vries}, \citenamefont {Harrison},\ and\ \citenamefont {Mendels}}]{zorko2008}%
    \BibitemOpen
    \bibfield  {author} {\bibinfo {author} {\bibfnamefont {A.}~\bibnamefont {Zorko}}, \bibinfo {author} {\bibfnamefont {S.}~\bibnamefont {Nellutla}}, \bibinfo {author} {\bibfnamefont {J.}~\bibnamefont {van~Tol}}, \bibinfo {author} {\bibfnamefont {L.~C.}\ \bibnamefont {Brunel}}, \bibinfo {author} {\bibfnamefont {F.}~\bibnamefont {Bert}}, \bibinfo {author} {\bibfnamefont {F.}~\bibnamefont {Duc}}, \bibinfo {author} {\bibfnamefont {J.-C.}\ \bibnamefont {Trombe}}, \bibinfo {author} {\bibfnamefont {M.~A.}\ \bibnamefont {de~Vries}}, \bibinfo {author} {\bibfnamefont {A.}~\bibnamefont {Harrison}},\ and\ \bibinfo {author} {\bibfnamefont {P.}~\bibnamefont {Mendels}},\ }\bibfield  {title} {\bibinfo {title} {{Dzyaloshinsky-{{Moriya}} Anisotropy in the Spin-1/2 {{Kagome}} Compound {{ZnCu}}$_3$({{OH}})$_6${{Cl}}$_2$}},\ }\href {https://doi.org/10.1103/PhysRevLett.101.026405} {\bibfield  {journal} {\bibinfo  {journal} {Phys. Rev. Lett.}\ }\textbf {\bibinfo {volume} {101}},\ \bibinfo {pages} {026405} (\bibinfo {year} {2008})}\BibitemShut {NoStop}%
  \bibitem [{\citenamefont {Jeschke}(2013)}]{jeschke2013}%
    \BibitemOpen
    \bibfield  {author} {\bibinfo {author} {\bibfnamefont {H.~O.}\ \bibnamefont {Jeschke}},\ \bibinfo {author} {\bibfnamefont {F.}~\bibnamefont {Salvat-Pujol}}\ and\ \bibinfo {author} {\bibfnamefont {R.}~\bibnamefont {Valent\'{\i}}},\ }
    \bibfield  {title} {\bibinfo {title} {First-principles determination of {{Heisenberg Hamiltonian}} parameters for the spin-1/2 kagome antiferromagnet {{ZnCu}}$_3$({{OH}})$_6${{Cl}}$_2$},\ }\href {https://doi.org/10.1103/PhysRevB.88.075106} {\bibfield  {journal} {\bibinfo  {journal} {Phys. Rev. B}\ }\textbf {\bibinfo {volume} {88}},\ \bibinfo {pages} {075106} (\bibinfo {year} {2013})}\BibitemShut {NoStop}%
  \bibitem [{\citenamefont {Rousochatzakis}\ \emph {et~al.}(2009)\citenamefont {Rousochatzakis}, \citenamefont {Manmana}, \citenamefont {L\"auchli}, \citenamefont {Normand},\ and\ \citenamefont {Mila}}]{Rousochatzakis2009}%
    \BibitemOpen
    \bibfield  {author} {\bibinfo {author} {\bibfnamefont {I.}~\bibnamefont {Rousochatzakis}}, \bibinfo {author} {\bibfnamefont {S.~R.}\ \bibnamefont {Manmana}}, \bibinfo {author} {\bibfnamefont {A.~M.}\ \bibnamefont {L\"auchli}}, \bibinfo {author} {\bibfnamefont {B.}~\bibnamefont {Normand}},\ and\ \bibinfo {author} {\bibfnamefont {F.}~\bibnamefont {Mila}},\ }\bibfield  {title} {\bibinfo {title} {{Dzyaloshinskii-Moriya anisotropy and nonmagnetic impurities in the $s = \frac12$ kagome system ZnCu$_3$(OH)$_6$Cl$_2$}},\ }\href {https://doi.org/10.1103/PhysRevB.79.214415} {\bibfield  {journal} {\bibinfo  {journal} {Phys. Rev. B}\ }\textbf {\bibinfo {volume} {79}},\ \bibinfo {pages} {214415} (\bibinfo {year} {2009})}\BibitemShut {NoStop}%
  \end{thebibliography}
\end{document}